\documentclass[12pt,letterpaper]{article}
\usepackage{jheppub}

\usepackage{graphicx}
\usepackage{bbm}
\usepackage{amsmath}
\usepackage{amssymb}
\usepackage{mathtools}
\usepackage{amsthm}
\usepackage{marvosym}
\usepackage{dsfont}
\usepackage[labelformat=simple]{subcaption}
\usepackage{enumitem} 
\usepackage[dvipsnames]{xcolor} 
\usepackage{rotating}
\allowdisplaybreaks
\usepackage{empheq}

  \definecolor{dark-gray}{gray}{0.20}
  \definecolor{gray}{gray}{0.30}
  \definecolor{light-gray}{gray}{0.80}
  \definecolor{dark-red}{rgb}{0.7,0,0}
  \definecolor{dark-green}{rgb}{0.1,0.4,0}
  \definecolor{dark-blue}{rgb}{0.3,0.3,0.7}
  \definecolor{light-blue}{rgb}{0.8,0.8,1}
      \definecolor{swamp}{RGB}{240, 199, 197}

\newcommand{\be}{\begin{equation}}
\newcommand{\ee}{\end{equation}}

\def\be{\begin{equation}}
\def\ee{\end{equation}}
\def\bea{\begin{eqnarray}}
\def\eea{\end{eqnarray}}

\newcommand{\beq}{\begin{equation}}  \newcommand{\eeq}{\end{equation}}
\newcommand{\bal}{\begin{aligned}}   \newcommand{\eal}{\end{aligned}}
\def\beqa{\begin{eqnarray}}
\def\eeqa{\end{eqnarray}}

\newcommand{\rmd}{\mathrm{d}}

\DeclareMathOperator{\cone}{cone}
\DeclareMathOperator{\polar}{polar}

\numberwithin{equation}{section}

\def\simleq{\; \raise0.3ex\hbox{$<$\kern-0.75em
      \raise-1.1ex\hbox{$\sim$}}\; }
   \def\simgeq{\; \raise0.3ex\hbox{$>$\kern-0.75em
      \raise-1.1ex\hbox{$\sim$}}\; }

\numberwithin{equation}{section}

\hypersetup{
	colorlinks=true,
	linkcolor=dark-blue,
	citecolor=dark-red,
	urlcolor=dark-green,
	linktoc=page
}

\theoremstyle{remark}

\newtheoremstyle{named}{}{}{\itshape}{}{\bfseries}{.}{.5em}{#3}
\theoremstyle{named}

\newtheorem{Definition}{Definition}

\title{\centering Taxonomy of Infinite Distance Limits}

\author{Muldrow Etheredge$^{1,2}$,}\author{Ben Heidenreich$^{1}$,}\author{Tom Rudelius$^{3}$,} \author{Ignacio Ruiz$^{4}$,} 
\author{Irene Valenzuela$^{4,5}$} 

\affiliation{$^{1}$Department of Physics, University of Massachusetts, Amherst, MA 01003, USA}

\affiliation{$^{2}$Kavli Institute for Theoretical Physics, University of California, Santa Barbara, CA 93106, USA}

\affiliation{$^{3}$Department of Mathematical Sciences, Durham University, Durham DH1 3LE, UK

}
\affiliation{$^{4}$Instituto de F\'isica Te\'orica UAM-CSIC and Departamento de F\'isica Te\'orica, Universidad Aut\'onoma de Madrid, Cantoblanco, 28049 Madrid, Spain}
\affiliation{$^{5}$CERN, Theoretical Physics Department, 1211 Meyrin, Switzerland}

\emailAdd{metheredge@umass.edu}
\emailAdd{bheidenreich@umass.edu}
\emailAdd{thomas.w.rudelius@durham.ac.uk}
\emailAdd{ignacio.ruiz@uam.es}
\emailAdd{irene.valenzuela@cern.ch}

\abstract{
The Emergent String Conjecture constrains the possible types of light towers in infinite-distance limits in quantum gravity moduli spaces. In this paper, we use these constraints to restrict the geometry of the scalar charge-to-mass vectors $(-\vec{\nabla}\log m)$ of the light towers and the analogous vector $(-\vec{\nabla}\log \Lambda_{\text{QG}})$ of the species scale. 
We derive taxonomic rules  that these vectors must satisfy in each duality frame.
Under certain assumptions, this allows us to classify the ways in which different duality frames can fit together globally  in the moduli space in terms of a finite list of polytopes. Many of these polytopes arise in known string theory compactifications, while others  suggest either undiscovered corners of the landscape or new  swampland constraints.
}

\begin{document}
\hypersetup{pageanchor=false}
\makeatletter
\let\old@fpheader\@fpheader
\renewcommand{\@fpheader}{\old@fpheader \vspace*{-0.1cm} 
\hfill ACFI-T24-04 \\ \vspace*{-0.1cm} 
\hfill CERN-TH-2024-067\\ \vspace*{-0.1cm} \hfill {IFT-UAM/CSIC-23-64}}
\maketitle
%{\small\tableofcontents}
\makeatother

%\setcounter{tocdepth}{2}

%\bigskip
%\pagebreak
\pagenumbering{roman}
\setcounter{page}{1}
\newpage
\section{Introduction}
\label{s.intro}
\pagenumbering{arabic}
\setcounter{page}{1}

In the realm of string theory and its low-energy effective field theory (EFT) descriptions, the values of all continuous parameters are determined by the vacuum expectation values of scalar fields, referred to as moduli when they are massless. In this context, perturbative regimes in the EFT correspond to infinite-distance limits in field space. Such limits have been studied extensively in recent years, and many of their features are now well understood.

Meanwhile, relatively little is known about the global properties of scalar field spaces in quantum gravity, largely due to the computational difficulties associated with strong coupling outside of asymptotic regimes.
However, asymptotic properties can sometimes provide information about global features of moduli spaces. In this paper, we will show how the microscopic nature of infinite-distance limits dictates how these different limits fit together in moduli space, and we will show how this constrains the different possible perturbative descriptions of a given theory, commonly known as duality frames. 

Central to this analysis are scalar charge-to-mass ratios, or ``$\zeta$-vectors," which are defined locally on moduli spaces. These $\zeta$-vectors encode how masses $m(\phi^i)$ of particles depend on the moduli $\phi^i$, and are defined as
	\begin{align}
		\vec{\zeta}=-\vec\nabla \log \frac{m}{M_\text{Pl,$d$}},
		\label{zetadef}
	\end{align}
	where the gradient is taken with respect to the moduli and $M_\text{Pl,$d$}$ is the Planck mass.
We will refer to $\zeta$-vectors of particle towers as \textbf{tower vectors}. At each point in a moduli space, one can consider the convex hull\footnote{This is done in analogy to the Weak Gravity Conjecture \cite{Palti:2017elp}, as for a scalar $\chi$ with mass $m(\phi)$ one can expand $\mathcal{L}\supseteq m(\phi)^2\chi^2=(m_0^2+2m_0\phi\partial_\phi m(\phi) )\chi^2+...$, with $\mu=\partial_{\phi}m$ measuring the scalar Yukawa charge induced by the moduli $\phi$, thereby the name 'scalar charge-to-mass ratio' \cite{Lee:2018spm, Gonzalo:2019gjp, DallAgata:2020ino,Andriot:2020lea,   Benakli:2020pkm}.} of these tower vectors for all of the particle towers. \emph{A priori}, this convex hull could take any of a wide variety of shapes and sizes: an effective field theorist could write down a set of particles whose masses depend on the moduli of the theory in any way they choose, and thereby generate a convex hull of arbitrary shape. However, as we will see, these convex hulls turn out to be highly constrained in the asymptotic, perturbative regimes of the theory. In particular, they are generated by infinite towers of states that emerge in these limits, and the microscopic nature of these towers fixes the value of $\vec\zeta$.

The existence of these towers of states is dictated by the Distance Conjecture \cite{Ooguri:2006in}, one of the most well-studied hypotheses of the swampland program \cite{Vafa:2005ui,Brennan:2017rbf,Palti:2019pca,vanBeest:2021lhn,Grana:2021zvf,Harlow:2022ich,Agmon:2022thq,VanRiet:2023pnx}. The Distance Conjecture proposes that whenever one travels a large geodesic distance $\Delta$ in the moduli space, one encounters a tower of light particles with exponentially-light characteristic masses $m\sim e^{-\alpha \Delta}$, for some positive constant $\alpha$, as $\Delta\rightarrow \infty$. This conjecture has been examined and verified in many string theory settings (see e.g.\cite{ Baume:2016psm,Klaewer:2016kiy, Blumenhagen:2017cxt, Grimm:2018ohb,Heidenreich:2018kpg, Blumenhagen:2018nts, Grimm:2018cpv, Buratti:2018xjt, Corvilain:2018lgw, Joshi:2019nzi,  Erkinger:2019umg, Marchesano:2019ifh, Font:2019cxq,  Gendler:2020dfp, Lanza:2020qmt, Klaewer:2020lfg, Rudelius:2023mjy,Ooguri:2024ofs,Aoufia:2024awo}), and it is linked to the famous duality web of string/M-theory. For a given infinite-distance geodesic with unit tangent vector $\hat t$, the exponential decay of the mass of the tower is given by $\alpha = \vec{\zeta} \cdot \hat t$. 

\begin{figure}[h]
\begin{center}
\begin{subfigure}{0.5\textwidth}
\center
\includegraphics[width=.9\textwidth]{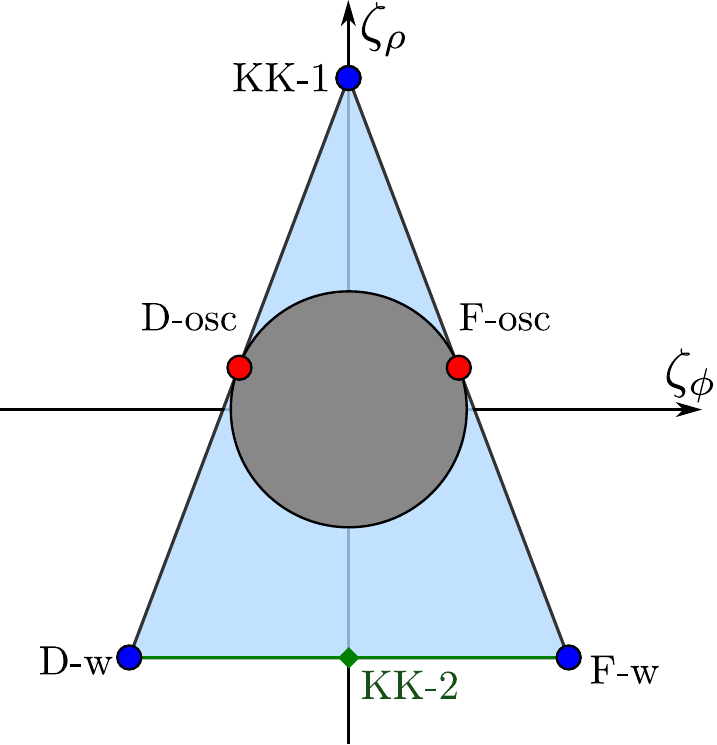}
\caption{Tower polytope example.} \label{f.alphas}
\end{subfigure}
\begin{subfigure}{0.45\textwidth}
\center
\includegraphics[width=.9\textwidth]{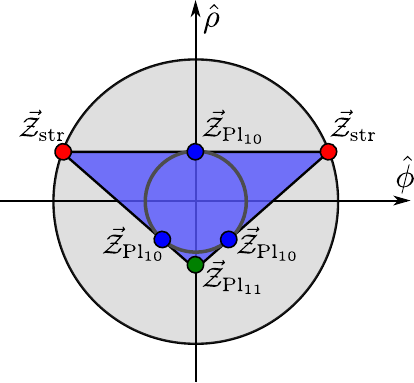}
\caption{Species polytope example} \label{f.alphasS}
\end{subfigure}
\caption{Example of a tower polytope and a species polytope. The tower polytope contains the disk of radius $\frac 1{\sqrt{d-2}}$, where $d$ is the number of spacetime dimensions. The species polytope is enclosed by the disk of radius $\frac 1{\sqrt{d-2}}$ and contains the disk of radius $\frac{1}{\sqrt{(d-1)(d-2)}}$. For $d=9$ this matches precisely with the polytopes obtained from Type IIB string theory compactified on a circle.
\label{f.lambdas}}
\end{center}
\end{figure}

In this work, we will be interested in understanding how different infinite-distance limits, and their associated towers, can be combined globally within a given moduli space. This information is encoded in the aforementioned convex hull of the tower vectors, since the towers generating the convex hull  provide the lightest towers in each of the infinite-distance limits. It has been observed that in some examples, including various 9d settings \cite{Etheredge:2022opl,Etheredge:2023odp,Calderon-Infante:2023ler,Etheredge:2023usk}, the convex hulls of these scalar charge-to-mass ratios for towers of particles are generated by rotations of polytopes, which we call \textbf{tower polytopes} (see Figure \ref{f.alphas}). Such polytopes dictate how the particle towers depend on the moduli and give information about the dualities of the theory. 

In this paper we show that these tower polytopes are tightly constrained by swampland conjectures about the asymptotic limits of moduli space. Under certain assumptions outlined in Section \ref{ss.assumpt}, we obtain a set of rules governing the tower vectors of the light towers in a generic infinite-distance limit, which enables us to derive a finite list of building blocks for the tower polytopes. Each such building block takes the form of a simplex in the scalar charge-to-mass space spanned by the tower vectors, and each such simplex is associated with a particular duality frame of the theory. If further properties of the moduli space are satisfied, we can glue these building blocks together across the different frames of the theory to find a finite list of tower polytopes. Comparing this list with polytopes that are known to arise from string theory compactifications, we reproduce many well-known cases and also obtain some potentially new ones.

The key ingredient for obtaining our taxonomic rules is the Emergent String Conjecture \cite{Lee:2019xtm,Lee:2019wij}, a refinement of the Distance Conjecture that specifies the microscopic nature of the towers of states. In particular, the Emergent String Conjecture holds that infinite-distance limits in the moduli space of a quantum gravity theory are either decompactification limits, in which the infinite tower of states is furnished by Kaluza-Klein modes, or emergent string limits, which feature a unique, emergent, critical, weakly coupled string with a tower of string oscillation modes. While its underlying motivation remains mysterious, the conjecture has been verified in many different flat space string compactifications\footnote{There is evidence, though, that the conjecture should be modified in the case of non-Einstein theories with AdS background to allow also for non-critical strings \cite{Baume:2020dqd,Perlmutter:2020buo, CalderonIV}.} \cite{Lee:2018urn,Lee:2019xtm,  Baume:2019sry, Xu:2020nlh,Lanza:2021udy,Castellano:2023jjt,Rudelius:2023odg}.

\begin{figure}[htp]
\begin{subfigure}{0.22\textwidth}
\centering
\includegraphics[width=\textwidth]{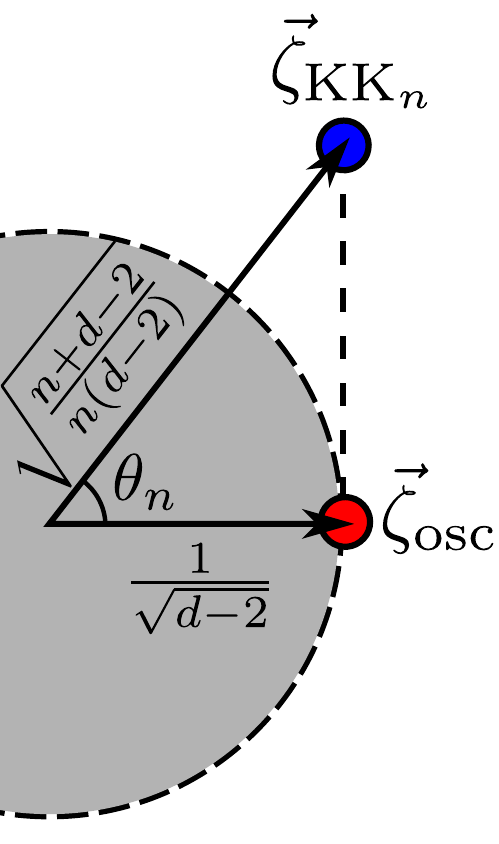}
\caption{\label{sf.sketch1}}
\end{subfigure}
\begin{subfigure}{0.27\textwidth}
\centering
\includegraphics[width=\textwidth]{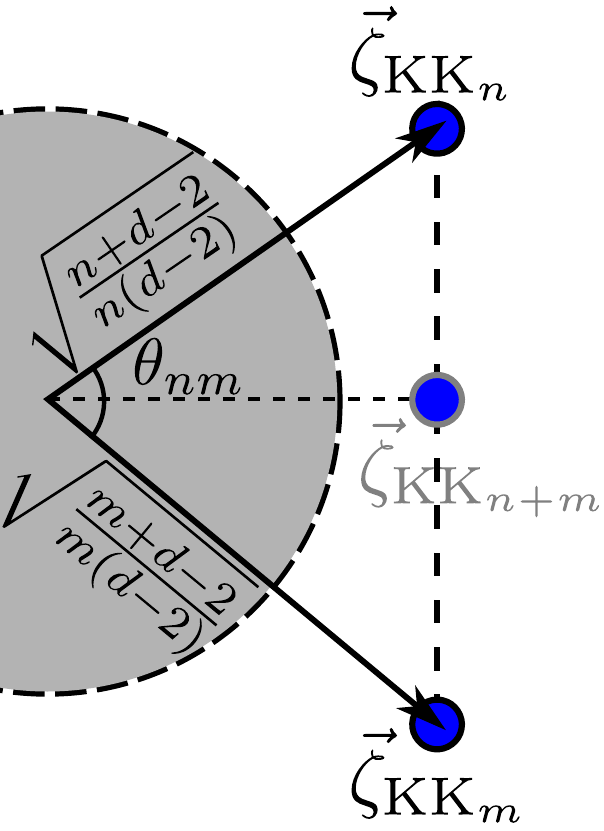}
\caption{\label{sf.sketch2}}
\end{subfigure}
\begin{subfigure}{0.50\textwidth}
\centering
\includegraphics[width=\textwidth]{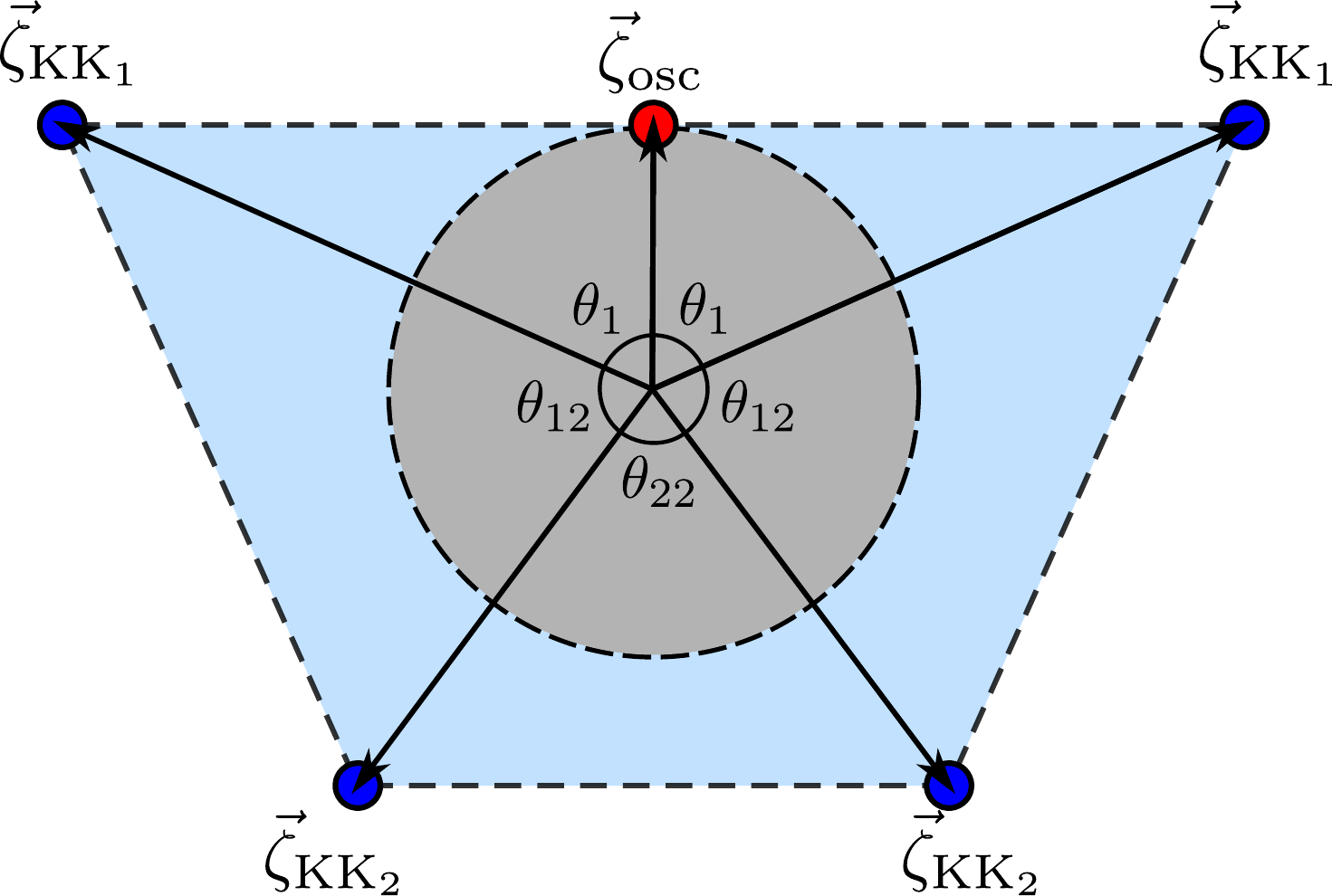}
\caption{\label{sf.sketch3}}
\end{subfigure}
\caption{Sketches of the different taxonomy rules.  Figure \subref{sf.sketch1} shows the tower vectors of a KK tower and a string tower, while  Figure \subref{sf.sketch2} represents two different KK towers decompactifying either $n$ or $m$ extra dimensions. Both the lengths and angles between the vectors are fixed by the nature of the tower. Figure \subref{sf.sketch3} depicts an example of a polytope that is consistent with the taxonomy rules. In the three figures the lower bound of $\frac{1}{\sqrt{d-2}}$ for  the exponential mass decay rate is depicted in gray.\label{f.sketch}}
\end{figure}

We will argue that in generic infinite-distance limits, the Emergent String Conjecture \cite{Lee:2019wij} constrains not only the lengths $|\vec{\zeta}|$ of the vectors generating the tower polytope, but also the angles between adjacent vectors. We illustrate this in Figures \ref{sf.sketch1} and \ref{sf.sketch2}, where the dots correspond to different towers of states that become light asymptotically. The length of each vector is fixed, and it depends on whether the vector corresponds to a KK tower or a tower of string oscillator modes. Similarly, the angle between two neighboring $\vec\zeta$-vectors is also fixed uniquely by the nature of the associated towers. These taxonomic rules allow us to build and classify the allowed tower polytopes, as illustrated in Figure \ref{sf.sketch3}.

These convex hulls are a useful tool in studying the Distance Conjecture in multi-dimensional moduli spaces \cite{Calderon-Infante:2020dhm} (see also \cite{Etheredge:2022opl, Etheredge:2023odp, Calderon-Infante:2023ler,Etheredge:2023usk}). As explained in \cite{Calderon-Infante:2020dhm}, the Distance Conjecture generically translates to the statement that the convex hull of the tower vectors of the light towers of states in each duality frame should lie outside a ball of radius $\alpha_\text{min}$, where $\alpha_\text{min} \sim O(1)$ is the minimum value of the exponential rate allowed by the Distance Conjecture\footnote{This was denoted in  \cite{Calderon-Infante:2020dhm} as the Convex Hull Distance Conjecture.}.
 We will see that this condition is indeed satisfied whenever the taxonomic rules derived in this paper hold, yielding  $\alpha_{min}=\frac{1}{\sqrt{d-2}}$, where $d$ is the number of spacetime dimensions, as expected by the Sharpened Distance Conjecture of \cite{Etheredge:2022opl} (see Figure \ref{sf.sketch1}-\ref{sf.sketch2}). However, as we will discuss, this does not necessarily guarantee  that  the decay rate of the lightest tower along \emph{any} geodesic satisfies $\alpha \geq \frac{1}{\sqrt{d-2}}$ unless additional assumptions are imposed.

Our analysis also produces restrictions on the asymptotic behavior of the moduli-dependent species scale $\Lambda_{\rm QG}$, which is the quantum gravity cut-off at which the EFT breaks down \cite{vandeHeisteeg:2023ubh, vandeHeisteeg:2023uxj,Calderon-Infante:2023ler, Castellano:2023stg,  Castellano:2023jjt}. The Emergent String Conjecture implies that in an asymptotic regime of moduli space, this species scale can be identified with either a string scale or a $(d+n)$-dimensional Planck scale associated with the decompactification of $n$ dimensions, depending on the duality frame.
It is convenient to introduce a \textbf{species vector} \cite{Calderon-Infante:2023ler} corresponding to the gradient of the logarithm of the species scale in a given duality frame of the theory,
\begin{align}
	\vec{\mathcal Z}=-\vec\nabla \log \frac{\Lambda_{\rm QG}}{M_{{\rm Pl},d}}\,.
\end{align}
This species vector parametrizes the variation of the species scale in moduli space, and it is useful for understanding the infinite-distance limits of the theory.
Notably, the species vector $\vec{\mathcal Z}$ plays a starring role in an intriguing ``pattern,'' observed first in
\cite{Castellano:2023jjt, Castellano:2023stg}, namely
\be
\vec{\zeta} \cdot \vec{\mathcal Z} = \frac{1}{d-2}\,,
\label{pattern1}
\ee
where $\vec \zeta$ is the tower vector of the lightest tower in a given infinite-distance limit.
As shown in \cite{Castellano:2023jjt, Castellano:2023stg} (see also \cite{Rudelius:2023spc}), this pattern holds in a vast array of string/M-theory compactifications.
In this work, we will see that it also follows from the Emergent String Conjecture under the assumptions outlined in Section \ref{ss.assumpt}. As such, it may be viewed as one of the taxonomic rules governing the geometry of the tower and species vectors.

In situations where our taxonomic rules can be applied globally across a suitable flat slice of the moduli space, one can further define a ``species polytope" \cite{Calderon-Infante:2023ler}, which is the convex hull of the set of species vectors in each of the infinite-distance limits of the theory. This species polytope is (up to normalization) the dual of the tower polytope (see Figure \ref{f.alphasS}). As a result, our taxonomic rules for tower polytopes immediately lead to taxonomic rules for species polytopes.

The structure of this paper is as follows. In Section \ref{ss.assumpt}, we present a brief summary of the rules governing the tower and species polytopes and the assumptions on which these rules rely. The detailed derivation of these rules is presented in Section \ref{s.rules}. In Section \ref{s.scope}, we discuss the scope of our analysis. In Section \ref{s.2dclassification}, we classify all two-dimensional slices of tower and species polytopes in dimensions 6-10, assuming that any decompactification limit gives a theory in at most eleven dimensions and that there are no strings in 11d. In Section \ref{s.fullpolytope}, we obtain all possible tower and species polytopes in dimensions 8-10 under the same assumptions, and we compare the results with the polytopes that are known to arise from certain string theory compactifications. We find that many of these polytopes appear in maximal and half-maximal supergravity, while other polytopes do not have known string theory realizations. We conclude in Section \ref{s.conc} with some final remarks, followed by a series of appendices. In Appendix \ref{a.MTheory}, we present a top-down derivation of the polytopes from string theory and detail the behavior of these polytopes under dimensional reduction. In Appendices  \ref{s.orthogonality} and \ref{a.GeodesicallyIncomplete}, we discuss the case where the tower are not constant but slide in moduli space. In Appendix \ref{a.GeodesicallyIncomplete}, we remark on the case of a geodesically-incomplete moduli space.

\subsection{Summary of results and assumptions \label{ss.assumpt}}

We now summarize the main results of our paper. Our taxonomy program proceeds in two steps. First, under one set of assumptions, we derive a set of ``taxonomic rules", which locally characterize the possible behavior of the light towers in a given duality frame. Second, under a more restrictive set of assumptions, we combine the results of different duality frames to classify the tower and species polytopes, which describe the global structure of the asymptotic regions of moduli space.

\subsubsection{Taxonomic rules}

We begin with step one. Our primary assumption is the Emergent String Conjecture, which implies the following conditions:
\begin{enumerate}
 \item The lightest tower in a given infinite-distance limit is either a KK tower or a tower of weakly coupled string oscillator modes. \label{assump1}
 \item The species scale in a given infinite-distance limit is either a higher-dimensional Planck scale or a string scale. \label{assump2}
\end{enumerate}
Consequently, we define a \textbf{principal tower} to be either (a) a tower of KK modes or (b) a tower of string oscillator modes.
We also impose the Emergent String Conjecture recursively to the higher dimensional theory that emerges upon decompactification. This latter assumption is stronger than it looks since it puts non-trivial constraints on the existence of bound states of the towers in lower dimensions, as explained in Section \ref{s.scope}.

To derive our taxonomic rules, we further restrict our attention to \textbf{regular} infinite-distance limits that satisfy the following assumptions, for simplicity:
\begin{enumerate}[resume]
 \item  In a decompactification limit, the endpoint of the decompactifying manifold is Ricci-flat except in regions of measure zero, such that the warp factor and varying field profiles dilute away in the limit.
  \label{assump3}
  \item The leading (i.e., lightest) principal tower is not degenerate, i.e., there are not multiple leading principal towers decaying at the same exponential rate.
   \label{assump4}
\item  Any decompactification limit corresponds, after decompactification, to an infinite-distance limit in a higher-dimensional theory which is also regular.
 \label{assump5}
\end{enumerate}
We will argue in Section \ref{subsec:generic} that regular infinite-distance limits are  \textbf{\emph{generic}} in the moduli space, in the sense that we expect irregular limits to occur only in regions of measure zero. In special limits where the assumption of regularity is violated, the rules may (but do not necessarily) break down, as we explain in Section \ref{s.scope}. Hence, the rules presented in this paper should be understood as a first step towards a taxonomy of infinite-distance limits.

We then consider an infinite-distance limit satisfying Assumptions \ref{assump1}-\ref{assump5} above, where some number $k$ of principal towers are lighter than the species scale.
Each principal tower is associated with a scalar field, either a volume field in the case of a KK tower or a dilaton in the case of a string oscillator tower, which span an $k$-dimensional slice of moduli space.
To each principal tower, we associate a tower vector
\be
\vec{\zeta} \equiv -\vec{\nabla} \log \frac{m}{M_{{\rm Pl},d}}\,,
\ee
where $M_{{\rm Pl},d}$ is the $d$-dimensional Planck scale.
Thus, the $k$ light principal towers give rise to $k$ tower $k$-vectors valued in the tangent bundle of this slice of moduli space. The convex hull of these vectors form the vertices of a $(k-1)$-simplex, which we call the \textbf{frame simplex}. In an infinite-distance limit in this slice of moduli space that satisfies Assumptions \ref{assump1}-\ref{assump5} above, the geometry  of the frame simplex (and in particular, is vertices, edges, and faces) are constrained to satisfy the following list of taxonomic rules.

Given any pair of tower vectors $\vec{\zeta}_a$, $\vec{\zeta}_b$, their dot product in the asymptotic limit satisfies
\be
\label{rule1}
\boxed{\vec{\zeta}_a \cdot \vec{\zeta}_b= \frac{1}{d-2} + \frac{1}{n_a}\delta_{ab}\,.}
\ee 
When considering the same tower (i.e., $a=b$), \eqref{rule1} fixes the lengths of the vertices, which are constrained to take values within a discrete set \cite{Etheredge:2022opl}:
\begin{align}
|\vec \zeta_{\text{KK}_n}|^2=\frac{n+d-2}{n(d-2)},\qquad  |\vec{\zeta}_\text{osc}|^2=\frac{1}{d-2}\,,
\label{vertexlengths}
\end{align}
where $d$ is the spacetime dimension, $\vec\zeta_{\text{KK}_n}$ is associated with the KK modes for a decompactification to $d+n$ dimensions, and $\vec{\zeta}_\text{osc}$ is associated with a tower of string oscillation modes (which formally can be recovered from setting $n=\infty$ in \eqref{rule1}).

The rule \eqref{rule1} also constrains the angles between the vertices of the frame simplex, when considering different towers (i.e. $a\neq b$). Namely, the angle $\theta_n$ between a string oscillator vertex and a KK$_n$ vertex is given by
\begin{equation}\label{e.angles intro}
   \cos \theta_{n} =\sqrt{\frac{n }{ n+d-2}}\,,
   \end{equation}
   while the angle between a KK$_m$ vertex and a KK$_n$ vertex is given by
   \begin{equation}
   \cos \theta_{m,n} =   \sqrt{\frac{n m}{ (n+d-2)(m+d-2)} } \,.
   \end{equation}
   Examples of these angles are shown in Figure \ref{sf.sketch3}. Equivalently, the lengths of the edges are constrained to be
   \be
|\vec \zeta_{\text{osc}} - \vec \zeta_{\text{KK}_n}|^2 = \frac{1}{n} \,,~~~~  |\vec \zeta_{\text{KK}_m} - \vec \zeta_{\text{KK}_n}|^2 = \frac{m+n}{mn} \,,
   \ee
   for an edge between (a) a string oscillator vertex and a KK$_n$ vertex and (b) a KK$_m$ vertex and a KK$_n$ vertex, respectively.

In a regular infinite-distance limit within the frame simplex satisfying Assumptions \ref{assump1}-\ref{assump5}, the value of the species scale $\Lambda_{\rm QG}$ is uniquely determined. Let $\vec{\mathcal Z} \equiv - \vec{\nabla} \log \frac{\Lambda_{\rm QG}}{M_{{\rm Pl},d}}$ denote the species vector as in \cite{Calderon-Infante:2023ler}. Then, the scalar product of the species vector $\vec {\mathcal Z}_\text{QG}$ and the tower vector $\vec \zeta_a$ of any of the vertices of the frame simplex satisfy asymptotically
  \begin{align}
	\boxed{\vec{\zeta}_a\cdot \vec{\mathcal Z}_\text{QG}=\frac 1{d-2},}
 \label{eq:pattern}
\end{align}
 which is precisely the pattern first observed in \cite{Castellano:2023jjt, Castellano:2023stg} relating the variation of the species scale and the lightest tower of states.
 
Consequently, the length of the species vector is also fixed:
\begin{align}
	\boxed{|\vec{\mathcal Z}_\text{QG}|^2=\frac{1}{d-2}-\frac{1}{D-2},}
\end{align}
where the species dimension $D$ is either the spacetime dimension in which the species scale equals the Planck scale (in a decompactification limit) or $D=\infty$ (in an emergent string limit).

For each face $\mathcal F$ of the frame simplex, spanned by tower-vectors $\vec \zeta_a$ for KK-modes decompactifying $n_1$,\dots,$n_k$ dimensions, the quantum gravity scale $\vec{\mathcal Z}_{\mathcal F}$ of this face is
\begin{align}
	\vec{\mathcal Z}_{\mathcal F}=\frac 1{D_{\mathcal F}-2}\sum_a n_a \vec\zeta_a,
\end{align}
where $D_{\mathcal F}=d+\sum_a n_a$ is the species dimension, which is the dimension that the theory decompactifies upon asymptotically traveling in the direction of the center of the face $\mathcal F$. For faces $\mathcal F_1$ and $\mathcal F_2$ within the same duality frame, the quantum gravity scales associated with each face satisfy the dot products
\begin{equation}
  \vec{\mathcal Z}_{\mathcal F}\cdot \vec{\mathcal Z}_{\mathcal F'}=\frac{1}{d-2}+\frac{D_{\mathcal F\cap\mathcal F'}-D_{\mathcal F}-D_{\mathcal F'}+2}{(D_{\mathcal F}-2)(D_{\mathcal F'}-2)},\label{e.pericenterproducts}
\end{equation}
where $D_{\mathcal F}$, $D_{\mathcal F'}$ and $D_{\mathcal F\cap \mathcal F'}$ are are the species dimensions of the frames
 $\mathcal F$, $\mathcal F'$, and $\mathcal F\cap \mathcal F'$.
 
\subsubsection{Classification of tower and species polytopes}

In progressing from step one to step two of the classification program, we make two further assumptions:

    \begin{enumerate}[resume]
     \item There is an asymptotically flat slice $\Sigma_k \cong_{\text{asymp}}
  {\mathbb{R}^k} $ of the moduli space $\mathcal{M}_d$, such that for every asymptotically straight line in $\Sigma_k$ there is a
  infinite-distance limit (geodesic ray) within $\mathcal{M}_d$ that
  asymptotically approaches it.
  \label{assump6}
  \item For a generic choice of asymptotically straight line in $\Sigma_k$, a subspace of the plane generated by the tower vectors of the frame simplex is asymptotically equal to the tangent space of $\Sigma_k$. Rules \eqref{rule1} and \eqref{eq:pattern} still apply to the vertices of the frame simplex and the species vector after projection to this subspace.
  \label{assump7}

\end{enumerate}

These are nontrivial assumptions and are not satisfied in many cases, as we will see below.
   When these assumptions do hold, however, then the frame simplices can be glued together globally to give a full tower polytope, which is necessarily generated by the towers becoming light at the different infinite-distance limits. When this is possible, then the pattern of \eqref{eq:pattern} implies that the dual polytope $P^\circ$
\begin{align}\label{e.dual pol}
	P^\circ=\left\{\vec{\mathcal{Z}}\ |\ \vec{\mathcal{Z}} \cdot \vec \zeta  \leq \frac{1}{d-2}\;\forall \,\vec \zeta\in P\right\}.
\end{align}
is equivalent to the species polytope, generated by the species vectors of the different duality frames. As a result, the angles between species vectors (which correspond to vertices of the species polytope) are also constrained.

The formula \eqref{e.pericenterproducts} describes dot products between pericenters of various facets of the species polytope. But, it does not describe dot products between vertices of the species polytopes. Suppose that two vertices $\vec{\mathcal Z}_\alpha$ and $\vec{\mathcal Z}_\beta$ of the species polytope are joined by an edge with pericenter $\vec{\mathcal Z}_{\mathcal F}$. Then the dot products between two vertices of the species polytope satisfy,
\begin{equation}
  \fbox{$\vec{\mathcal Z}_{\alpha}\cdot \vec{\mathcal Z}_\beta=\frac{1}{d-2}-\frac 1{D_{\alpha \beta}-2}\left[1+\sqrt{\frac{(D_\alpha-D_{\alpha\beta})(D_\beta-D_{\alpha\beta})}{(D_\alpha-2)(D_\beta-2)}}\right]$,}
\end{equation}
where $D_{\alpha}$, $D_\beta$, and $D_{\alpha\beta}$ are the species dimensions associated to $\vec{\mathcal Z}_\alpha$, $\vec{\mathcal Z}_\beta$, $\vec{\mathcal Z}_{\mathcal F}$. This constrains the angles between adjacent vertices, which are uniquely determined in terms of the vertex types.

In Section \ref{s.Applications} we will use the taxonomic rules \eqref{rule1} and \eqref{eq:pattern} to build tower and species polytopes that are consistent with the above assumptions.

\section{Taxonomy rules}\label{s.rules}

Consider the moduli space $\mathcal{M}_d$ of a
$d$-dimensional quantum gravity theory (QGT), endowed with a natural Riemannian metric
$\mathsf{G}_{i j} (\phi)$ defined by the Planck-normalized kinetic terms of
the moduli:
\begin{equation}
  S_{\text{kin}} \supset - \frac{1}{2 \kappa_d^2} \int \textsf{} \mathsf{G}_{i
  j} (\phi) d \phi^i \wedge \star d \phi^j .
\end{equation}
For simplicity, we use vector symbols such as $\vec{\zeta}$ to denote
tangent/cotangent vectors on this Riemannian space, where we freely (and
silently) convert between the two using the metric $\mathsf{G}_{i j}$. In an
abuse of notation, we also use the same vector symbols to denote
tangent/cotangent vectors on naturally defined subspaces of the moduli space
that will arise during the discussion. Whenever we write these vectors in
components, we choose a convenient orthonormal frame $\mathsf{G}_{i j} =
\mathsf{e}_i^a \mathsf{e}_j^b \delta_{a b}$ to do so. $\vec{\nabla} f$ will
denote the moduli space gradient, the components of which are partial
derivatives $(\vec{\nabla} f)_i = \frac{\partial f}{\partial \phi^i}$ in a
coordinate basis, but which involve the inverse vielbein $(\vec{\nabla} f)_a =
\mathsf{e}_a^i \frac{\partial f}{\partial \phi^i}$ in an orthonormal basis. With these conventions in mind, we rarely need to invoke either the metric
$\mathsf{G}_{i j}$ or the vielbein $\mathsf{e}_i^a$ explicity.

\subsection{The structure of a regular infinite-distance limit}\label{s.inf dist str}

Consider an infinite-distance limit in the moduli space $\mathcal{M}_d$ of a
$d$-dimensional QGT. To be precise, by this we mean a semi-infinite path
$\phi^i (s)$, $s \geqslant 0$, through $\mathcal{M}_d$ such that the shortest
route between $\phi^i (0)$ and $\phi^i (s)$ is along the path itself. This
implies that (1) the path is a geodesic travelling to infinite distance and
(2) it does so ``as quickly as possible'' (without meandering).\footnote{For
example, if $\mathcal{M}_d$ is a flat cylinder then helical paths winding
around the cylinder are not infinite-distance limits---even though they are
geodesics that go to infinite distance---because the ``straight'' paths that
do not wind are shorter.} Such a path is known as a \textbf{geodesic ray} in
the mathematical literature. We declare geodesic rays that asymptotically
approach each other to define equivalent limits so that, e.g., the choice of
starting point is not part of specifying the infinite-distance limit.

According to the Distance Conjecture, one or more particle towers become light
in our chosen infinite-distance limit. In general, this collection of towers
may be quite complicated. To simplify things, we develop a notion of a more
tractable, ``regular'' infinite-distance limit (definition \ref{def:regular}
below). We then classify the possible towers in regular limits, and use our
understanding of these limits to better understand general infinite-distance limits.

\subsubsection{Deriving the rules: one tower scale}

Let $m (\phi)$ be the mass scale of the lightest tower in an infinite-distance limit. The moduli dependence
the tower scale relative to the $d$-dimensional Planck scale $M_{{\rm Pl,}d}$ is
characterized by the \textbf{tower vector} $\vec{\zeta} \equiv -
\vec{\nabla} \log \frac{m}{M_{{\rm Pl,}d}}$. In general, there might be
\emph{multiple} leading towers becoming light at the same rate, each with
different tower vectors $\vec{\zeta}$, $\vec{\zeta}'$, $\dots$. In this
case, we say that the leading towers are \textbf{degenerate}. To avoid this
complication, let us assume for the time being that the leading tower is
non-degenerate.

Per the Emergent String Conjecture, this tower is either (1) a KK tower
associated to decompactification to a $(D = d + n)$-dimensional theory, or (2) a
tower of oscillator modes of a perturbative fundamental string.

Consider the case where the leading tower is a KK tower, and let us further
assume that the theory decompactifies along an ``empty" Ricci-flat manifold
$X_n$. The moduli of the $d$-dimensional theory consist of those of the
$D$-dimensional theory together with the overall volume and shape moduli of
the compact manifold $X_n$ and the axions arising from the $p$-form gauge
fields of the $D$-dimensional theory reduced along $p$-cycles of $X_n$.
Expressed in this basis, the tower vector of the leading tower takes the form:
\begin{equation}
  \vec{\zeta}_{\text{KK$_n$}} = \left(\begin{array}{cccc}
    \vec{\zeta}^{(D)}_{\text{KK$_n$}} \; ; &
    \zeta^{(\text{vol})}_{\text{KK$_n$}} \; ; &
    \vec{\zeta}^{(\text{shape})}_{\text{KK$_n$}} \; ; &
    \vec{\zeta}^{(\text{axion})}_{\text{KK$_n$}}
  \end{array}\right) = \left(\begin{array}{cccc}
    \vec{0} \hspace{0.8em} ; & \sqrt{\frac{n + d - 2}{n (d - 2)}}
    \hspace{0.8em} ; & \vec{\zeta}^{(\text{shape})}_{\text{KK$_n$}} \; ; &
    \vec{0}
  \end{array}\right) .
\end{equation}
Here we have temporarily left open the possibility that the KK scale depends
on the shape moduli. This is because $X_n$ may become ``long and narrow'' in
some limits of moduli space, making some KK modes lighter than the
overall-volume KK scale and others heavier. Thus, if
$\vec{\zeta}^{(\text{shape})}_{\text{KK$_n$}} \neq \vec{0}$ we would expect
multiple towers with different values of
$\vec{\zeta}^{(\text{shape})}_{\text{KK$_n$}}$. However, since by assumption
the leading tower is non-degenerate, we conclude that
$\vec{\zeta}^{(\text{shape})}_{\text{KK$_n$}} = \vec{0}$,\footnote{For
instance, in the case of a torus $X_n = T^2$, the KK modes indeed depend on
the complex structure modulus $\tau$ as well as the overall volume. Then
either (1) $\tau$ is frozen, e.g., by a discrete quotient of the form $(T^2
\times Y_m) /\mathbb{Z}_k$ $(k = 3, 4, 6)$, or else (2) in a
\emph{regular} infinite-distance limit one cycle of the torus
decompactifies before the other, hence there are two separate KK scales and
only the $S^1$ KK modes appear in the leading tower. Since $S^1$ has no shape
moduli, this agrees with $\vec{\zeta}^{(\text{shape})} = \vec{0}$.} i.e.,
\begin{equation}
  \vec{\zeta}_{\text{KK$_n$}} = \left(\begin{array}{cccc}
    \vec{0} \hspace{0.8em} ; & \sqrt{\frac{n + d - 2}{n (d - 2)}} \; ; &
    \vec{0} \; ; & \vec{0}
  \end{array}\right) \qquad \Rightarrow \qquad \left|
  \vec{\zeta}_{\text{KK$_n$}} \right|^2 = \frac{n + d - 2}{n (d - 2)} =
  \frac{1}{d - 2} + \frac{1}{n} . \label{eqn:KKzeta}
\end{equation}
Thus, the tower vector of the leading tower has a fixed length, determined by
the spacetime dimension $d$ and the number of dimensions $n$ that decompactify
at the tower scale.

Before proceeding, we revisit the assumption that the theory decompactifies
along an empty, Ricci flat manifold $X_n$. More generally, the
decompactification along $X_n$ may involve branes, fluxes and moduli gradients
with their associated warping and/or Ricci curvature
\cite{Maldacena:2000mw}. In such cases, our conclusions still
follow if the warped, Ricci-curved regions associated to these sources grow
parametrically more slowly than the overall volume of $X_n$, resulting in an
\textbf{asymptotically empty} geometry in the decompactification limit.

Note, however, that a new class of ``brane moduli'' can appear in such
asymptotically empty scenarios. These moduli control the positions of warped /
Ricci-curved regions and/or degrees of freedom that are localized in these
regions. However, the KK modes are determined by the bulk geometry of $X_n$,
hence $\vec{\zeta}^{(\text{brane})}_{\text{KK$_n$}} = \vec{0}$ and the above
argument is unmodified.

If there are no other light towers beyond the leading KK tower, the species
scale $\Lambda_{\text{QG}} (\phi)$ is reduced from the $d$-dimensional Planck
scale $M_{{\rm Pl,}d}$ down to the $D$-dimensional Planck scale $M_{{\rm Pl,}D} \ll M_{{\rm Pl,}d}$. The moduli
dependence of $\Lambda_{\text{QG}} (\phi)$ is characterized by the
\textbf{species vector} $\vec{\mathcal{Z}} \equiv - \vec{\nabla} \log
\frac{\Lambda_{\text{QG}}}{M_{{\rm Pl,}d}}$, equal to
\begin{equation}
  \vec{\mathcal{Z}}_{\text{Pl}_D} = - \vec{\nabla} \log \frac{M_{{\rm Pl,}D}}{M_{{\rm Pl,}d}} =
  \left(\begin{array}{cccc}
    \vec{0} \; ; & \sqrt{\frac{n}{(n + d - 2) (d - 2)}} \; ; & \vec{0} \; ; &
    \vec{0}
  \end{array}\right), \label{eqn:PlanckZvector}
\end{equation}
in this case. Notice that
\begin{equation}
  | \vec{\mathcal{Z}}_{\text{Pl}_D} |^2 = \frac{n}{(n + d - 2) (d - 2)} =
  \frac{1}{d - 2} - \frac{1}{D - 2}, \qquad \vec{\zeta}_{\text{KK$_n$}} \cdot
  \vec{\mathcal{Z}}_{\text{Pl}_D} = \frac{1}{d - 2}, \label{eqn:PlanckZ}
\end{equation}
where the latter equality is an example of the tower-species pattern
discovered in \cite{Castellano:2023jjt, Castellano:2023stg}. We generalize our discussion later to allow
for additional light towers between the KK scale and $\Lambda_{\text{QG}}
(\phi)$.

Now consider the case where the leading tower consists of oscillator modes of
a perturbative fundamental string. The tension $T_s (\phi)$ of the fundamental
string is controlled by a dilaton $\phi_s$, which is a universal part of the
string spectrum much like the graviton. The universal dilaton coupling fixes
the tower vector of the oscillator modes appearing at the string scale $m_s \sim 
\sqrt{T_s}$ to be:
\begin{equation}
  \vec{\zeta}_{\text{osc}} = \left(\begin{array}{cc}
    \zeta^{\phi_s}_{\text{osc}} \; ; & \vec{\zeta}_{\text{osc}}^{\left(
    \text{other} \right)}
  \end{array}\right) = \left(\begin{array}{cc}
    \frac{1}{\sqrt{d - 2}} \; ; & \vec{0}
  \end{array}\right) \qquad \Rightarrow \qquad \left| \vec{\zeta}_{\text{osc}}
  \right|^2 = \frac{1}{d - 2} . \label{eqn:oscZeta}
\end{equation}
Thus, the tower vector of the leading tower again has a fixed length, this
time determined by the spacetime dimension $d$ alone.

Because the density of oscillator modes grows exponentially, in this case the
species scale is parametrically the same as the string scale,
$\Lambda_{\text{QG}} \sim m_s$, up to corrections that are subexponential in
the moduli. The species vector is therefore:
\begin{equation}
  \vec{\mathcal{Z}}_{\text{str}} = \left(
    \mathcal{Z}_{\text{str}}^{\phi_s} ,\mathcal{Z}_{\text{str}}^{\left(
    \text{other} \right)}
 \right) = \left(\begin{array}{cc}
    \frac{1}{\sqrt{d - 2}} \quad  \vec{0}
  \end{array}\right) = \vec{\zeta}_{\text{osc}} \quad \Rightarrow \quad
  \left| \vec{\mathcal{Z}}_{\text{str}} \right|^2 = \frac{1}{d - 2}, \quad
  \vec{\mathcal{Z}}_{\text{str}} \cdot \vec{\zeta}_{\text{osc}} = \frac{1}{d -
  2}, \label{eqn:stringZ}
\end{equation}
again consistent with the tower-species pattern. Note that formally
\eqref{eqn:oscZeta}, \eqref{eqn:stringZ} are special cases of
\eqref{eqn:KKzeta}, \eqref{eqn:PlanckZ} with $n \rightarrow \infty$, so that a
string oscillator tower is formally analogous to KK tower for $n = \infty$
decompactifying dimensions, and likewise for the associated species scales. We
make repeated use of this analogy for notational convenience throughout our
paper.

Note that even when the leading tower is degenerate, if one of the degenerate
towers is a tower of string oscillator modes, then the above reasoning can
still be applied to the oscillator tower, and the rules \eqref{eqn:oscZeta},
\eqref{eqn:stringZ} are still satisfied. For now, we simply ignore the tower
vectors of the remaining, degenerate towers. (Note that these typically have
subexponential density---consisting, e.g., of KK and/or winding modes---and they lie parametrically at the species scale.)

By contrast, when KK towers degenerate there is no single, dominant tower with
a fixed tower vector. Instead, there will be multiple towers with various
values of $\vec{\zeta}^{(\text{shape})} \neq \vec{0}$, all parametrically
below the species scale. It is useful to keep this case separate from the
simpler, non-degenerate scenario considered above, and we defer further
consideration of it until Section \ref{subsec:degenerate}.

\subsubsection{Deriving the rules: multiple tower scales\label{sec:rules}}

In the case of a decompactification limit, we recover a higher-dimensional QGT
parametrically above the KK scale. Projecting onto the higher-dimensional
moduli space $\mathcal{M}_{(D)}$, the original infinite-distance limit
$\phi_{(d)}^i (s)$ lifts to a path $\phi_{(D)}^I (s)$ through
$\mathcal{M}_{(D)}$. This path cannot have any shortcuts along it, because if
it does then there will be corresponding shortcuts along the path
$\phi_{(d)}^i (s)$ through the complete moduli space $\mathcal{M}_{(d)}$,
contradicting the assumption that we are considering an infinite-distance limit. Therefore, either (1) $\phi_{(D)}^I (s) = \phi_{(D)}^I (0)$ is a single
point in $\mathcal{M}_D$, or (2) $\phi_{(D)}^I (s)$ is itself an
infinite-distance limit of $\mathcal{M}_{(D)}$.

In the first case, there is (parametrically) only one tower scale, which is
covered by the discussion above. In the second case, we refocus our attention
on the infinite-distance limit $\phi_{(D)}^I (s)$ in the $D$-dimensional
theory. If this limit satisfies the same assumptions as above, we can reason
in a recursive manner. The required assumptions are encapsulated in the
following regularity conditions:

\begin{Definition}[Definition]
  A \textbf{regular} infinite-distance limit is one with
  either\label{def:regular}
  \begin{enumerate}
    \item A leading string oscillator tower, {\underline{or}}
    
    \item A leading KK tower, such that
    \begin{enumerate}
      \item The tower is {\textbf{non-degenerate}} (so that there are not several leading  towers decaying at the same rate, i.e., the limit is characterized by a single
      tower vector $\vec{\zeta} \equiv - \vec{\nabla} \log \frac{m}{M_{{\rm Pl,}d}}$)
      {\underline{and}}
      
      \item The decompactification manifold is {\textbf{asymptotically empty}}
      (Ricci flat with vanishing background fields, except in regions of
      measure zero) {\underline{and}}
      
      \item After decompactification, the lift of the infinite-distance limit
      to the higher dimensional theory is also regular.
    \end{enumerate}
  \end{enumerate}
\end{Definition}

As discussed above, we impose non-degeneracy for KK towers (which occur
parametrically below the species scale), but not for string oscillator towers
(which occur parametrically at the species scale). The final condition is
recursive: after each decompactification, the same regularity conditions are
applied in the new description.

Given a regular infinite-distance limit, we obtain a parametric hierarchy of
tower scales up to the species scale by applying the following steps
recursively, starting at $a = 1$ with the original $d_1 \equiv d$ dimensional
theory:
\begin{enumerate}
  \item Let $m_a$ be the mass scale of the leading tower in a regular
  infinite-distance limit of a $d_a$-dimensional theory.
  
  \item If this is a KK tower associated to the decompactification of $n_a$
  dimensions then we consider the lift of the infinite-distance limit to the
  $d_{a + 1} = d_a + n_a$ dimensional decompactified theory.
  \begin{enumerate}
    \item If the lift is an infinite-distance limit of the $d_{a +
    1}$-dimensional theory, then we return to step $1$ for this infinite
    distance limit in the decompactified theory, incrementing $a \rightarrow a
    + 1$.
    
    \item If the lift is a single point in the moduli space of the $d_{a +
    1}$-dimensional theory, then $\Lambda_{\text{QG}}$ is parametrically equal
    to the Planck scale $M_{{\rm Pl,}d_{a+1}}$ of this theory, and there are no other
    towers parametrically below this scale, so we stop here.\footnote{There
    can be light towers between the KK scale and the higher-dimensional Planck
    scale but with these assumptions their masses are fixed in Planck units,
    so they are not parametrically separated from the Planck scale.}
  \end{enumerate}
  \item If this is a string oscillator tower, then $\Lambda_{\text{QG}}$ is
  parametrically equal to $m_a$, so we stop here.
\end{enumerate}
The end result is a parametric hierarchy of tower mass scales below the
species scale,
\begin{equation}
  m_1 \ll m_2 \ll \cdots \ll m_k \lesssim \Lambda_{\text{QG}},
\end{equation}
where $k$ is the \textbf{rank} of the limit in question and the first $(k -
1)$ scales $m_a$ are KK scales with $n_a < \infty$ dimensions decompactifying,
and the last scale $m_k$ is either a KK scale (in which case $m_k \ll
\Lambda_{\text{QG}}$) or a string scale (in which case $m_k \simeq
\Lambda_{\text{QG}}$, up to subexponential corrections). Associated to these
scales, we have a collection of tower and species vectors:
\begin{equation}
  \vec{\zeta}_1 \equiv - \vec{\nabla} \log \frac{m_1}{M_{{\rm Pl,}d}}, \qquad \ldots,
  \qquad \vec{\zeta}_k \equiv - \vec{\nabla} \log \frac{m_k}{M_{{\rm Pl,}d}}, \qquad
  \vec{\mathcal{Z}}_{\text{QG}} \equiv - \vec{\nabla} \log
  \frac{\Lambda_{\text{QG}}}{M_{{\rm Pl,}d}} .
\end{equation}
One of the main results of this paper is that the geometry of these vectors is constrained by the following
\emph{taxonomy rules}:
\begin{equation}
  \fbox{$\vec \zeta_a\cdot \vec \zeta_b=\frac{1}{d-2}+\frac 1{n_a}\delta_{ab},\qquad \vec \zeta_a\cdot \vec{\mathcal Z}_\text{QG}=\frac{1}{d-2},\qquad |\vec{\mathcal{Z}}_\text{QG}|^2=\frac{1}{d-2}-\frac{1}{D-2},$}
  \label{eqn:taxonomyRules}
\end{equation}
where for notational compactness we formally set $n_k = \infty$ when $m_k$ is
a string scale and $D \equiv d + \sum_a n_a$ is the \textbf{species
dimension}. Note that these rules once again include the tower-species pattern
of \cite{Castellano:2023stg, Castellano:2023jjt}. This is not an extra input, but rather a consequence
of our starting assumptions. 

The proof of \eqref{eqn:taxonomyRules} is inductive in the rank of the limit.
We have already seen that it holds for rank $k = 1$ limits, see
\eqref{eqn:KKzeta}, \eqref{eqn:PlanckZ}, \eqref{eqn:oscZeta},
\eqref{eqn:stringZ}. Now assume that the rules hold for the rank  ($k - 1$)
-limit in the $d_2 = d + n_1$ dimensional theory obtained from decompactifying
$n_1$ dimensions at the leading KK scale $m_1 (\phi)$. Thus,
\begin{equation}
  \vec{\zeta}_a^{(d_2)} \cdot \vec{\zeta}_a^{(d_2)} = \frac{1}{d_2 - 2} +
  \frac{1}{n_a} \delta_{a b}, \qquad \vec{\zeta}_a^{(d_2)} \cdot
  \vec{\mathcal{Z}}_{\text{QG}}^{(d_2)} = \frac{1}{d_2 - 2}, \qquad |
  \vec{\mathcal{Z}}_{\text{QG}}^{(d_2)} |^2 = \frac{1}{d_2 - 2} - \frac{1}{D -
  2}, \label{eqn:inductiveRules}
\end{equation}
for $a, b > 1$, where $\vec{\zeta}^{(d_2)},
\vec{\mathcal{Z}}_{\text{QG}}^{(d_2)}$ refer to the tower and species vectors
in the $d_2$-dimensional theory and $D = d + \sum_a n_a = d_2 + \sum_{a > 1}
n_a$ is the same before and after compactification. Since
\begin{equation}
  \vec{\zeta}_a^{(d)} = - \vec{\nabla} \log \frac{m_a}{M_{{\rm Pl,}d}} = - \vec{\nabla}
  \log \frac{m_a}{M_{{\rm Pl,}d_2}} - \vec{\nabla} \log \frac{M_{{\rm Pl,}d_2}}{M_{{\rm Pl,}d}} =
  \vec{\zeta}^{(d_2)}_a + \vec{\mathcal{Z}}_{\text{Pl}_{d_2}} \qquad \text{for
  $a > 1$,}
\end{equation}
and likewise $\vec{\mathcal{Z}}_{\text{QG}}^{(d)} =
\vec{\mathcal{Z}}_{\text{QG}}^{(d_2)} + \vec{\mathcal{Z}}_{\text{Pl}_{d_2}}$,
using \eqref{eqn:KKzeta}, \eqref{eqn:PlanckZvector} we find:\footnote{Note
that $\vec{\zeta}_a$, $a > 1$ describes the moduli dependence of the
\emph{KK scale} at which the $n_a$ dimensions in question decompactify,
not the moduli dependence of the mass of an individual KK mode. The latter may
be more complicated, depending, e.g., on the axions, but this dependence is
irrelevant to our argument.}
\begin{align}
  \vec{\zeta}_a^{(d)} = & \left\{ \begin{array}{ccc}
    \left(\begin{array}{cccc}
      0 \; ; & \sqrt{\frac{d_2 - 2}{n_1 (d - 2)}} \; ; & \vec{0} \; ; &
      \vec{0}
    \end{array}\right), &  & a = 1,\\
    \left(\begin{array}{cccc}
      \vec{\zeta}^{(d_2)}_a \; ; & \sqrt{\frac{n_1}{(d_2 - 2) (d - 2)}} \; ; &
      \vec{0} \; ; & \vec{0}
    \end{array}\right), &  & a > 1,
  \end{array} \right. \nonumber\\
  \vec{\mathcal{Z}}_{\text{QG}}^{(d)} = & \left(\begin{array}{cccc}
    \vec{\mathcal{Z}}_{\text{QG}}^{(d_2)} \; ; & \sqrt{\frac{n_1}{(d_2 - 2) (d
    - 2)}} \; ; & \vec{0} \; ; & \vec{0}
  \end{array}\right), 
\end{align}
in the same basis as before. Taking the dot products of these vectors, it is
straightforward to verify \eqref{eqn:taxonomyRules} assuming
\eqref{eqn:inductiveRules}, completing the inductive proof.

\subsubsection{The frame simplex}

To understand the implications of the taxonomy rules
\eqref{eqn:taxonomyRules}, note that they fix the Gram matrix (matrix of dot
products) of the set of vectors $\vec{\zeta}_1$, $ \ldots$ $\vec{\zeta}_k$, $\vec{\mathcal{Z}}_{\text{QG}}$. Up to an overall rotation, a set of vectors is
completely determined by its Gram matrix, so the taxonomy rules completely fix
the geometry of the vectors $\vec{\zeta}_1$, $\ldots$ $\vec{\zeta}_k$, $\vec{\mathcal{Z}}_{\text{QG}}$. We now summarize this geometry.

One can show that the $(k + 1) \times (k + 1)$ Gram matrix specified by
\eqref{eqn:taxonomyRules} is positive semi-definite (as required for any Gram
matrix), with rank $k$. Thus, the Gram matrix has a single null eigenvector,
corresponding to a single linear relation between the vectors $\vec{\zeta}_1,
\ldots, \vec{\zeta}_k, \vec{\mathcal{Z}}_{\text{QG}}$:
\begin{equation}
  \sum_a n_a \vec{\zeta}_a = (D - 2)  \vec{\mathcal{Z}}_{\text{QG}} .
  \label{eqn:zetaZreln}
\end{equation}
In other words, the tower vectors are linearly independent, and together they
determine the species vector.

\begin{figure}[h]
\begin{center}
\begin{subfigure}{0.54\textwidth}
\center
\includegraphics[width=.95\textwidth]{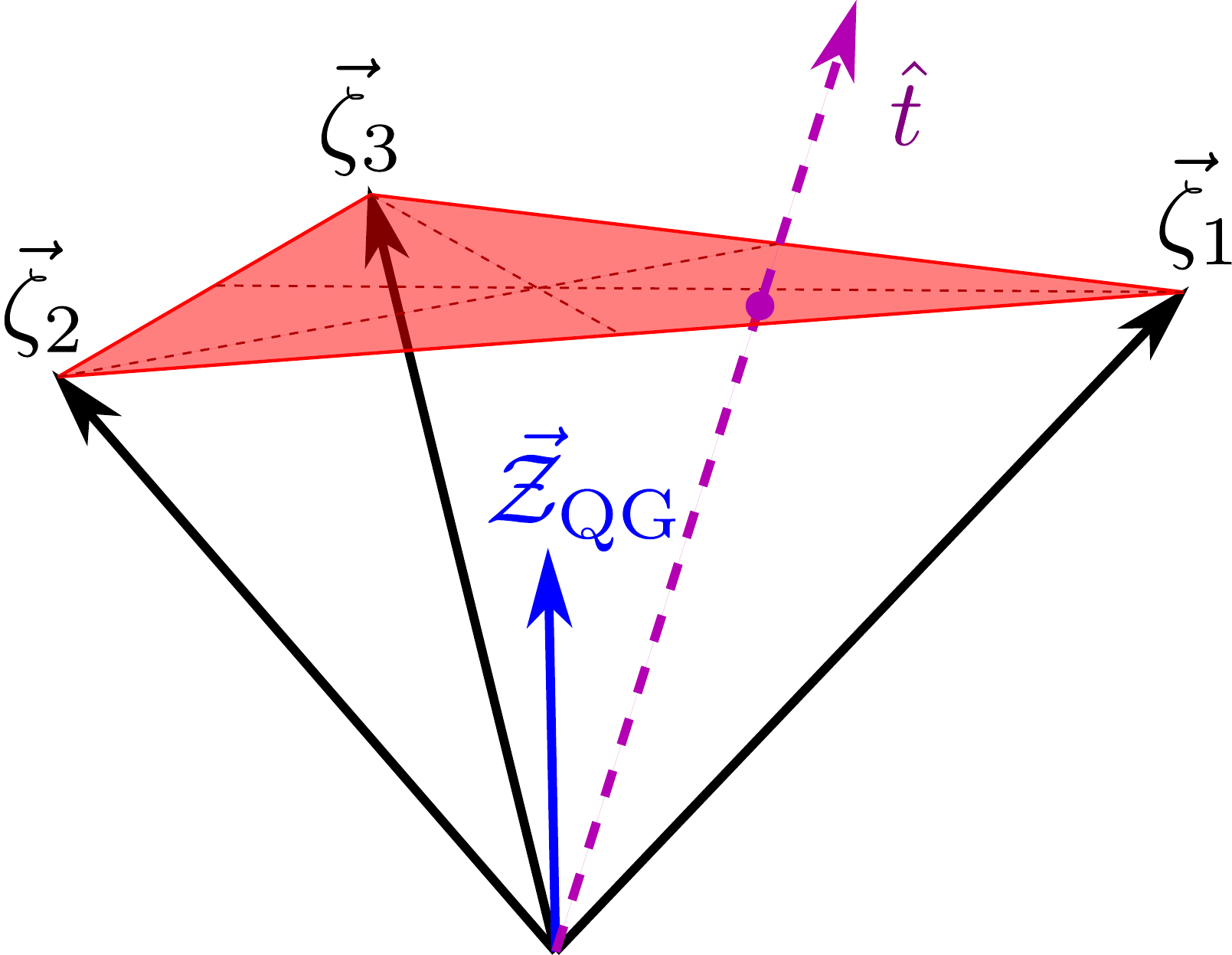}
\caption{Planckian phase.} \label{f.3dFRAMEch1}
\end{subfigure}
\begin{subfigure}{0.45\textwidth}
\center
\includegraphics[width=.95\textwidth]{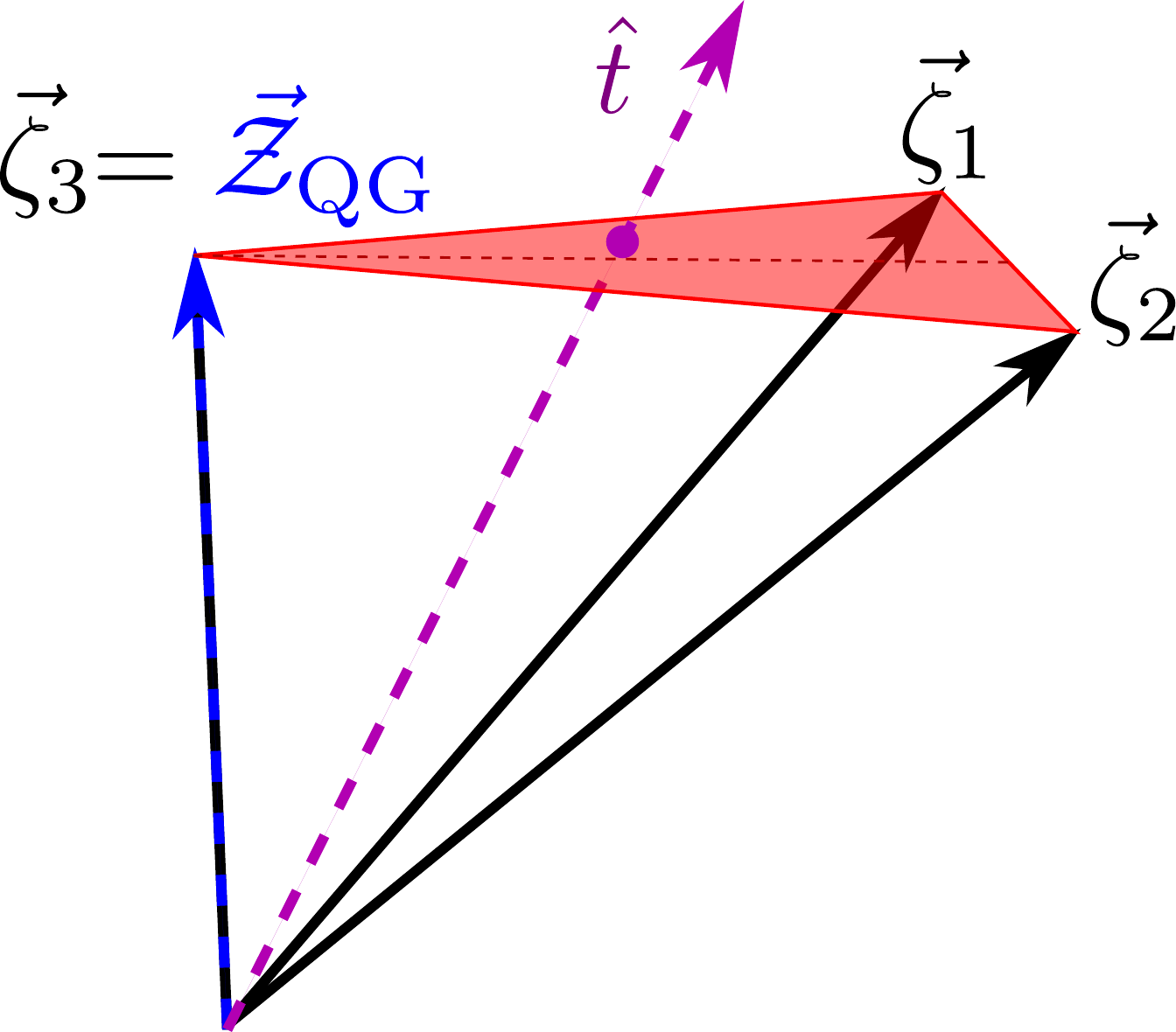}
\caption{Stringy phase} \label{f.3dFRAMEch2}
\end{subfigure}
\end{center}
\caption{The tower vectors $\{\vec{\zeta}_1,\vec\zeta_2,\vec{\zeta}_3\}$ (black arrows), frame simplex (red triangle), species vector $\vec{\mathcal{Z}}_{\rm QG}$ (blue arrow) and direction vector $\hat{t}$ (purple arrow, see \S\ref{subsec:dir}) for two examples of regular infinite-distance limits. Both limits are rank 3 (with 3 tower vectors spanning a 3d principal plane), where \subref{f.3dFRAMEch1} is a Planckian phase and \subref{f.3dFRAMEch2} is a stringy phase. The directions over which several towers degenerate are depicted in dashed lines.
}
\label{f.3dFRAME}
\end{figure}

Therefore, the tower vectors $\vec{\zeta}_1, \ldots, \vec{\zeta}_k$ span a
$k$-plane  in the moduli tangent space, which we call the \textbf{principal
plane}. This is just the radion-radion-\dots-radion or
radion-...-radion-dilaton plane that arose naturally from the overall volume
moduli and/or the dilaton in our derivation above, but the existence of this
plane follows from the rules \eqref{eqn:taxonomyRules} independent of the
derivation. Within the principal plane, the convex hull of the tower vectors
$\vec{\zeta}_1, \ldots, \vec{\zeta}_k$ is a $(k - 1)$-simplex, which we call
the \textbf{frame simplex}, $\Delta = \{ \vec{\zeta}_1, \ldots,
\vec{\zeta}_k \}$.\footnote{The frame simplex has not only a size and shape
but also a specified location relative to the origin. One can think of it as
the base of the cone generated by the tower vectors.} The vertices of the
frame simplex are the tower vectors, and the species vector
$\vec{\mathcal{Z}}_{\text{QG}}$ is orthogonal to the simplex since
$(\vec{\zeta}_a - \vec{\zeta}_b) \cdot \vec{\mathcal{Z}}_{\text{QG}} = 0$ from
\eqref{eqn:taxonomyRules} as noticed in \cite{Castellano:2023jjt}. Some examples of frame simplices are shown in Figure \ref{f.3dFRAME}.

The \textbf{pericenter} of the frame simplex, where it comes closest the
origin, is a point of special interest. One finds
\begin{equation}
  \vec{\zeta}_{\text{pc}} = \frac{1}{\sum_b n_b}  \sum_a n_a \vec{\zeta}_a =
  \frac{D - 2}{\sum_b n_b}  \vec{\mathcal{Z}}_{\text{QG}} .\label{e:zeta min}
\end{equation}
This is the definition of the effective tower given in \cite{Castellano:2022bvr, Castellano:2021mmx}. There are two cases to consider (illustrated in Figure \ref{f.3dFRAME}):
\begin{enumerate}
  \item When $\sum_a n_a < \infty$, i.e., in a \textbf{}\textbf{Planckian
  phase} where $\Lambda_{\text{QG}} \sim M_{{\rm Pl,}D}$ is (parametrically) a Planck
  scale in the species dimension $D$, the pericenter lies in the
  \emph{interior} of the frame simplex, with
  \begin{equation}
    | \vec{\zeta}_{\text{pc}} | = \sqrt{\frac{\left( \sum_a n_a + d - 2
    \right)}{\left( \sum_a n_a \right) (d - 2)}},
  \end{equation}
  just like a tower vector for the decompactification of $\sum_a n_a$
  dimensions.
  
  \item When $\sum_a n_a = \infty$ ($n_k = \infty$), i.e., in a
  \textbf{stringy phase} where $\Lambda_{\text{QG}} \sim m_s$ is
  (parametrically) a string scale and the species dimension $D = \infty$, the
  pericenter lies on the \emph{boundary} of the frame simplex,
  \begin{equation}
    \vec{\zeta}_{\text{pc}} = \vec{\zeta}_k \qquad \Rightarrow \qquad |
    \vec{\zeta}_{\text{pc}} | = \frac{1}{\sqrt{d - 2}},
  \end{equation}
  which is the tower vector for the string oscillator modes.
\end{enumerate}
In either case, $| \vec{\zeta}_{\text{pc}} | \geqslant \frac{1}{\sqrt{d -
2}}$, suggesting a connection to the Sharpened Distance Conjecture \cite{Etheredge:2022opl}. This will be made more precise in
Section \ref{subsec:SDC}.

\subsubsection{The direction of the infinite-distance limit} \label{subsec:dir}

The above taxonomy rules hold asymptotically once a particular regular infinite-distance limit is chosen.
While it does not appear in the preceeding rules \eqref{eqn:taxonomyRules},
the \textbf{direction vector} $\hat{t}^i = \frac{d \phi^i}{d s}$
of the infinite-distance limit with respect to the tower vectors is also constrained. For instance, it must be
the case that $\hat{t} \cdot \vec{\zeta}_a > 0$ for each tower vector since
the corresponding tower becomes light in the infinite-distance limit in
question. In fact, since all the towers in question appear at or below the
species scale by assumption, the stronger constraint $\hat{t} \cdot
\vec{\zeta}_a \geqslant \hat{t} \cdot \vec{\mathcal{Z}}_{\text{QG}}$ must
hold. To analyze this constraint, assume for now that $\hat{t}$ lies entirely
within the principal plane. Then:
\begin{equation}
  \hat{t} = \sum_b \lambda_b  \vec{\zeta}_b,
\end{equation}
for some constants $\lambda_a$, $a = 1, \ldots, k$. Applying this ansatz to
the constraint $\hat{t} \cdot (\vec{\zeta}_a - \vec{\mathcal{Z}}_{\text{QG}})
\geqslant 0$ and using the taxonomy rules, we obtain:
\begin{equation}
  \frac{1}{n_a} \lambda_a \geqslant 0 .
\end{equation}
In a Planckian phase, all the towers lie strictly below the top-dimensional
Planck scale and also $n_a < \infty$ for all $a$, hence the $\lambda_a$ are
all positive, implying that $\hat{t}$ is a positive linear combination of the
principal tower vectors, i.e., it lies in the interior of $\text{cone} (\Delta) =
\text{cone} (\vec{\zeta}_1, \ldots, \vec{\zeta}_k)$.

In a stringy phase, we likewise obtain $\lambda_a > 0$ for all $a < k$, but
the above argument does not constrain $\lambda_k$ since $\vec{\zeta}_k =
\vec{\mathcal{Z}}_{\text{QG}}$. However, there is another constraint: since
the string theory is weakly coupled by assumption, the string scale must like
below the Planck scale in the top, $d_k$-dimensional theory. The vector
$\vec{\mathcal{Z}}_{\text{Pl}}$ associated to the latter scale satisfies the
taxonomy rules:
\begin{equation}
  \vec{\mathcal{Z}}_{\text{Pl}} \cdot \vec{\zeta}_a = \frac{1}{d - 2} -
  \frac{1}{\sum_{a < k} n_a + d - 2} \delta_{a k}, \qquad |
  \vec{\mathcal{Z}}_{\text{Pl}} |^2 = \frac{1}{d - 2} - \frac{1}{\sum_{a < k}
  n_a + d - 2} . \label{eqn:ZPl}
\end{equation}
These rules can be proven inductively as before, starting with
$\vec{\mathcal{Z}}_{\text{Pl}} = \vec{0}$ in the case $k = 1$ (since the
Planck scale is fixed in Planck units). The requirement that the string is
parametrically weakly coupled is then $\hat{t} \cdot
\vec{\mathcal{Z}}_{\text{QG}} > \hat{t} \cdot \vec{\mathcal{Z}}_{\text{Pl}}$.
This works out to:
\begin{equation}
  \hat{t} \cdot (\vec{\mathcal{Z}}_{\text{QG}} -
  \vec{\mathcal{Z}}_{\text{Pl}}) = \frac{1}{\sum_{a < k} n_a + d - 2}
  \lambda_k > 0 \qquad \Rightarrow \qquad \lambda_k > 0,
\end{equation}
so we conclude that $\hat{t}$ lies in the interior of $\text{cone} (\Delta)$,
as before.

If $\hat{t}$ does not lie in the principal plane then the above argument still
applies to its projection $\hat{t}_{\parallel}$ onto this plane. Thus, we
conclude that
\begin{equation}
  \fbox{$\begin{matrix}\text{Projected onto the principal plane, the  direction}\\\text{ vector $\hat t^i\equiv \frac {d\phi^i}{ds}$ lies in the interior of cone$(\Delta)$.}\end{matrix}$}
  \label{eqn:dirrule}
\end{equation}
This constrains the orientation of the frame simplex with respect to
$\hat{t}_{\|}$, completing the taxonomy rules. Note that the component of
$\hat{t}$ perpendicular to the principal plane is not fixed by the rules (in
either magnitude or direction).

\subsection{Connection with the Sharpened Distance
Conjecture}\label{subsec:SDC}

We can now connect our results to the Sharpened Distance Conjecture \cite{Etheredge:2022opl}. The
exponential rate at which each tower becomes light is $\alpha_a =
\vec{\zeta}_a \cdot \hat{t}$. If $\hat{t}$ lies within the principal plane,
then since it is inside $\text{cone} (\{ \vec{\zeta}_a \})$, simple geometric
considerations lead to:
\begin{equation}
  \alpha_{\max} = \max (\{ \alpha_a \}) \geqslant | \vec{\zeta}_{\text{pc}} |,
\end{equation}
where the inequality is saturated when $\hat{t} \propto
\vec{\zeta}_{\text{pc}}$. Since $| \vec{\zeta}_{\text{pc}} | \geqslant
\frac{1}{\sqrt{d - 2}}$ as noted above, this would imply the Sharpened Distance Conjecture in any regular infinite-distance limit.

However, if $\hat{t}$ does not lie within the principal plane then we obtain
the weaker constraint:
\begin{equation}\label{e.theta}
  \alpha_{\max} \geqslant \frac{\cos \theta}{\sqrt{d - 2}},
\end{equation}
where $\theta \in [0, \pi / 2)$ is the angle between $\hat{t}$ and the
principal plane. This creates a danger of violating the Sharpened Distance
Conjecture, especially in a stringy phase where $| \vec{\zeta}_{\text{pc}} | =
\frac{1}{\sqrt{d - 2}}$; then if $\theta \neq 0$ for $\hat{t} \propto
\vec{\zeta}_{\text{pc}}$, the Sharpened Distance Conjecture would be violated.

It is plausible that any infinite-distance limit in the landscape has $\theta = 0$, so that any limit with $\theta \neq 0$ resides in the swampland. However, this is difficult to prove rigorously, as we discuss in greater detail in Section \ref{subsec:parallel}. If $\theta \neq 0$ occurs in the landscape, then
the Sharpened Distance Conjecture does not follow from the Emergent String
Conjecture, even in the regular limits we have been considering.\footnote{Even
if $\theta \neq 0$ occurs in the landscape, the Sharpened Distance Conjecture
may still be satisfied, depending on the details. However, in this case the
\emph{connection} between the Sharpened Distance Conjecture and the ESC
becomes more tenuous, even in the absence of other complications such as
non-asymptotically empty decompactifications.}

\subsection{The structure of a duality frame}\label{subsec:dualityframe}

So far we have focused on a single, fixed infinite-distance limit, imposing
regularity conditions to simplify the physics. We now allow the infinite
distance limit to vary continuously. To be precise, two infinite-distance limits $\phi_0$, $\phi_1$ are \emph{continuously connected} if there is a
continuous family of paths $\phi^i_t (s), t \in [0, 1]$ such that $\phi^i_t
(s)$ is an infinite-distance limit for each value of $t$. As discussed in
Section \ref{subsec:generic}, we expect that a \emph{generic}
infinite-distance limit is regular, i.e., any irregular limit $\phi_0$ should
sit inside some continuous family $\phi_t$, $t \in [0, 1]$ such that $\phi_t$
is regular for $t > 0$. In other words, we expect irregular limits to be of measure zero in this continuous family of paths. If so, the space of continuously connected infinite
distance limits in a given theory can be understood by piecing together
continuous families of regular infinite-distance limits.

Thus, we consider what happens as a regular infinite-distance limit is
continuously varied. Each regular limit in the continuous family is
characterized by a frame simplex $\Delta$, species vector
$\vec{\mathcal{Z}}_{\text{QG}}$, and direction vector $\hat{t}$ satisfying the
taxonomy rules \eqref{eqn:taxonomyRules}, \eqref{eqn:dirrule}. However, with
the exception of the direction vector, the taxonomy rules are
\emph{rigid}, not allowing for continuous variations in, e.g., the shape
of the frame simplex. Thus, as the limit varies continuously, the frame
simplex and species vector remain fixed as long as the identification of the set of light towers remains the same,\footnote{In principle, the
tower/species vectors can rotate within the principal plane (and the principal
plane can rotate within the full tangent space) while still respecting the
taxonomy rules. However, such a rotation can be removed (up to possible
monodromies, if the continuous family of infinite-distance limits is not
simply connected) by a convenient choice of frame on the tangent space.} with
only the direction vector $\hat{t}$ continuously varying.

The variation $\delta \hat{t}$ in the direction vector can be decomposed into
components both (1) parallel and (2) perpendicular to the principal plane.
Note that, while we expect the direction vector $\hat{t}$ to lie wholly within
the principal plane, see Section \ref{subsec:parallel}, this does
\emph{not} imply that $\delta \hat{t}$ lies within the principal plane.
This is possible because the top-dimensional theory in the chain of
decompactifications may have moduli that can be turned on (such as NSNS moduli
in a stringy phase), generating a \emph{new} tower scale in the hierarchy,
and adding a dimension to the principal plane (increasing the rank by one).
Thus, regular limits with different ranks can be continuously connected.
However, because the moduli space is finite dimensional, there is some
\emph{maximum} size for the principal plane after which no additional
infinite-distance limits remain in the top-dimensional theory, and the rank
cannot increase further upon small variations in the direction of the limit. In other words, there is a maximum size for the principal plane for which the identification of the set of light towers remains fixed.
Such \textbf{full rank} limits are (locally) generic in the space of regular
limits and are described by the same \textbf{duality frame}, since they have the same microscopic identification of the species scale $\vec{\mathcal{Z}}_{\text{QG}}$.

Starting with a full-rank limit, we can vary $\hat{t}$ within the principal
plane. Since the moduli space metric is (asymptotically) flat in this plane,
geodesics are straight lines, and we can vary $\hat{t}$ in an arbitary
direction within the plane. This continues until one of the following failure
modes occurs:
\begin{enumerate}
  \item One of the decompactification limits is no longer asymptotically
  empty, i.e., the warped / Ricci-curved regions begin to grow at the same
  rate as the overall volume of the manifold.
  
  \item Two or more towers that are parametrically lighter than
  $\Lambda_{\text{QG}}$ degenerate.
  
  \item $\hat{t}$ reaches the edge of $\text{cone} (\Delta)$.
\end{enumerate}
The first two of these indicate a breakdown in regularity, which is largely
beyond the scope of the present paper. Nonetheless, when multiple KK towers
degenerate it is \emph{sometimes} possible to continue past the degenerate
locus to reach another continuous family of regular infinite-distance limits.
When this preserves the frame simplex, up to the natural reordering of the
hierarchy of tower scales due to the change in the direction vector, we say
that the degeneration is \textbf{ignorable}. For instance, this occurs for
compactifications on empty, direct product manifolds $X_m \times Y_n$ when the
hierarchy between the sizes of the two manifolds reverses. (Note that even
ignorable degenerations come with interesting additional physics, as discussed
in Section \ref{subsec:degenerate}.)

What happens when $\hat{t}$ reaches the edge of $\text{cone} (\Delta)$? In
this case, one or more of the tower (corresponding to the tower vectors
that do not lie on the edge in question) will get heavier than the species
scale.\footnote{In the case of a tower of string oscillator modes, this means
that the string coupling will go to $1$, i.e., the string scale will disappear
into the higher-dimensional Planck scale.} Then (up to ignorable
degenerations) we obtain another regular infinite-distance limit, but with a
lower-dimensional frame simplex/principal plane. This is precisely the
reverse of the process, discussed above, by which the principal plane can grow
in dimension.

\subsubsection{The species star}

We now suppose that no irregular infinite-distance limits (besides ignorable
degenerations) appear as we scan the direction vector $\hat{t}$ across the
interior of the frame simplex $\Delta$. As noted above, all of these limits
share the same underlying species-scale physics and can be thought of as
residing in a single duality frame. Approaching a boundary of $\Delta$, one or
more of the tower scales merges with the species scale, reducing the frame
simplex to one of its faces $\mathcal{F} \subset \Delta$,\footnote{In what
follows, a $p$-simplex $\Delta$ is represented as the set of its $p + 1$
linearly independent vertices, hence a $q$-face $\mathcal{F} \subseteq \Delta$
is a subset consisting of $q + 1$ of these vertices, where by convention we
exclude the ``$(- 1)$-simplex'' $\mathcal{F}= \emptyset$ from consideration.
For convenience, we use the notations $\vec{\zeta}_a \in
\mathcal{F}$ and $a \in \mathcal{F}$ interchangeably when the meaning is clear
from the context.} which is itself a lower-rank simplex generated by the
vectors of the towers that remain parametrically below the species scale. In
particular, $\mathcal{F}$ is the face of $\Delta$ in whose interior the
boundary point that we are approaching lies. On this boundary, the species
vector changes to
\begin{equation}
  \vec{\mathcal{Z}}_{\mathcal{F}} = \frac{1}{D_{\mathcal{F}} - 2} \sum_{a \in
  \mathcal{F}} n_a \vec{\zeta}_a, \label{eqn:Zface}
\end{equation}
where $D_{\mathcal{F}} = d + \sum_{b \in \mathcal{F}} n_b$ is the
corresponding species dimension, which is either the spacetime dimension in
which the corresponding Planck scale occurs or $D_{\mathcal{F}} = \infty$ if
the species scale is a perturbative string scale. Note that
$\vec{\mathcal{Z}}_{\mathcal{F}} = \vec{\mathcal{Z}}_{\text{QG}}$ if
$\mathcal{F}$ includes a string oscillator tower, whereas otherwise it is easy
to see that $\vec{\mathcal{Z}}_{\mathcal{F}}$ is distinct for each distinct
face $\mathcal{F} \subseteq \Delta$ of the frame simplex due to the linear
independence of the tower vectors.

Because the species vector changes there, one can think of each boundary of
the frame simplex $\Delta$ as representing a new duality frame, or perhaps
more accurately, the \emph{onset} of a new duality frame. For instance, in
the Planckian phase associated to M-theory on a rectangular two-torus, the
boundaries of the frame simplex correspond to infinite-distance limits in
which the nine-dimensional theory decompactifies to M-theory on a circle of
fixed radius. While one might call this the ``same'' duality frame, depending
on the radius of the circle this might better be thought of as decompactifying
to type IIA string theory at fixed string coupling. From either viewpoint,
these boundary limits are Planckian\footnote{This is true even in the type IIA
description because the string coupling is fixed in this limit, not going parametrically to
zero.} with species dimension $D_{\mathcal{F}} = 10$, reduced from species
dimension $D = 11$ for generic limits in the interior of $\Delta$.

The structure of these duality ``onsets'' is described by the set of species
vectors $\vec{\mathcal{Z}}_{\mathcal{F}}$ corresponding to the faces
$\mathcal{F} \subseteq \Delta$ of the frame simplex, including as a special
case the original species vector $\vec{\mathcal{Z}}_{\text{QG}} =
\vec{\mathcal{Z}}_{\Delta}$. These vectors are the vertices of a geometric object, which we will refer to as the \textbf{species star} $\Sigma$. We now state the properties of this object, later sketching the proofs of these statements.

\begin{figure}[h]
\begin{center}
\begin{subfigure}{0.54\textwidth}
\center
\includegraphics[width=.95\textwidth]{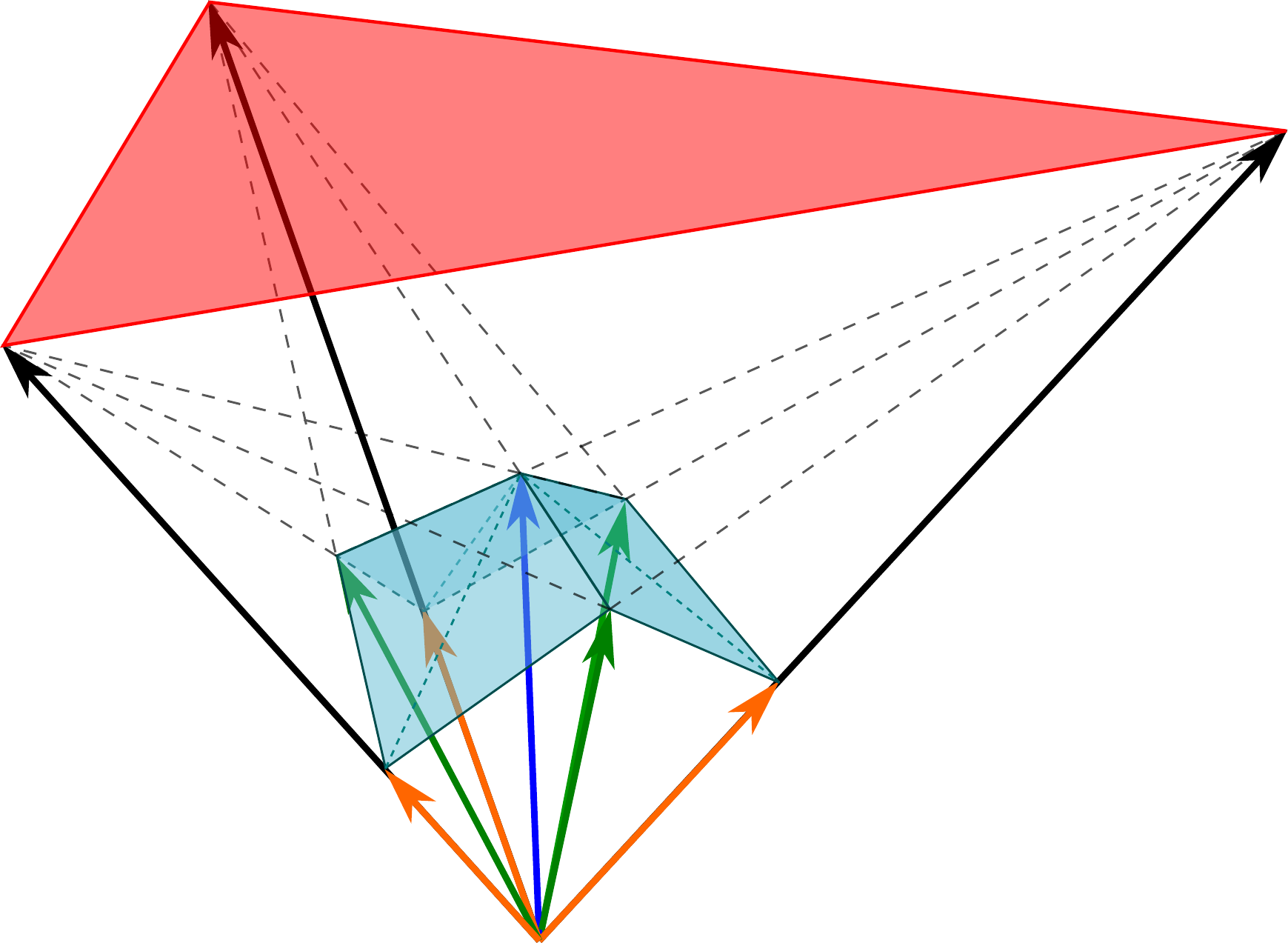}
\caption{Plankian phase.} \label{f.3dSTARch1}
\end{subfigure}
\begin{subfigure}{0.45\textwidth}
\center
\includegraphics[width=.95\textwidth]{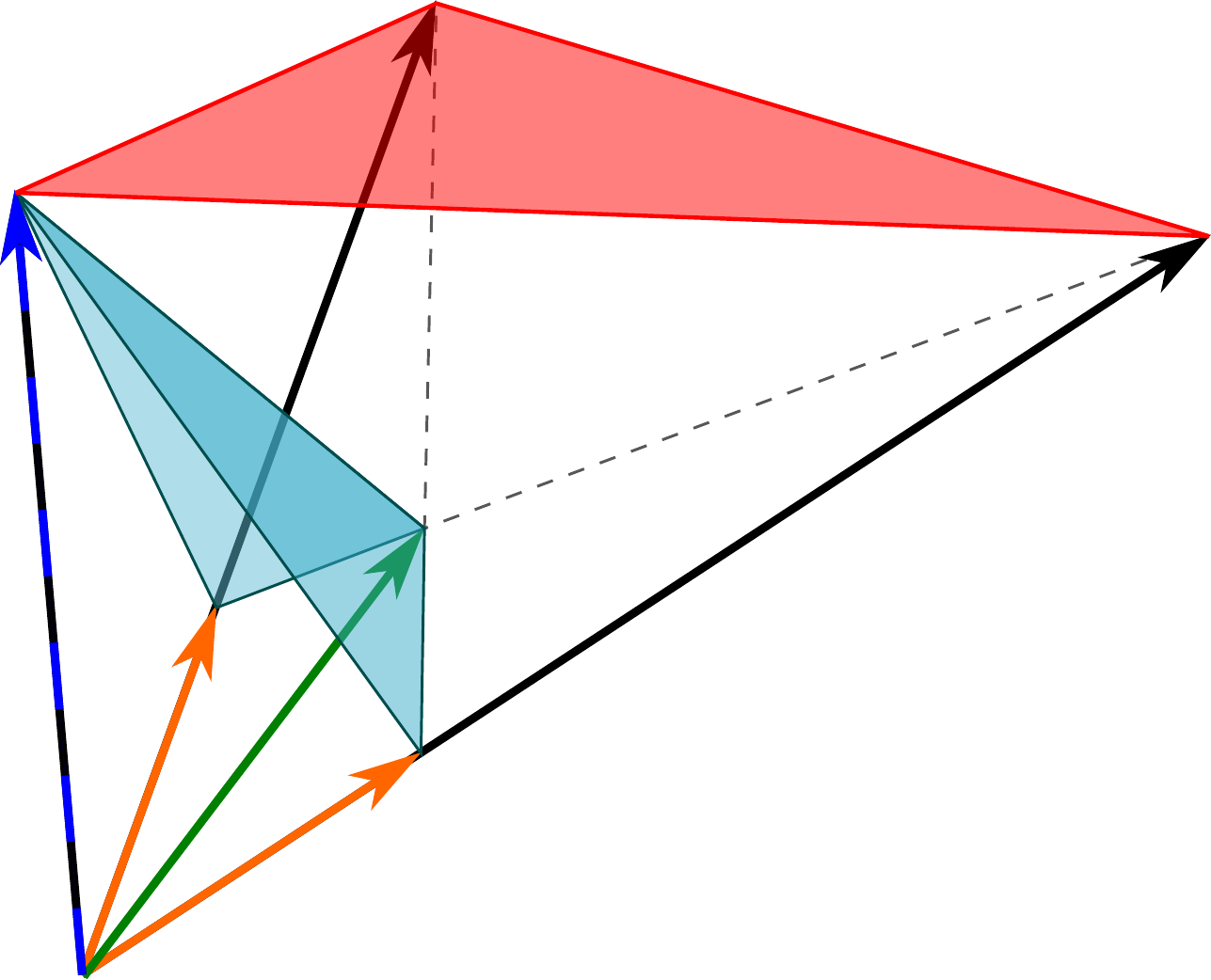}
\caption{Stringy phase} \label{f.3dSTARch2}
\end{subfigure}
\end{center}
\caption{The species stars (in light blue) for the rank-3 Planckian and stringy phases whose frame simplices are depicted in Figure \ref{f.3dFRAME}. The species vectors $\vec{\mathcal{Z}}_{\mathcal{F}}$ associated to the $0$, $1$, and $2$ faces of the frame simplex are depicted in orange, green, and blue, respectively. Note that in the stringy phase, $\vec{\mathcal{Z}}_{\text{Pl}}$ is the green arrow (associated to the facet of $\Delta$ omitting the string oscillator tower). The dashed lines illustrate the polar duality between the frame simplex and the species star.}
\label{f.3dSTAR}
\end{figure}

In a Planckian phase, the species star\footnote{In a simplicial complex, the
\emph{star} of a vertex consists of every simplex sharing that vertex. The
faces of the species star are not simplicial, but it admits a natural
triangulation with a simplex $\{ \vec{\mathcal{Z}}_{\mathcal{F}_1}, \ldots,
\vec{\mathcal{Z}}_{\mathcal{F}_p} \}$ for each inclusion sequence
$\mathcal{F}_1 \subset \mathcal{F}_2 \subset \cdots \subset \mathcal{F}_p$ of
faces of $\Delta$ such that $\mathcal{F}_{p - 1}$ does not include
a string oscillator tower. With this
triangulation, the species star is indeed the star of the vertex
$\vec{\mathcal{Z}}_{\text{QG}}$ (or, more correctly in a stringy phase, the star of the edge
joining $\vec{\mathcal{Z}}_{\text{QG}}$ with $\vec{\mathcal{Z}}_{\text{Pl}}$).}
consists of $k$ facets meeting
at their common vertex $\vec{\mathcal{Z}}_{\text{QG}}$ and ending on the
boundaries of $\cone(\Delta)$. In a stringy phase, the species star consists of $k - 1$
facets meeting along their common edge joining $\vec{\mathcal{Z}}_{\text{QG}}
= \vec{\zeta}_{\text{osc}}$ with $\vec{\mathcal{Z}}_{\text{Pl}} =
\frac{\sum_{a < k} n_a \vec{\zeta}_a}{d + \sum_{a < k} n_a - 2}$ (see~\eqref{eqn:ZPl}). Both cases are illustrated in Figure~\ref{f.3dSTAR}.

The geometry of the species star is such that its faces intersect the
boundaries of $\cone(\Delta)$ perpendicuarly, hence the pericenter of
each face (besides $\vec{\mathcal{Z}}_{\text{QG}}$) is also a vertex of
$\Sigma$ lying on the boundary of $\cone(\Delta)$. These vertices are the
species vectors $\vec{\mathcal{Z}}_{\mathcal{F}}$, where for a $p$-face
$\mathcal{F} \subseteq \Delta$ of the frame simplex not containing a string
oscillator tower, $\vec{\mathcal{Z}}_{\mathcal{F}}$ is the pericenter of a $(k
- p - 1)$-face of $\Sigma$. (When $\mathcal{F}$ contains a string oscillator
tower then $\vec{\mathcal{Z}}_{\mathcal{F}} = \vec{\mathcal{Z}}_{\text{QG}}$
as noted previously.)

This gives $\Sigma$ a structure that is combinatorically dual to $\Delta$:
each $p$-face of $\mathcal{F}$ of $\Delta$ (not containing a string oscillator
tower) corresponds to the $(k - p - 1)$-face of $\Sigma$ with pericenter
$\vec{\mathcal{Z}}_{\mathcal{F}}$. Indeed, geometrically this is just a
standard (polar) duality in disguise. In particular, consider the polar set of
the frame simplex:
\begin{equation}
  \polar(\Delta) \equiv \biggl\{ \vec{\mathcal{Z}} \biggm| \vec{\mathcal{Z}}
  \cdot \vec{\zeta}_a \leqslant \frac{1}{d - 2}, a = 1, \ldots, k \biggr\} .
\end{equation}
This region is bounded by $k$ semi-infinite facets meeting at their common
vertex $\vec{\mathcal{Z}}_{\text{QG}}$. The species star $\Sigma$ is precisely
the portion of this boundary that lies within $\cone(\Delta)$:
\begin{equation}
  \Sigma = \partial (\polar(\Delta)) \cap \cone(\Delta) .
  \label{eqn:speciesstar}
\end{equation}
Note that the reason $\Sigma$ has only $k - 1$ facets in a stringy phase is
because in this case one of the facets of $\partial (\polar(\Delta))$
lies wholly outside $\cone(\Delta)$.

The aforementioned properties of the species star can be proven using the following taxonomy rules:
\begin{align}
  \vec{\mathcal{Z}}_{\mathcal{F}} \cdot \vec{\zeta}_a &= \frac{1}{d - 2} -
  \frac{1}{D_{\mathcal{F}} - 2} \delta_{a \notin \mathcal{F}}, 
  \label{eqn:genpattern}\\
  \vec{\mathcal{Z}}_{\mathcal{F}} \cdot \vec{\mathcal{Z}}_{\mathcal{F}'} &=
  \frac{1}{d - 2} - \frac{D_{\mathcal{F} \cup \mathcal{F}'} -
  2}{(D_{\mathcal{F}} - 2) (D_{\mathcal{F}'} - 2)},  \label{eqn:speciesstarRule}
\end{align}
which follow by direct calculation from \eqref{eqn:Zface} and the taxonomy
rules~\eqref{eqn:taxonomyRules}. Here $\mathcal{F} \cup \mathcal{F}'$
denotes the face of $\Delta$ whose vertex set is the union of those of
$\mathcal{F}$ and $\mathcal{F}'$ (such a face always exists because $\Delta$
is a simplex) and $D_{\mathcal{F} \cup \mathcal{F}'} = d + \sum_{b \in
\mathcal{F} \cup \mathcal{F}'} n_b$ is the associated species dimension.

We now sketch a few details of the proof, as they naturally introduce the
subject of \emph{recursion}, to be discussed in~\S\ref{sec:framerecursion}. Define the species star as
$\Sigma [\Delta] \equiv \partial (\polar(\Delta)) \cap \cone(\Delta)$. To show that its vertices are indeed the species vectors $\{
\vec{\mathcal{Z}}_{\mathcal{F}} |\mathcal{F} \subseteq \Delta \}$, consider
the related object $\tilde{\Sigma} [\Delta] \equiv \polar(\Delta) \cap
\cone(\Delta)$, a portion of whose boundary is $\Sigma [\Delta]$. One
can show by induction on the rank $k$ of $\Delta$ that the vertices of
$\tilde{\Sigma} [\Delta]$ are $\{ \vec{0} \} \cup \{
\vec{\mathcal{Z}}_{\mathcal{F}} |\mathcal{F} \subseteq \Delta \}$, where the
base case $k = 1$ is easy to check and the inductive step proceeds by noting
that each vertex lies on one or more facets, then considering each of the
facets in turn. The facets of $\tilde{\Sigma} [\Delta]$ are of two types:
\begin{enumerate}
  \item ``Inner'' facets, which are facets of $\cone(\Delta)$,
  intersected with $\polar(\Delta)$, and
  \item ``Outer'' facets, which are facets of $\polar(\Delta)$,
  intersected with $\cone(\Delta)$.
\end{enumerate}
In the first (inner) case, there is a facet for each vertex $\vec{\zeta}_a \in
\Delta$, defined by the equations:
\begin{equation}
  \vec{\mathcal{Z}} \in \cone(\Delta_{\bar{a}}) \qquad \text{and}
  \qquad \vec{\mathcal{Z}} \cdot \vec{\zeta}_b \leqslant \frac{1}{d - 2},
  \quad \forall b,
\end{equation}
where $\Delta_{\bar{a}} = \Delta \setminus \{ \vec{\zeta}_a \}$ is the
rank-$(k - 1)$ frame simplex obtained by omitting the tower vector
$\vec{\zeta}_a$. One can check that the condition $\vec{\mathcal{Z}} \cdot
\vec{\zeta}_a \leqslant \frac{1}{d - 2}$ is a consequence of the other
conditions, at which point these equations reduce to the equations defining
$\tilde{\Sigma} [\Delta_{\bar{a}}]$. Thus, by the inductive assumption, the
vertices of this facet are $\{ \vec{0} \} \cup \{
\vec{\mathcal{Z}}_{\mathcal{F}} |\mathcal{F} \subseteq \Delta \setminus \{
\vec{\zeta}_a \} \}$.

In the second (outer) case, there is a facet for each vertex $\vec{\zeta}_a
\in \Delta$ that is not a string oscillator tower,\footnote{Although
$\polar(\Delta)$ does have a facet associated to a string oscillator
tower $\vec{\zeta}_a = \vec{\zeta}_{\text{osc}}$, this facet intersects
$\cone(\Delta)$ at a single point. This can be shown by following the
same steps as below, resulting in a frame simplex $\Delta_a'$ that is formally
in infinite spacetime dimension $D_a = \infty$. As a result, $\tilde{\Sigma}
[\Delta_a']$ is simply the intersection between $\cone(\Delta)$ and its
polar cone, which is a single point, $\tilde{\Sigma} [\Delta_a'] = \{ \vec{0}
\}$, implying that the facet of $\polar(\Delta)$ in question intersects
$\cone(\Delta)$ at the single point $\vec{\mathcal{Z}}_a =
\vec{\mathcal{Z}}_{\text{QG}}$.} now defined by the equations:
\begin{equation}
  \vec{\mathcal{Z}} \in \cone(\Delta) \qquad \text{and} \qquad
  \vec{\mathcal{Z}} \cdot \vec{\zeta}_a = \frac{1}{d - 2} \qquad \text{and}
  \qquad \vec{\mathcal{Z}} \cdot \vec{\zeta}_b \leqslant \frac{1}{d - 2},
  \quad \forall b \neq a. \label{eqn:outerFacet}
\end{equation}
Let $\vec{\mathcal{Z}}_a \equiv \frac{n_a}{d + n_a - 2}  \vec{\zeta}_a =
\vec{\mathcal{Z}}_{\{ \vec{\zeta}_a \}}$ be the species vector associated to
the vertex $\vec{\zeta}_a \in \Delta$, with associated species dimension $D_a
\equiv d + n_a$. Non-trivially, applying \eqref{eqn:genpattern},
\eqref{eqn:speciesstarRule} one can rewrite the conditions \eqref{eqn:outerFacet}
as
\begin{equation}
  \vec{\mathcal{Z}} - \vec{\mathcal{Z}}_a \in \cone(\Delta_a') \qquad
  \text{and} \qquad (\vec{\mathcal{Z}} - \vec{\mathcal{Z}}_a) \cdot
  (\vec{\zeta}_b - \vec{\mathcal{Z}}_a) \leqslant \frac{1}{D_a - 2}, \quad
  \forall b \neq a, \label{eqn:facetRecurse}
\end{equation}
where $\Delta_a' \equiv \{ \vec{\zeta}_b - \vec{\mathcal{Z}}_a |b \neq a \}$
is a particular $(k - 2)$-simplex and again one of the original conditions
(part of the set $\vec{\mathcal{Z}} \in \cone(\Delta)$) turns out to be
redundant and has been dropped. Applying \eqref{eqn:genpattern},
\eqref{eqn:speciesstarRule} once again, we see that the shifted tower and species
vectors $\vec{\zeta}_b' \equiv \vec{\zeta}_b - \vec{\mathcal{Z}}_a$ for $b
\neq a$ and $\vec{\mathcal{Z}}_{\text{QG}}' \equiv
\vec{\mathcal{Z}}_{\text{QG}} - \vec{\mathcal{Z}}_a$ satisfy the taxonomy
rules~\eqref{eqn:taxonomyRules} associated to the simplex $\Delta_a'
= \{ \vec{\zeta}_b' |b \neq a \}$ in spacetime dimension $D_a$:
\begin{equation}
  \vec{\zeta}_b' \cdot \vec{\zeta}_c' = \frac{1}{D_a - 2} + \frac{1}{n_b}
  \delta_{b c}, \qquad \vec{\zeta}_b' \cdot \vec{\mathcal{Z}}_{\text{QG}}' =
  \frac{1}{D_a - 2}, \qquad |\vec{\mathcal{Z}}_{\text{QG}}'|^2 = \frac{1}{D_a - 2} - \frac{1}{D - 2}.
  \label{eqn:towerRecurse}
\end{equation}
Moreover, \eqref{eqn:facetRecurse} is equivalent to
\begin{equation}
  \vec{\mathcal{Z}} \in \tilde{\Sigma} [\Delta_a'] + \vec{\mathcal{Z}}_a .
\end{equation}
By the inductive assumption, the vertices of $\tilde{\Sigma} [\Delta_a']$ are
$\{ \vec{0} \} \cup \{ \vec{\mathcal{Z}}'_{\mathcal{F}'} |\mathcal{F}'
\subseteq \Delta_a' \}$. Examining \eqref{eqn:Zface}, one finds that:
\begin{equation}
  \vec{\mathcal{Z}}'_{\mathcal{F}'} + \vec{\mathcal{Z}}_a =
  \vec{\mathcal{Z}}_{\mathcal{F}},
\end{equation}
where $\mathcal{F}$ is the face of $\Delta$ with vertices $\vec{\zeta}_a$ plus
$\vec{\zeta}_b = \vec{\zeta}_b' + \vec{\mathcal{Z}}_a$ for each vertex
$\vec{\zeta}_b' \in \mathcal{F}'$. Likewise, $\vec{0} + \vec{\mathcal{Z}}_a =
\vec{\mathcal{Z}}_{\mathcal{F}}$ where $\mathcal{F}= \{ \vec{\zeta}_a \}$.
Thus, the vertices of the facet of $\tilde{\Sigma} [\Delta]$ in question are
$\{ \vec{\mathcal{Z}}_{\mathcal{F}} | \{ \vec{\zeta}_a \} \subseteq
\mathcal{F} \subseteq \Delta \}$.

Combining the inner and outer cases, we conclude that the vertices of
$\tilde{\Sigma} [\Delta]$ are $\{ \vec{0} \} \cup \{
\vec{\mathcal{Z}}_{\mathcal{F}} |\mathcal{F} \subseteq \Delta \}$. Since the
species star $\Sigma [\Delta] \equiv \partial (\polar(\Delta)) \cap
\cone(\Delta)$ consists of the outer facets of $\tilde{\Sigma}
[\Delta]$, its vertices are $\{ \vec{\mathcal{Z}}_{\mathcal{F}} |\mathcal{F}
\subseteq \Delta \}$ as claimed.

Other properties of the species star follow more immediately from the rules
\eqref{eqn:genpattern}, \eqref{eqn:speciesstarRule}. For instance, let
$\mathcal{F}$ be a $p$-face of $\Sigma$ not containing a string oscillator
tower. As a special case of \eqref{eqn:speciesstarRule} we obtain:
\begin{equation}
  \vec{\mathcal{Z}}_{\mathcal{F}} \cdot \vec{\mathcal{Z}}_{\mathcal{F}'} =
  \frac{1}{d - 2} - \frac{1}{D_{\mathcal{F}} - 2} \qquad \text{when} \qquad
  \mathcal{F} \subseteq \mathcal{F}' . \label{eqn:subfaceRule}
\end{equation}
Equivalently, $\vec{\mathcal{Z}}_{\mathcal{F}} \cdot
(\vec{\mathcal{Z}}_{\mathcal{F}'} - \vec{\mathcal{Z}}_{\mathcal{F}}) = 0$ when
$\mathcal{F}' \supseteq \mathcal{F}$, which implies that the species vectors
$\vec{\mathcal{Z}}_{\mathcal{F}'}$ with $\mathcal{F}' \supseteq \mathcal{F}$
all lie in the plane with pericenter $\vec{\mathcal{Z}}_{\mathcal{F}}$.
Indeed, retracing the above inductive argument, these are the remaining
vertices of a $(k - p - 1)$-face of $\Sigma$. Since the entire face lies in
the $\vec{\mathcal{Z}}_{\mathcal{F}}$-pericenter plane, the pericenter of this
face is $\vec{\mathcal{Z}}_{\mathcal{F}}$, as claimed.

\subsubsection{Partial decompactification and recursion} \label{sec:framerecursion}

The modified frame simplex $\Delta_a' = \{ \vec{\zeta}_b - \vec{\mathcal{Z}}_a
|b \neq a \}$ that appeared in the above inductive arugment has a simple
physical interpretation: it is the frame simplex after \emph{partial
decompactification}, where we send the corresponding KK scale $m_a$ to zero in
$(D_a = d + n_a)$-dimensional Planck units.

To be precise, choose a direction vector that is nearly parallel to
$\vec{\zeta}_a$, $\hat{t} \propto \hat{\zeta}_a + \varepsilon \hat{t}'$ for
$\hat{t}' \cdot \hat{\zeta}_a = 0$. Then the rate at which each tower becomes
light in $d$-dimensional Planck units is
\begin{equation}
  \vec{\zeta}_b \cdot \hat{t} = \frac{1}{| \vec{\zeta}_a |}  \biggl( \frac{1}{d
  - 2} + \frac{1}{n_a} \delta_{a b} \biggr) + \varepsilon \vec{\zeta}_b \cdot
  \hat{t}' + O (\varepsilon^2), \qquad \vec{\mathcal{Z}}_{\text{QG}} \cdot
  \hat{t} = \frac{1}{(d - 2) | \vec{\zeta}_a |} + \varepsilon
  \vec{\mathcal{Z}}_{\text{QG}} \cdot \hat{t}' + O (\varepsilon^2) .
\end{equation}
Thus the towers and the species scale all become light in this limit, but the
KK tower $\vec{\zeta}_a$ becomes light more quickly than the others. We now
rewrite this in $(D_a = d + n_a)$-dimensional Planck units. Since
\begin{equation}
  \vec{\zeta}_b' \equiv - \vec{\nabla} \frac{m_b}{M_{\text{Pl}, D_a}} = -
  \vec{\nabla} \frac{m_b}{M_{\text{Pl}, d}} + \vec{\nabla} \frac{M_{\text{Pl},
  D_a}}{M_{\text{Pl}, d}} = \vec{\zeta}_b - \vec{\mathcal{Z}}_a, \quad
  \text{and} \quad \vec{\mathcal{Z}}_{\text{QG}}' =
  \vec{\mathcal{Z}}_{\text{QG}} - \vec{\mathcal{Z}}_a,
\end{equation}
we find:
\begin{equation}
  \vec{\zeta}_b' \cdot \hat{t} = \frac{1}{n_a | \vec{\zeta}_a |} \delta_{a b}
  + \varepsilon \vec{\zeta}_b \cdot \hat{t}' + O (\varepsilon^2), \qquad
  \vec{\mathcal{Z}}_{\text{QG}}' \cdot \hat{t} = \varepsilon
  \vec{\mathcal{Z}}_{\text{QG}} \cdot \hat{t}' + O (\varepsilon^2) .
\end{equation}
Thus, in $(D_a = d + n_a)$-dimensional Planck units only the tower
$\vec{\zeta}_a$ becomes light quickly, whereas the other towers and the
species scale become light much more slowly when $\varepsilon \ll 1$ (or not
at all, depending on the choice of $\hat{t}'$).

This allows us to separate scales, partially decompactifying to $D_a = d +
n_a$ dimensions while keeping track of a further, ``slow'' infinite-distance
limit in the resulting theory. This ``slow'' limit will be regular if the full
infinite-distance limit we started with is regular, with tower vectors
$\vec{\zeta}_b' = \vec{\zeta}_b - \vec{\mathcal{Z}}_a$, $b \neq a$ and species
vector $\vec{\mathcal{Z}}_{\text{QG}}' = \vec{\mathcal{Z}}_{\text{QG}} -
\vec{\mathcal{Z}}_a$. Thus, $\Delta_a' = \{ \vec{\zeta}_b' |b \neq a \}$ is
the resulting tower simplex. Referring to \eqref{eqn:towerRecurse}, we see that
\begin{equation}
  \vec{\zeta}_b' \cdot \vec{\zeta}_c' = \frac{1}{D_a - 2} + \frac{1}{n_b}
  \delta_{b c}, \qquad \vec{\zeta}_b' \cdot \vec{\mathcal{Z}}_{\text{QG}}' =
  \frac{1}{D_a - 2}, \qquad |\vec{\mathcal{Z}}_{\text{QG}}'|^2 = \frac{1}{D_a - 2} - \frac{1}{D - 2},
\end{equation}
so $\Delta_a'$ satisfies the taxonomy rules~\eqref{eqn:taxonomyRules} in $D_a$ spacetime dimensions.\footnote{One
can also check that $\hat{t}' \in \cone(\Delta_a')$ as a result of
$\hat{t} \in \cone(\Delta)$.}

More generally, one can consider a direction vector that is infinitesimally
close to a point in the interior of a face $\mathcal{F} \subset \Delta$ of the
full frame simplex $\Delta$. As above, such a limit has both ``fast'' and
``slow'' components. The fast component is described by the frame simplex
$\mathcal{F}$ with tower vectors $\vec{\zeta}_a$, $a \in \mathcal{F}$ and
species vector $\vec{\mathcal{Z}}_{\mathcal{F}}$, as measured in
$d$-dimensional Planck units. This limit describes the decompactification to
$D_{\mathcal{F}} = d + \sum_{a \in \mathcal{F}} n_a$ dimensions. Following
this decompactification, the slow component remains, which is described by the
frame simplex $\Delta_{\mathcal{F}}'$ with tower vectors $\vec{\zeta}_b' =
\vec{\zeta}_b - \vec{\mathcal{Z}}_{\mathcal{F}}$, $b \notin \mathcal{F}$ and
species vector $\vec{\mathcal{Z}}_{\text{QG}}' = \vec{\mathcal{Z}}_{\text{QG}}
- \vec{\mathcal{Z}}_{\mathcal{F}}$ as measured in
$D_{\mathcal{F}}$-dimensional Planck units. As in the above special case, one
can verify that $\Delta_{\mathcal{F}}'$ satisfies the taxonomy rules
\eqref{eqn:taxonomyRules} in $D_{\mathcal{F}}$ spacetime dimensions.
Indeed, such \emph{recursion relations}\footnote{These relations have a
physical origin, but correspond to interesting geometric facts. For instance,
$\Delta$ can be assembled from its ``fast'' and ``slow'' components
$\mathcal{F}$ and $\Delta_{\mathcal{F}}'$ by orienting the two simplices in
orthogonal planes and then translating the origin of plane containing
$\Delta_{\mathcal{F}}'$ to the point $\vec{\mathcal{Z}}_{\mathcal{F}}$ in the
plane containing $\mathcal{F}$.} are built into the derivation of the taxonomy
rules. We revisit these recursion relations in \S\ref{sec:polytoperecursion}.

\subsection{Connection with the tower-species pattern}

The taxonomic rule \eqref{eq:pattern} corresponds to the tower species pattern between the light tower of states and the species scale observed in \cite{Castellano:2023stg}. Evidence from plethora of string theory compactifications was provided in \cite{Castellano:2023jjt}. Here, we have re-derived it from bottom-up under the assumptions\footnote{The assumptions used in this paper to derive the pattern from bottom-up are analogous to the bottom-up conditions already given in \cite{Castellano:2023jjt}. The relation between the condition on the existence of bound states \cite{Castellano:2023jjt} and the Emergent String Conjecture is explained in Section \ref{s.bound}, while the other two conditions  \cite{Castellano:2023jjt} are analogous to the definition of a regular limit in this paper.}  outlined in Section \ref{s.inf dist str}. However, it could be that this pattern applies more generally (for instance, some of the examples in  \cite{Castellano:2023jjt} included irregular limits for which several KK towers degenerate due to non-trivial dependence on the complex structure moduli of the compactification). Interestingly, this pattern is not completely independent from the taxonomic rule of the towers as we explain in the following.

A special case of \eqref{eqn:genpattern} is
\begin{equation}
  \vec{\mathcal{Z}}_{\mathcal{F}} \cdot \vec{\zeta}_a = \frac{1}{d - 2},
  \qquad \text{when $a \in \mathcal{F}$.} \label{eqn:pattern1}
\end{equation}
This extends the tower-species pattern \cite{Castellano:2023stg, Castellano:2023jjt} to every face of the
frame simplex. Recall \eqref{eqn:Zface} as well:
\begin{equation}
  \vec{\mathcal{Z}}_{\mathcal{F}} = \frac{1}{D_{\mathcal{F}} - 2} \sum_{a \in
  \mathcal{F}} n_a \vec{\zeta}_a . \label{eqn:pattern2}
\end{equation}
Physically, this can be understood as dictating how the species scale is
controlled by the relevant tower scales.

Interestingly, \eqref{eqn:pattern1}, \eqref{eqn:pattern2} together imply the
taxonomy rules \eqref{eqn:taxonomyRules}. To see this, first consider the case
where $\mathcal{F}$ is a single vertex $a$. Then,
\begin{equation}
  \vec{\mathcal{Z}}_a = \frac{n_a  \vec{\zeta}_a}{d + n_a - 2}, \qquad
  \vec{\mathcal{Z}}_{\mathcal{F}} \cdot \vec{\zeta}_a = \frac{1}{d - 2} \qquad
  \Rightarrow \qquad | \vec{\zeta}_a |^2 = \frac{d + n_a - 2}{n_a  (d - 2)} =
  \frac{1}{d - 2} + \frac{1}{n_a} .
\end{equation}
Next, let $\mathcal{F}$ be the edge between vertices $a$ and $b$, in which
case:
\begin{equation}
  \vec{\mathcal{Z}}_{a b} = \frac{n_a  \vec{\zeta}_a + n_b  \vec{\zeta}_b}{d +
  n_a + n_b - 2}, \qquad \vec{\mathcal{Z}}_{a b} \cdot \vec{\zeta}_a =
  \frac{1}{d - 2} \qquad \Rightarrow \qquad  \vec{\zeta}_a \cdot \vec{\zeta}_b
  = \frac{1}{d - 2} .
\end{equation}
Finally, letting $\mathcal{F}$ be the entire frame simplex, we find that
$\vec{\mathcal{Z}}_{\text{QG}} = \vec{\mathcal{Z}}_{\mathcal{F}}$ satisfies:
\begin{equation}
  \vec{\mathcal{Z}}_{\text{QG}} = \frac{1}{D - 2} \sum_a n_a \vec{\zeta}_a
  \qquad \Rightarrow \qquad \vec{\mathcal{Z}}_{\text{QG}} \cdot \vec{\zeta}_a
  = \frac{1}{d - 2}, \qquad | \vec{\mathcal{Z}}_{\text{QG}} |^2 = \frac{1}{d -
  2} - \frac{1}{D - 2},
\end{equation}
which completes the rederivation of the taxonomy rules
\eqref{eqn:taxonomyRules}. 

More intuitively, we can explain the above derivation as follows. The definition of the species scale in terms of the light towers of states implies that the tower vector $\vec \zeta_b$ with $b \not\in \mathcal{F}$ projects into the individual species $\vec{\mathcal{Z}}_\mathcal{F}$ when moving along the direction of a tower in $a \in \mathcal{F}$ (see dotted black lines in Figure \ref{f.3dSTAR}). This is because  the states from the tower $\vec \zeta_b$ must enter the EFT and become lighter than the species scale $\vec{\mathcal{Z}}_a$ as we move away from the facet $\mathcal{F}$, such that its contribution lowers the species scale to yield $\vec{\mathcal{Z}}_{\text{QG}}$.  One therefore gets that $\vec{\zeta}_a\cdot\vec{\zeta}_b=\vec{\zeta}_a\cdot\vec{\mathcal{Z}}_{\mathcal{F}}=\frac{1}{d-2}$.

Thus, the tower-species pattern combined with the definition of the species scale in terms of the towers, as encoded in
\eqref{eqn:Zface}, implies the rest of the taxonomy rules. Note, however, that this
presupposes the ability to explore infinite-distance limits directed along
every boundary of the frame simplex, so this reasoning does not go through if
we cannot reach every boundary due, e.g., to the appearance of irregular
infinite-distance limits (other than ignorable degenerations). Our taxonomy
rules, however, apply to any regular infinite-distance limit regardless of
whether irregular limits appear for other directions within the frame simplex,
so in this sense they contain more information than the pattern.

\subsection{Combining duality frames\label{sec:dualities}}

We now consider what happens when the direction vector $\hat{t}$ moves
\emph{outside} of the original frame simplex. This corresponds in many
examples to the familiar concept of a \emph{duality},\footnote{Dualities
may also occur when moving through a locus where the infinite-distance limit
becomes irregular, see Section \ref{subsec:degenerate} for further
discussion.} since this will bring us to consider a new frame simplex associated to a different duality frame where the nature of the species scale may change. In what follows, we describe how in certain cases these frame simplices can be glued together so that the moduli space is divided into subregions corresponding to different perturbative descriptions of the theory. At the interface, the different descriptions are related to each other by duality transformations.

As we have seen, when $\hat{t}$ reaches a boundary of the frame simplex
$\Delta$, the principal plane reduces in dimension, where the new frame
simplex is the face $\mathcal{F} \subset \Delta$ of the original frame simplex
in whose interior the boundary point in question lies. Physically, this
corresponds to ``turning off'' the portion of the infinite-distance limit that
lies in the moduli space $\mathcal{M}_{d_i}$ of a higher-dimensional theory in
the chain of decompactifications, so that on the boundary of the frame simplex
we decompactify to a fixed point in the moduli space of this
higher-dimensional theory.\footnote{Note that this ``switching off'' process
is indeed continuous in the space of infinite-distance limits of the
lower-dimensional theory. It amounts to reducing the speed at which the limit
is taken in the moduli space of the higher-dimensional theory relative to the
speed at which the intervening dimensions decompactify until this speed
reaches 0, at which point the limit goes to a fixed point in the moduli space
of the higher-dimensional theory.}

\begin{figure}[h]
\begin{center}
\begin{subfigure}{0.52\textwidth}
\center
\includegraphics[width=.95\textwidth]{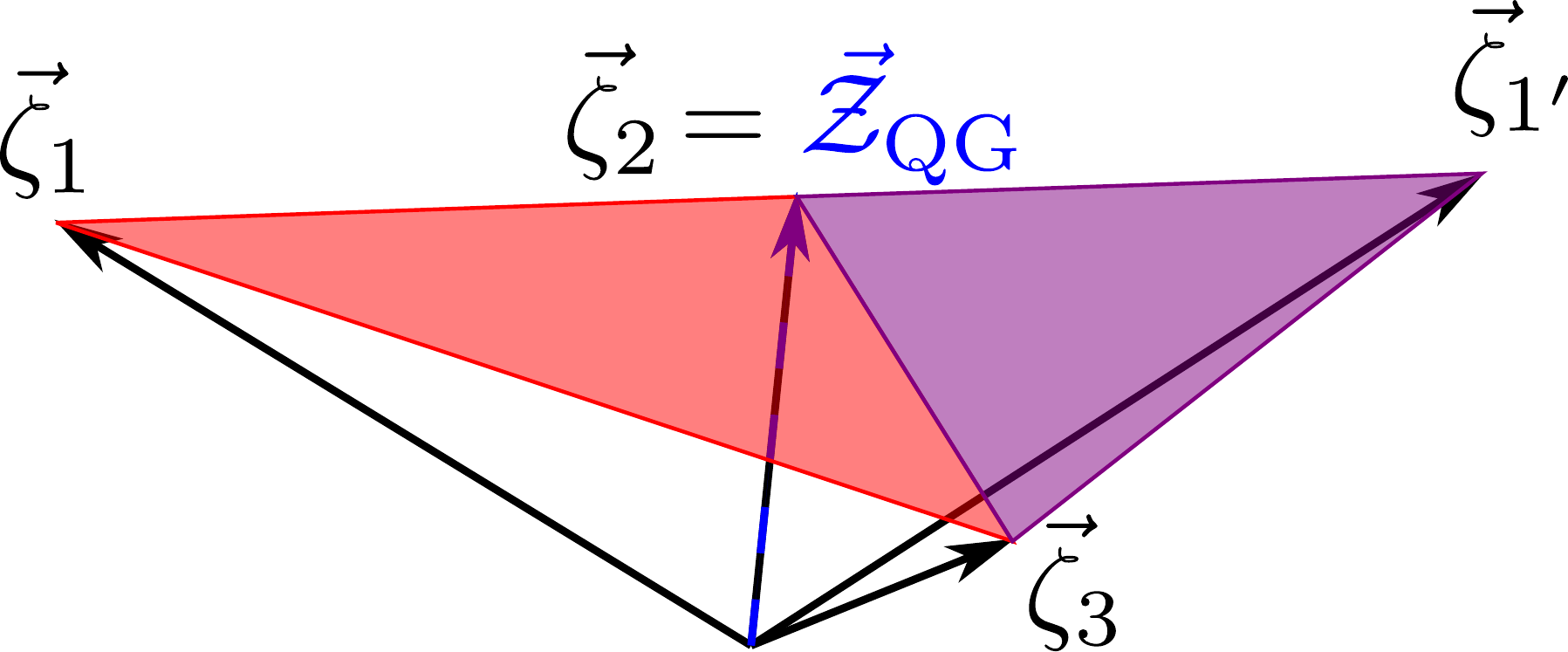}
\vspace{0.3cm}
\caption{Two duality frames with the same $\vec{\mathcal{Z}}_{\rm QG}$.} \label{f.glue2}
\end{subfigure}
\begin{subfigure}{0.47\textwidth}
\center
\includegraphics[width=.95\textwidth]{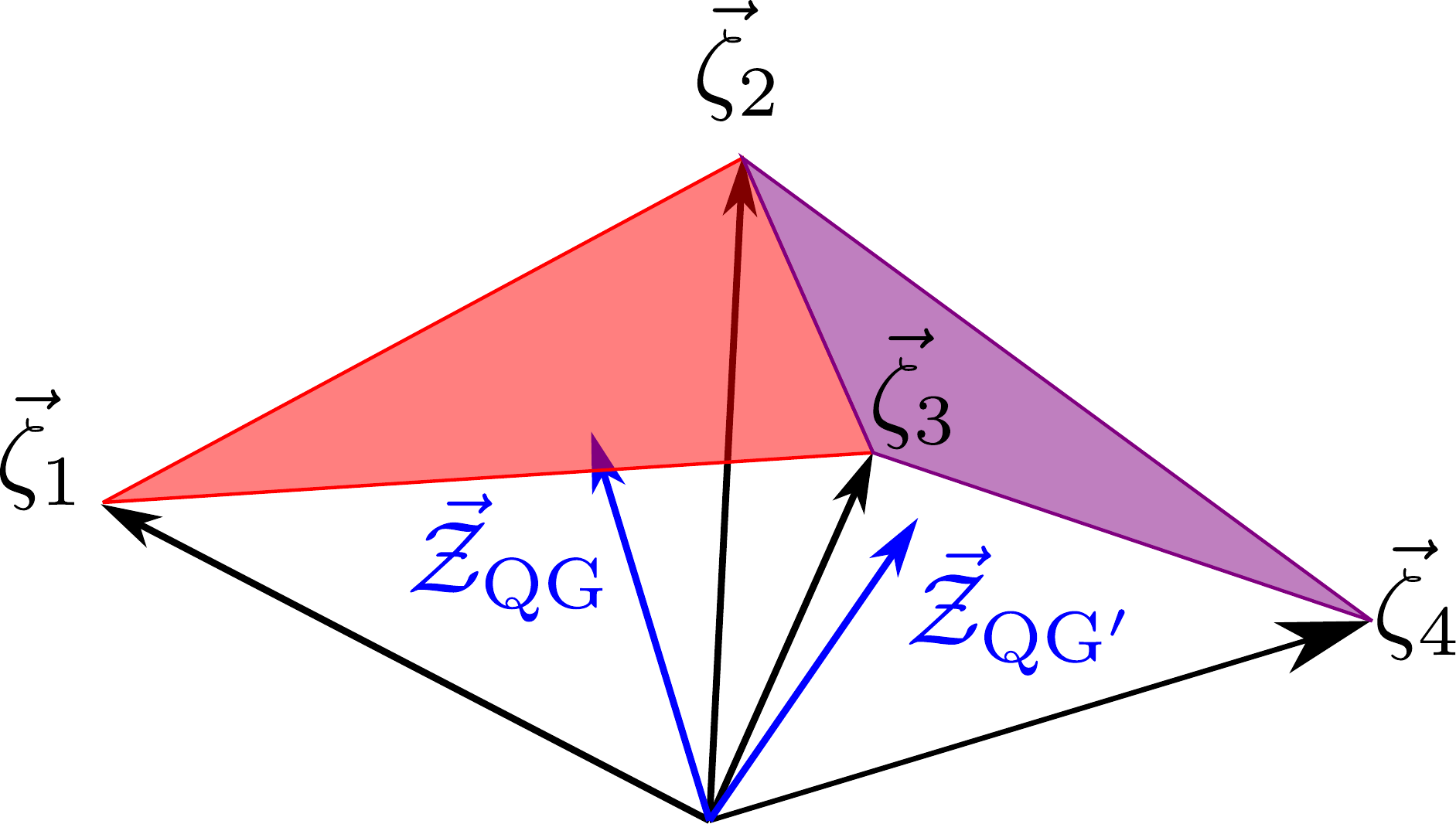}
\caption{Two duality frames with different $\vec{\mathcal{Z}}_{\rm QG}$.} \label{f.glue1}
\end{subfigure}
\end{center}
\caption{A duality between two different duality frames can be represented by ``gluing'' the frame simplices for these frames together. The two frame simplices can feature \subref{f.glue2} the same species vector or \subref{f.glue1} two different species vectors. The former only occurs in stringy phases and corresponds to a T-duality of the perturbative string in question. The latter can be thought of as a generalized S-duality; in particular, one of the towers reaches the Planck scale on the boundary between the two phases, suggesting the onset of strong coupling on general emergence grounds. 
}
\label{f.glue}
\end{figure}

To proceed ``through'' the boundary of the original frame simplex, we simply turn on a \emph{different} infinite-distance limit in the moduli space $\mathcal{M}_{d_i}$. Assuming that the resulting infinite-distance limit is
regular (as expected for a generic infinite-distance limit, see
Section \ref{subsec:generic}), we obtain another frame simplex $\Delta'$
of which $\mathcal{F}$ \emph{is also a face}. One can then imagine
embedding both $\Delta$ and $\Delta'$ in the same principal plane, such
that they are correctly oriented relative to one another to join along
$\mathcal{F}$, forming a geometric simplicial complex. Assuming that
$\mathcal{F}$ is a facet of both $\Delta$ and $\Delta'$, this
\emph{rigidly} locks the two simplices together. In this way, we can start
to piece duality frames together for form a larger \textbf{frame complex}
that encodes multiple duality frames and the dualities relating them. This is depicted in Figures \ref{f.glue1} and \ref{f.glue2} for two frame simplices, respectively featuring the same and different asymptotic species scales. Interestingly, the former corresponds to a T-duality while the latter can be thought of as a generalized S-duality.

Ideally, one would like to continue this process until every frame simplex is
glued to another frame simplex along each one of its facets, such that the
frame complex encodes every possible duality, presenting a global picture of
the infinite-distance limits, duality frames, and dualities of a given QGT.
Unfortunately, this is not generally possible, for several reasons. (1) We
gave no prescription for how to choose another infinite-distance limit in
$\mathcal{M}_{d_i}$; there might be multiple options, or no options at all.
(2) Even after specifying a choice of dualities to trace, one can encounter
\emph{monodromies} in the complex of frame simplices. For instance, if the
principal plane is two-dimesional then locally the frame simplices glue
together like the faces of a polygon, but upon passing 360$^{\circ}$ around
the origin, the candidate polygon may not close. Likewise, if the principal
plane is three-dimensional then locally the frame simplices glue together like
the faces of a polyhedron, but upon circling one of the vertices by
360$^{\circ}$, the faces of the candidate polyhedron may not mesh. Similar
issues can occur for a principal plane of any dimension $k > 1$.

\subsubsection{The tower polytope}\label{ss.towerpolytope}

To circumvent these complications, we impose some global structure on the
moduli space that will allow us to glue the different frame simplices into a global polytope. Although this procedure cannot be directly applied to any moduli space, it will serve as a proof of principle of how the taxonomy rules can be used to constrain how different infinite-distance limits can globally fit together in the moduli space. 

First, let us assume that the moduli space has an asymptotically
flat slice, where ``asymptotically flat'' means that the asymptotic boundary
is globally isometric to $\mathbb{R}^k$ (i.e., the moduli space curvature on
the slice goes asymptotically to zero \emph{and} any
asymptotically-visible global differences from $\mathbb{R}^k$ such as a
deficit angle are absent). We would like to asymptotically identify generic
straight lines on this slice with regular infinite-distance limits whose
principal plane is the tangent space to the slice. However, this may fail for
several reasons. (1) Straight lines within the slice may not be geodesic rays
within the entire moduli space, and thus fail to meet our criteria for an
infinite-distance limit (which can lead to a violation of the taxonomy rules).
For example, within the flat slice $C_0 = (\text{const})$ in the type IIB
moduli space, the path $g_s \rightarrow \infty$ is not a geodesic ray when the
fixed value of $C_0$ is not rational. (2) Even if generic lines within our
slice are geodesic rays, the principal plane for such a limit may include
directions outside the slice.

\begin{figure}[h]
\begin{center}
\includegraphics[width=\textwidth]{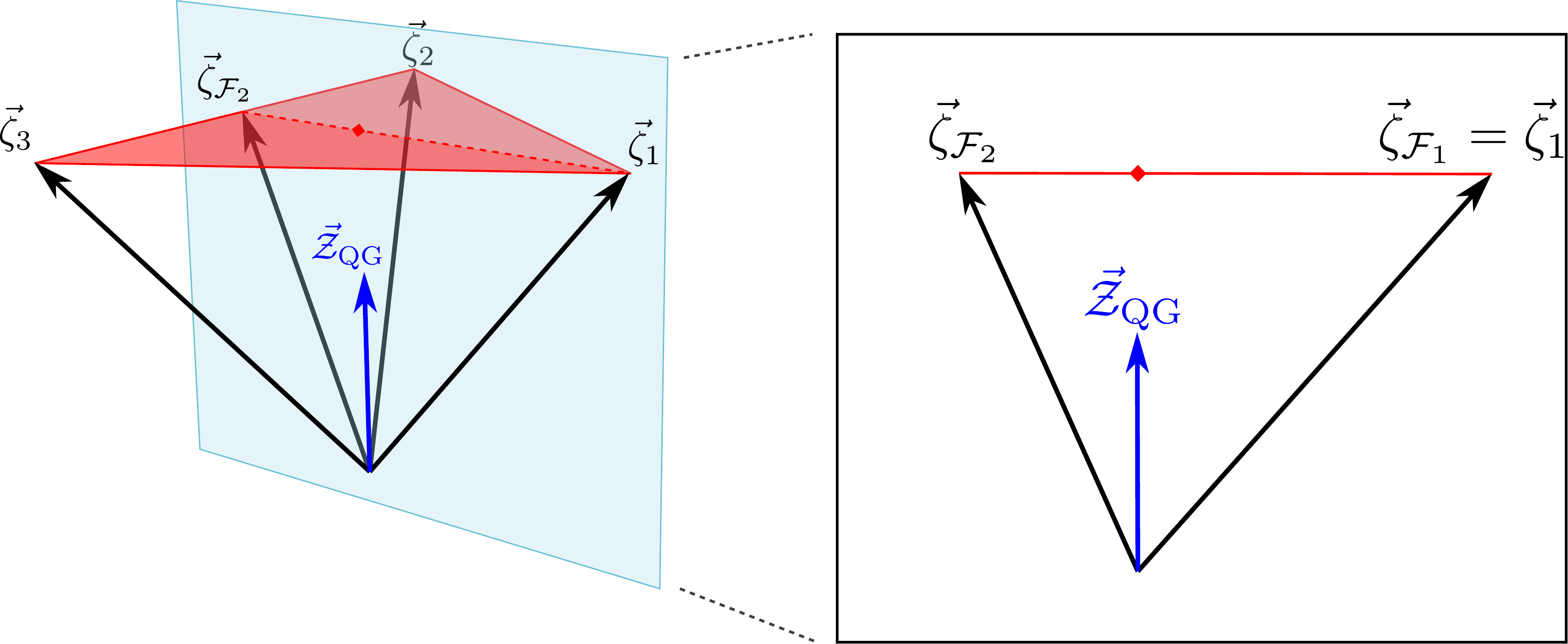}
\end{center}
\caption{A ``good'' projection of a rank 3 frame simplex down to a rank 2 ``effective'' frame simplex. To perform the projection, the tower vectors $\{\vec\zeta_1,\,\vec\zeta_2,\,\vec\zeta_3\}$ are partitioned into disjoint faces $\mathcal{F}_1 = \{ \vec\zeta_1 \}$, $\mathcal{F}_2 = \{ \vec\zeta_2, \vec\zeta_3 \}$ and the frame simplex is projected onto the plane generated by the face pericenters $\vec\zeta_{\mathcal{F}_1}$,  $\vec\zeta_{\mathcal{F}_2}$, which are the tower vectors of the effective frame simplex. Note that the intersection of the frame simplex with this plane is equal to its projection onto the plane, which is an equivalent way to define a ``good'' projection.
}
\label{f.projecting}
\end{figure}

It is not necessarily fatal if the principal plane has directions
outside our chosen slice, provided that some ``effective'' version
of it reduces to the tangent space of the slice. To be precise, consider
grouping the $k$ tower vectors, $\vec{\zeta}_a$, $a = 1, \ldots, k$ into $k_0$
faces $\mathcal{F}_1, \ldots, \mathcal{F}_{k_0}$, such that every vertex
belongs to exactly one face. Now consider the pericenters of these faces:
\begin{equation}
  \vec{\zeta}_{\mathcal{F}_{\alpha}} \equiv \frac{1}{n_{\mathcal{F}_{\alpha}}}
  \sum_{a \in \mathcal{F}_{\alpha}} n_a \vec{\zeta}_a, \qquad
  n_{\mathcal{F}_{\alpha}} \equiv \sum_{a \in \mathcal{F}_{\alpha}} n_a .
\end{equation}
One finds that:
\begin{equation}
  \vec{\zeta}_{\mathcal{F}_{\alpha}} \cdot \vec{\zeta}_{\mathcal{F}_{\beta}} =
  \frac{1}{d - 2} + \frac{1}{n_{\mathcal{F}_{\alpha}}} \delta_{\alpha \beta},
  \qquad \vec{\zeta}_{\mathcal{F}_{\alpha}} \cdot
  \vec{\mathcal{Z}}_{\text{QG}} = \frac{1}{d - 2},
\end{equation}
which reproduces the taxonomy rules with the ``effective'' tower vectors
$\vec{\zeta}_{\mathcal{F}_{\alpha}}$. This \textbf{good projection} of the frame
simplex physically corresponds to artificially freezing some of the moduli,
resulting in a lower-dimensional effective principal plane, see Figure \ref{f.projecting} for an illustration, so that we only move along directions perpendicular to the ``frozen'' one.

Thus, to succeed in defining a global notion of the principal plane, we
require the following assumptions.
\begin{enumerate}
  \item There is an asymptotically flat slice $\Sigma_k \cong_{\text{asymp}}
  {\mathbb{R}^k} $ of the moduli space $\mathcal{M}_d$.
  
  \item For every asymptotically straight line in $\Sigma_k$ there is a
  infinite-distance limit (geodesic ray) within $\mathcal{M}_d$ that
  asymptotically approaches it.
  
  \item For a generic choice of asymptotically straight line in $\Sigma_k$,
  the frame simplex of the associated infinite-distance limit admits a good projection with principal plane asymptotically equal to the tangent space of
  $\Sigma_k$.
\end{enumerate}
As discussed in \cite{Etheredge:2023odp, Etheredge:2023usk}, there are several good rules
of thumb for obtaining such a slice in a specific QGT. For now let us assume
that we have done so.

With the slice $\Sigma_k$ in hand, (1) we unambiguously fix which new infinite
distance limit of $\mathcal{M}_{d_i}$ to explore when passing through the
facets of the frame simplex and (2) we eliminate the possibility of monodromy,
since the principal plane is globally defined, and returning to the same
direction vector $\hat{t}$ brings us back to the same infinite-distance limit.\footnote{To be precise, this is true up to the impact parameter of the
asymptotically straight line. However, this impact parameter has no effect on
the frame simplex in a regular infinite-distance limit.}

\begin{figure}[h]
\begin{center}
\begin{subfigure}{0.52\textwidth}
\center
\includegraphics[width=.9\textwidth]{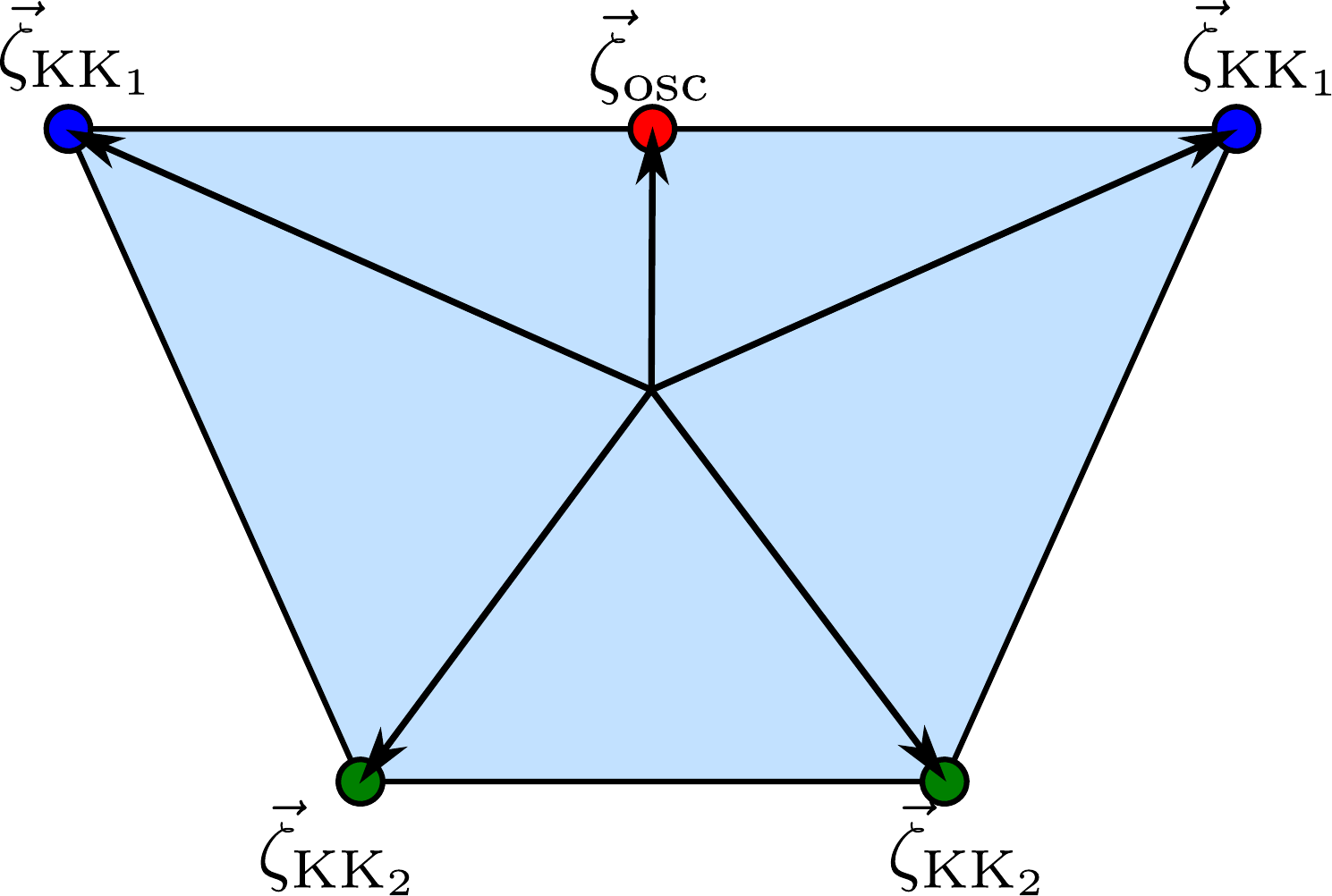}
\caption{Rank-2 tower polytope in $d=7$.} \label{f.sketch4}
\end{subfigure}
\begin{subfigure}{0.46\textwidth}
\center
\includegraphics[width=.9\textwidth]{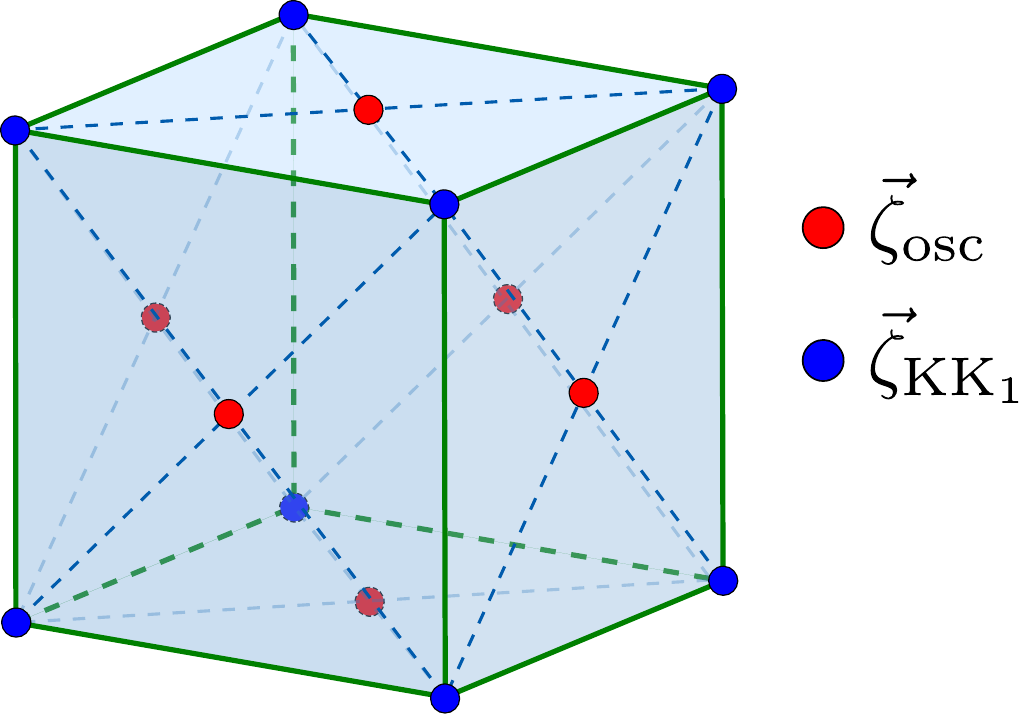}
\caption{Rank-3 tower polytope in $d=4$.} \label{f.cube4dTOW}
\end{subfigure}
\end{center}
\caption{Examples of tower polytope with different number of moduli and $d$. For \subref{f.cube4dTOW} the triangulation into simplices is shown in dashed blue lines. The depicted polytopes are realized as a rank 2 and 3 slices of the full polytope resulting from compactifying M-theory on $T^4$ and $T^7$, respectively.
}
\label{f.towerPol}
\end{figure}

Thus, after choosing such a slice we can complete our program of gluing frame
simplices together, resulting in a closed frame complex enclosing the origin,
the \textbf{tower polytope} for the slice $\Sigma_k$ in the QGT in question, as showed in Figure \ref{f.towerPol} for two specific examples.
Since the taxonomy rules for the tower vectors are rigid, there is a discrete
set of allowed tower polytopes. With some extra input---such as an upper bound
on the spacetime dimension after decompactification---this set becomes finite,
allowing for a classification program.

\begin{figure}[h]
\begin{center}
\begin{subfigure}{0.43\textwidth}
\center
\includegraphics[width=.95\textwidth]{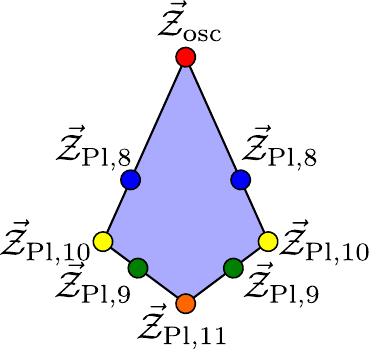}
\caption{Rank-2 species polytope in $d=7$.} \label{f.sketch4sp}
\end{subfigure}
\begin{subfigure}{0.56\textwidth}
\center
\includegraphics[width=.95\textwidth]{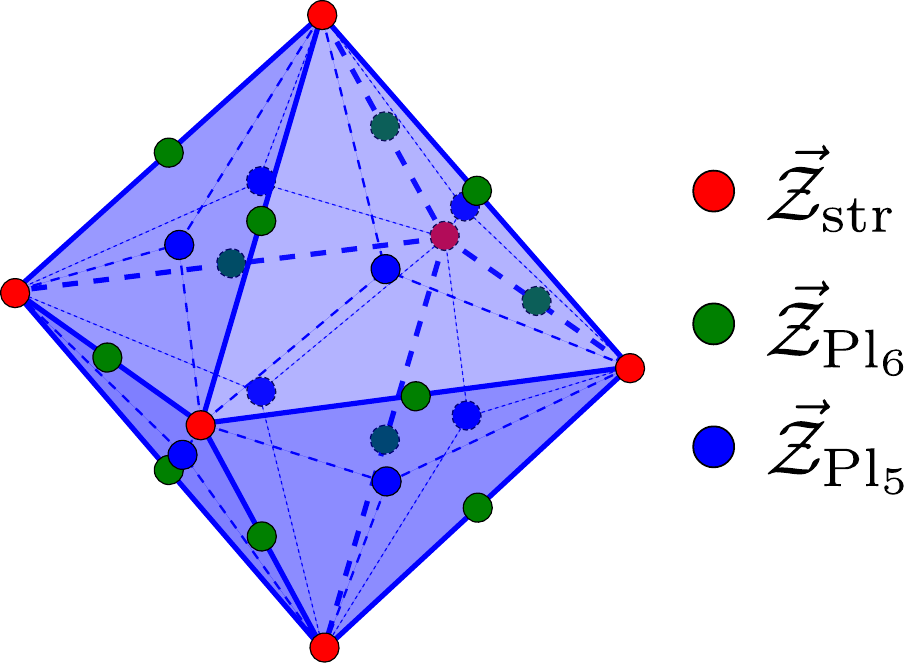}
\caption{Rank-3 species polytope in $d=4$.} \label{f.cube4dSP}
\end{subfigure}
\end{center}
\caption{Example of species polytope (dual to the tower polytopes from Figures \ref{f.sketch4} and \ref{f.cube4dTOW}) with two and three moduli in $d=7$ and 4. Also pictured is the triangulation into simplices for \subref{f.cube4dSP}. These species polytopes appear as a rank-2 and 3 slices of the full species polytopes of M-theory on and $T^4$ and $T^7$.}
\label{f.spPOL}
\end{figure}

Note that while the surface of the tower polytope is triangulated by the frame
complex, the \emph{geometric} facets, vertices, etc., of the tower polytope are not
identical to those of the complex. In particular, every tower vector is a
geometric vertex of the tower polytope \emph{except} for string oscillator
towers, which are the pericenters of geometric facets. Thus, the stringy faces
of the tower polytope are in general nonsimplicial (specifically, they are
\emph{cross-polytopes}) and are triangulated by multiple frame simplices,
whereas the Planckian faces are simplicial.

Let us remark that the above procedure of gluing the different frame simplices to build the tower polytope does not allow for having two string towers connected to each other without having a KK vertex in between. This is consistent with the fact that the Emergent String Conjecture requires a \emph{unique} string becoming light in the emergent string limits. We did not need to impose this as an input, but rather it emerged as an output of the taxonomic rules.

\subsubsection{The species polytope}

Just as dualities glue frame simplices together into a complex, which form a
closed polytope given a suitable slice $\Sigma_k \subseteq \mathcal{M}_d$,
likewise species stars naturally glue together into a larger geometric object. In particular, since the pericenter $\vec{\mathcal{Z}}_\mathcal{F}$ of every face of a species star lies on its boundary, where it matches the pericenter of faces of adjoining species stars, the faces of neighboring species stars combine into larger faces that span multiple duality frames. This is unlike the frame complex, where each geometric facet represents a distinct duality frame (in a Planckian phase) or at most a collection of T-dual frames (in a stringy phase).

Given a suitable slice $\Sigma_k \subseteq \mathcal{M}_d$ satisfying the above
assumptions and gluing together the ``projected'' species stars (constructed
in the obvious way from the projected frame simplices), one obtains the
\textbf{species polytope}, the vertices of which are the species vectors
$\vec{\mathcal{Z}}_{\text{QG}}$ for each duality frame.\footnote{Note that in
the case of stringy phases, multiple duality frames (related by, e.g.,
T-duality) share the same species vector $\vec{\mathcal{Z}}_{\text{QG}} =
\vec{\zeta}_{\text{osc}}$, controlled by the common string scale.} See Figure \ref{f.spPOL} for the species polytopes associated to the examples from Figure \ref{f.towerPol}.

 Due to the
relationship $\vec{\mathcal{Z}}_{\text{QG}} \cdot \vec{\zeta}_a = \frac{1}{d -
2}$ for each frame simplex (the tower-species pattern \cite{Castellano:2023stg, Castellano:2023jjt}), the
species polytope is precisely the polar dual of the tower polytope, where the
polar dual of a set $P \subseteq \mathbb{R}^k$ is defined as:
\begin{equation}
  P^{\circ} = \biggl\{ \vec{\mathcal{Z}} \biggm| \forall \vec{\zeta} \in P,
  \vec{\mathcal{Z}} \cdot \vec{\zeta} \leqslant \frac{1}{d - 2} \biggr\},
\end{equation}
with the $\frac{1}{d - 2}$ normalization chosen in accordance with the
tower-species pattern. Thus, for instance, each KK tower vector (geometric
vertex) in the tower polyope is dual to a facet of the species
polytope, and likewise each geometric facet of the tower polytope is dual to a
vertex in the species polytope, etc.

Just as the geometry of the tower polytope vertices are rigidly fixed by the
taxonomy rules \eqref{eqn:taxonomyRules}, it is interesting to ask whether
analogous rules directly fix the geometry of the vertices of the species polytope.
The rule \eqref{eqn:speciesstarRule} applies only within a single duality frame,
so it does not directly address this question. However, consider two vertices
$\vec{\mathcal{Z}}_{\alpha}$, $\vec{\mathcal{Z}}_{\beta}$ of the species
polytope that are joined by an edge. The pericenter of the edge is
$\vec{\mathcal{Z}}_{\mathcal{F}}$, where $\mathcal{F} \subset
\Delta_{\alpha} \cap \Delta_{\beta}$ is a common facet of the corresponding
frame simplices $\Delta_{\alpha}, \Delta_{\beta}$. We can then compute the dot
product between the two vertices by decomposing each into components parallel
and perpendicular to the pericenter-$\vec{\mathcal{Z}}_{\mathcal{F}}$ plane.
We find:
\begin{equation}
  \vec{\mathcal{Z}}_{\alpha}^{\perp} = \vec{\mathcal{Z}}_{\beta}^{\perp} =
  \vec{\mathcal{Z}}_{\mathcal{F}}, \qquad \vec{\mathcal{Z}}_{\alpha}^{\|} =
  \vec{\mathcal{Z}}_{\alpha} - \vec{\mathcal{Z}}_{\mathcal{F}}, \qquad
  \vec{\mathcal{Z}}_{\beta}^{\|} = \vec{\mathcal{Z}}_{\beta} -
  \vec{\mathcal{Z}}_{\mathcal{F}},
\end{equation}
hence
\begin{equation}
  \vec{\mathcal{Z}}_{\alpha} \cdot \vec{\mathcal{Z}}_{\beta} =
  \vec{\mathcal{Z}}_{\alpha}^{\perp} \cdot \vec{\mathcal{Z}}_{\beta}^{\perp} +
  \vec{\mathcal{Z}}_{\alpha}^{\|} \cdot \vec{\mathcal{Z}}_{\beta}^{\|} = |
  \vec{\mathcal{Z}}_{\mathcal{F}} |^2 - | \vec{\mathcal{Z}}_{\alpha}^{\|} |  |
  \vec{\mathcal{Z}}_{\beta}^{\|} |,
\end{equation}
where we use the fact that $\vec{\mathcal{Z}}_{\alpha}^{\|}$ is antiparallel
to $\vec{\mathcal{Z}}_{\beta}^{\|}$ since $\vec{\mathcal{Z}}_{\mathcal{F}}$
lies on the line between them. Thus, using \eqref{eqn:subfaceRule} we find
\begin{equation}
  \vec{\mathcal{Z}}_{\alpha} \cdot \vec{\mathcal{Z}}_{\beta} = \frac{1}{d - 2}
  - \frac{1}{D_{\alpha \beta} - 2} - \sqrt{\left( \frac{1}{D_{\alpha \beta} -
  2} - \frac{1}{D_{\alpha} - 2} \right) \left( \frac{1}{D_{\alpha \beta} - 2}
  - \frac{1}{D_{\beta} - 2} \right)},
\end{equation}
where $D_{\alpha}, D_{\beta}$ and $D_{\alpha \beta}$ are the species
dimensions associated to $\vec{\mathcal{Z}}_{\alpha}$,
$\vec{\mathcal{Z}}_{\beta}$ and $\vec{\mathcal{Z}}_{\mathcal{F}}$.
Simplifying, we find:
\begin{equation}
  \fbox{$\vec{\mathcal Z}_{\alpha}\cdot \vec{\mathcal Z}_\beta=\frac{1}{d-2}-\frac 1{D_{\alpha \beta}-2}\left[1+\sqrt{\frac{(D_\alpha-D_{\alpha\beta})(D_\beta-D_{\alpha\beta})}{(D_\alpha-2)(D_\beta-2)}}\right]$.}\label{eqn:speciesPolytopeRule}
\end{equation}
This rule applies to any two vertices of the species polytope that
are joined by an edge. More generally, it applies to the pericenters $\vec{\mathcal{Z}}_{\alpha}$, $\vec{\mathcal{Z}}_{\beta}$ of \emph{any} pair of
faces of the species polytope,
provided that the pericenter $\vec{\mathcal{Z}}_{\alpha \beta}$ of the line between them
is the pericenter of face of the polytope.
For example, in the special case where $\vec{\mathcal{Z}}_{\beta}$ lies in the
$\vec{\mathcal{Z}}_{\alpha}$-pericenter plane, we have $D_{\alpha \beta} =
D_{\alpha}$, so that
\begin{equation}
  \vec{\mathcal{Z}}_{\alpha} \cdot \vec{\mathcal{Z}}_{\beta} = |
  \vec{\mathcal{Z}}_{\alpha} |^2 = \frac{1}{d - 2} - \frac{1}{D_{\alpha} - 2},
  \qquad \text{when $\vec{\mathcal{Z}}_{\beta}$ lies in the
  $\vec{\mathcal{Z}}_{\alpha}$-pericenter plane,}
  \label{eqn:speciesPolytopeSpecRule}
\end{equation}
which is a restatement of \eqref{eqn:subfaceRule}.

The rule \eqref{eqn:speciesPolytopeRule} rigidly fixes the geometry of the
species polytope in an analogous fashion to the taxonomy rules
\eqref{eqn:taxonomyRules} applied to the tower polytope.

\subsubsection{Recursion of polytopes} \label{sec:polytoperecursion}

Just as with the frame simplex and the species star, the tower and species
polytopes can be built up recursively in the rank. The nature of this
recursion is somewhat easier to explain in the case of the species polytope,
which we discuss first.

Consider a facet of the species polytope, with pericenter
$\vec{\mathcal{Z}}_1$. Then for any two species vectors
$\vec{\mathcal{Z}}_{\alpha}$, $\vec{\mathcal{Z}}_{\beta}$ on this facet,
\eqref{eqn:speciesPolytopeRule} implies that
\begin{equation}
  (\vec{\mathcal{Z}}_{\alpha} - \vec{\mathcal{Z}}_1) \cdot
  (\vec{\mathcal{Z}}_{\beta} - \vec{\mathcal{Z}}_1) = \frac{1}{D_1 - 2} -
  \frac{1}{D_{\alpha \beta} - 2}  \left[ 1 + \sqrt{\frac{(D_{\alpha} -
  D_{\alpha \beta}) (D_{\beta} - D_{\alpha \beta})}{(D_{\alpha} - 2)
  (D_{\beta} - 2)}} \right], \label{eqn:speciesPolyRecurs}
\end{equation}
where we use the special case \eqref{eqn:speciesPolytopeSpecRule} to compute
$\vec{\mathcal{Z}}_{\alpha} \cdot \vec{\mathcal{Z}}_1 =
\vec{\mathcal{Z}}_{\beta} \cdot \vec{\mathcal{Z}}_1 = | \vec{\mathcal{Z}}_1
|^2$. The rule \eqref{eqn:speciesPolyRecurs} has the same form as
\eqref{eqn:speciesPolytopeRule} with $d \rightarrow D_1$. Thus, \emph{each facet of the species polytope is itself a species polytope} in
spacetime dimension equal to the species dimension $D$ of the pericenter of
the facet. More generally, this applies to any $p$-face of the
species polytope, not just to its facets.

The physical interpretation of this is the same as in
Section \ref{subsec:dualityframe}: the pericenter of the facet
correponds to a KK tower vector $\vec{\zeta}_1$. Taking the limit where this
KK tower becomes exponentially light in $D_1 = d + n_1$ dimensional Planck
units, we recover a $D_1$-dimensional theory with an inherited asymptotically
flat slice $\Sigma_{k - 1}$, etc., such that the species polytope of this
theory is the facet of the original species polytope that we began with.

We now study the same limit in the tower polytope. Per the taxonomy rules
\eqref{eqn:taxonomyRules} and \eqref{eqn:pattern1}, we have:
\begin{equation}
  \vec{\zeta}_a^{(\text{adj})} \cdot \vec{\zeta}_1 = \frac{1}{d - 2}, \qquad
  \vec{\mathcal{Z}}_1 \cdot \vec{\zeta}_1 = \frac{1}{d - 2} \qquad \Longrightarrow
  \qquad (\vec{\zeta}_a^{(\text{adj})} - \vec{\mathcal{Z}}_1) \cdot
  \vec{\zeta}_1 = \vec{0},
\end{equation}
for any tower vector $\vec{\zeta}_a^{(\text{adj})}$ joined to $\vec{\zeta}_1$
by an edge in the frame complex. Recall that $(\vec{\zeta}_a^{(\text{adj})} - \vec{\mathcal{Z}}_1)$ is equivalent to the tower vector with the mass written in the higher dimensional Planck units. By comparison $(\vec{\zeta}_1 -
\vec{\mathcal{Z}}_1) \cdot \vec{\zeta}_1 = \frac{1}{n_1} > 0$, whereas it is
not hard to see that, since the frame complex is convex, for any
\emph{other} tower vector $\vec{\zeta}_{\text{far}}$ in the frame complex,
$\vec{\zeta}_a^{(\text{far})} \cdot \vec{\zeta}_1 < \frac{1}{d - 2}$ and hence
$(\vec{\zeta}_a^{(\text{far})} - \vec{\mathcal{Z}}_1) \cdot \vec{\zeta}_1 <
0$. Thus, in the limit $\hat{t} \propto \vec{\zeta}_1$, the KK tower
corresponding to $\vec{\zeta}_1$ becomes exponentially light (as expected),
whereas the towers joined to $\vec{\zeta}_1$ by an edge in the frame complex
remain at a fixed scale in $D_1$-dimensional Planck units and all other towers
become heavy.

As already shown in \eqref{eqn:towerRecurse}, the tower vectors
$\vec{\zeta}_a^{(\text{adj})} - \vec{\mathcal{Z}}_1$ indeed satisfy the
taxonomy rules in $D_1$ spacetime dimensions. These vectors are precisely the
vertices of the link of the tower vector $\vec{\zeta}_1$ within the frame
complex. Thus, \emph{the link of each geometric vertex in the tower
polytope is itself a tower polytope}, again in spacetime dimension $D_1 = d +
n_1$. Note that in this case it is important that the link is computed in the
frame complex, which includes the string oscillator towers as vertices.
Geometrically, the link can also be thought of as the \emph{vertex figure}
of the vertex in question. Concrete examples illustrating these recursion relations of polytopes are shown in Section \ref{s.applications.recursionillustration} (see Figures \ref{fig: 10from9} to \ref{fig: 10from8}).

\section{Scope of the taxonomy rules}\label{s.scope}

The taxonomic rules derived in the previous section hold under certain assumptions, as outlined there and in Section \ref{ss.assumpt}. Some of these assumptions are believed to be universal features of quantum gravity (for instance, some can be motivated by the Emergent String Conjecture) while others are assumptions about the geometry of the moduli space that do not hold universally. As a result, we emphasize that our rules are \emph{not} universal: they do not apply at all points in quantum gravity moduli spaces (not even in all the asymptotic limits). In this section, we investigate these assumptions in more detail, exploring the conditions under which they are satisfied and explaining how the taxonomic rules can break down when they are violated.

\subsection{Emergent String Conjecture and bound states\label{s.bound}}

The derivation of the taxonomic rules in Section \ref{s.rules} relied on the assumption that the Emergent String Conjecture (ESC) holds in any effective field theory consistent with a UV quantum gravity completion (Assumptions \ref{assump1} and  \ref{assump2} in Section \ref{ss.assumpt}). Moreover, we assumed that the ESC can be applied recursively to the higher-dimensional theory that emerges upon decompactification. In this subsection, we emphasize that this a stronger condition than merely imposing the ESC in the original lower dimensional theory, and we highlight the crucial implication of this assumption: the existence of bound states of neighboring principal towers. Such bound states are necessary to avoid pathologies that would otherwise violate the taxonomy rules.

For purposes of illustration, consider a 2-dimensional moduli space with a frame simplex generated by the tower vectors of two KK principal towers, associated with the decompactification of $n$ dimensions and $m$ dimensions, respectively. Then, following the arguments of the previous section, a generic infinite-distance limit in this duality frame should correspond to a decompactification of $m+n$ dimensions, with a species scale given by the $(d+m+n)$-dimensional Planck scale. However, if the KK modes of the two towers do not form bound states, then the the total number of states contributing to the species scale will be given simply by the sum of the light modes of the two towers, $N\sim N_1+N_2$, which is too small. As a result, the species scale will be too large, resulting in a violation of the pattern, $\vec{\zeta} \cdot \vec{\mathcal Z} \neq \frac{1}{d-2}$ \cite{Castellano:2021mmx,Castellano:2023jjt}. In contrast, if the KK modes do form bound states that populate a (sub-)lattice of their KK charges, then the total number of light species is multiplicative, $N\sim N_1 N_2$, which leads to the expected scaling of the species and the orthogonality of the species vector with the convex hull of the tower vectors, in concordance with our taxonomic rules.

\begin{figure}[h]
\begin{center}
\includegraphics[width = 85mm]{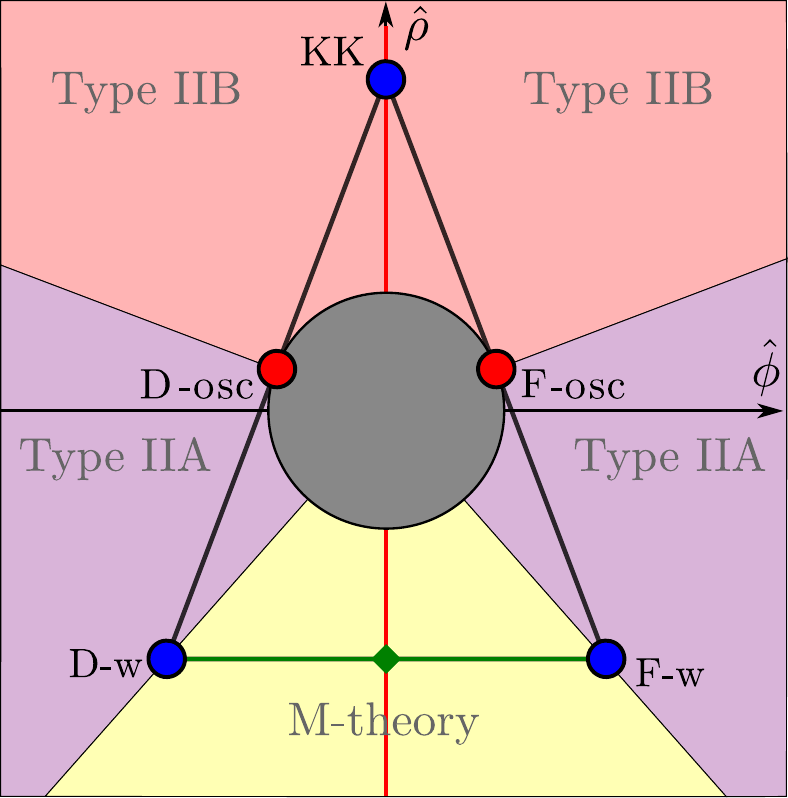}
\end{center}
\caption{Tower polytope for Type IIB string theory compactified on $S^1$ to $d=9$, with the different KK (in blue) and string oscillator towers depicted, as well as the different duality frame descriptions. The moduli correspond to the canonically normalized 10d dilaton and the $S^1$ radius. The sphere of radius $\frac{1}{\sqrt{d-2}}=\frac{1}{\sqrt{7}}$ is shown in gray and the Type IIB self-dual line in red. See \cite{Etheredge:2023odp} for more details on the limits of the duality frames and the expression of the different towers tower vectors.}
\label{f.CHIIB}
\end{figure}

From this, we conclude that any two principal KK towers that are connected by an edge of the tower polytope (so that there is one direction along which both decay at the same rate) must form bound states and, therefore, can be described microscopically as KK towers from the perspective of the same duality frame. This means that if two KK vertices of the tower polytope are connected by an edge and their KK modes do not form bound states (which occurs if they are interpreted as KK towers in different duality frames, e.g. KK and winding modes), then the interior of this edge must also contain a string oscillator vertex. This string oscillator vertex separates the two KK towers into distinct duality frames.

An example of this occurs in the tower polytope of Type IIB string theory on $S^1$ (see Figure \ref{f.CHIIB}), where the edge connecting the KK mode vertex and the F-string winding mode vertex is separated into two distinct frame simplices by the F-string oscillator vertex. As a result, the KK and winding modes are never simultaneously lighter than the species scale. Moreover, in this case the neighboring vertices (i.e. the KK and string vertices) again form bound states comprising the Kaluza-Klein replicas of the string modes, since the KK tower can be described as perturbative states from the string worldsheet perspective. Without such bound states, the taxonomy rules of the previous section could be violated.

In \cite{Castellano:2023jjt} it was noted that this condition on the formation bound states was essential for the tower-species pattern to hold. Here we point out that this condition follows from imposing the ESC recursively in the higher dimensional theory that emerges upon decompactification, so that all the light towers below the species scale can be described as perturbative states under the same duality frame.

\subsection{Regular vs. irregular infinite-distance limits}

In Section \ref{s.rules}, we focused on deriving the taxonomy rules for regular infinite-distance limits (see the definition at the beginning of Section \ref{sec:rules}). Unlike the assumption of the Emergent String Conjecture, however, this regularity assumption is violated in known examples of infinite-distance limits, and in such cases the taxonomy rules can be violated. In what follows, we will argue that generic infinite-distance limits are regular, and we will briefly discuss what happens when a limit becomes irregular, although a more systematic analysis is left for future work.

\subsubsection{Generic infinite-distance limits are
regular}\label{subsec:generic}

Let us start by arguing that---provided that the ESC holds---infinite
distance limits should be generically regular. For instance, if there are
degenerate towers in some particular infinite-distance limit, a slight
variation in the \emph{direction} of the limit (as measured, e.g., within
the tangent space of some fixed reference point) will produce parametric
splittings between these towers. While this argument assumes that a continuous
family of infinite-distance limits with a suitably varying direction exists,
this is true in every example we know of involving degenerate towers.

Infinite-distance limits involving decompactifications that are not
asymptotically empty are subtler. Let us temporarily turn the problem around
and consider what happens as the direction of the asymptotic limit is varied
for an asymptotically empty (but not strictly empty) decompactification. As
this direction is varied, the exponential rates controlling both the growth of
the overall volume as well as the size of the warped / Ricci-curved regions
will vary. In special directions, these rates will coincide and the warped /
Ricci-curved regions will grow at the same rate as the overall volume in the
decompactification limit. The relative size of these regions compared to that
of $X_n$ can still be adjusted by adjusting the limit, but typically this does
not change the \emph{direction} of the limit (which controls the
exponential rates), only its \emph{impact parameter}. Adjusting this
impact parameter towards one extreme, the strongly warped / Ricci-curved
regions grow to fill all of $X_n$.

Continuing beyond such a special direction, the warping / Ricci-curvature
will naively blow up, and it is no longer possible to view the theory as
decompactifying along $X_n$. The Emergent String Conjecture then demands that
there is a ``dual'' description of this limit which is again either a
decompactification limit along some new manifold $Y_n$ or an emergent string
limit. Either way, we conclude that a non-asymptotically-empty
decompactification can only occur in specific directions in which various
exponential rates coincide, hence varying the direction of the infinite
distance limit should generically produce a regular infinite-distance limit.
This agrees with the examples studied in \cite{Etheredge:2023odp}, as explained in Section \ref{s:sliding}.

Thus, we expect that regularity is generic amongst infinite-distance limits.
If so, we can understand the collection of \emph{all} infinite-distance limits by starting with regular limits and allowing them to vary continuously.

\subsubsection{Irregular limits: degenerate towers}\label{subsec:degenerate}

We now briefly consider what happens when a regular infinite-distance limit
becomes irregular as the direction vector is varied. This can happen in one of
two ways, either because some of the tower scales
parametrically below $\Lambda_{\text{QG}}$ degenerate, or because we reach a decompactification limit that is not asymptotically
empty. Let us first focus on the first case, which is potentially more benign.

\begin{figure}[h]
\begin{center}
\includegraphics[width=.55\textwidth]{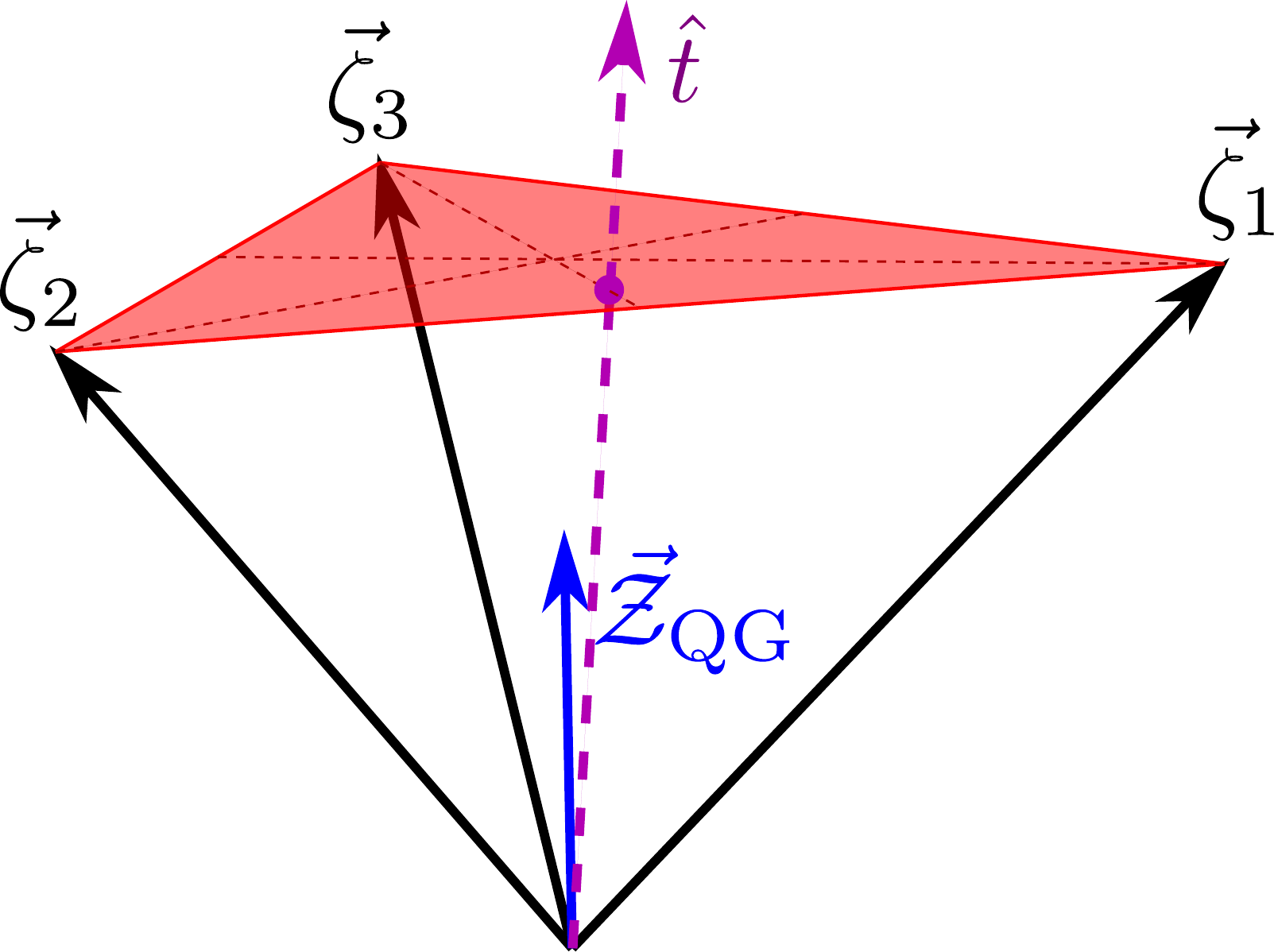}
\end{center}
\caption{Example of a direction $\hat{t}$ over which which several towers (in this case those associated to $\vec{\zeta}_1$ and $\vec{\zeta}_2$) become degenerate in a way that we can ``ignore''. The rank 3 Planckian phase is the same as depicted in Figure \ref{f.3dFRAMEch1}.
}
\label{f.3dFRAMEdg}
\end{figure}

The case of degenerate towers correspond to having several leading towers parametrically below $\Lambda_{\text{QG}}$ that decay at the same rate, see Figure \ref{f.3dFRAMEdg}.
If this occurs, these towers are necessarily
KK scales, and their degeneration corresponds to a previously hierarchical
decompactification on $X_m$ followed by $Y_n$ with $\text{vol} (X_m) \gg
\text{vol} (Y_n)$ morphing into the decompactification of some manifold $Z_{m
+ n}$ without hierarchical KK scales. In some cases, we can take another limit
of this manifold where we again obtain a hierarchical decompactification, but
now on $Y_n$ followed by $X_m$ (so that $\text{vol} (Y_n) \gg \text{vol}
(X_m)$). When this occurs we can \emph{continue through} the degenerate
locus and reach another regular infinite-distance limit. Although the sequence
of the number of decompactifying dimensions is now altered (by exchanging $n$
with $m$), this only affects the taxonomy rules \eqref{eqn:taxonomyRules} by
permuting the tower vectors, so with an apropriate choice of conventions we can
again consider the frame simplex to be the same as before the degeneration,
with only $\hat{t}$ having changed. This is what we have termed an ``ignorable
degeneration'' in Section \ref{subsec:dualityframe}.
Two important comments are in order. Firstly, it is not necessarily the case
that $Z_{m + n} = X_m \times Y_n$ (for instance, there may be a fibration).
When this is \emph{not} true, the reversed limit discussed above may not
exist, so it is not always possible to continue through a degenerate locus in
the manner described above.  In that case, we might only get a portion of the frame simplex, as in the M-theory on K3 example of Appendix \ref{app:K3}. Secondly, there may be \emph{other}
(previously undetected) hierarchical limits of $Z_{m + n}$ that open up in the
degeneration limit. Thus, the principal plane may connect to the principal
plane of another family of regular infinite-distance limits at points of
degeneration, even if the degeneration is ignorable. A simple example of this
is in the moduli space of 9d maximal SUGRA, where the limit in which
$M$-theory decompactifies along a $T^2$ of fixed shape is such a degenerate
limit. This limit admits continuous deformations where the $T^2$ shape modulus
goes off to any of the various infinite-distance limits of Teichm\"uller space.
(These are of course related by dualities, but they correspond to
\emph{different} principal planes in the tangent space of moduli space at
this point).

\subsubsection{Irregular limits: sliding vertices and warped geometries \label{s:sliding}}

The other regularity condition we assumed in our derivation of the the taxonomy rules in Section \ref{s.rules} was asymptotic emptiness; namely, we assumed that the decompactification manifold approaches a Ricci-flat manifold with background fields that vanish outside of regions of measure zero. This assumption may be violated, however, in decompactification limits that involve a significant amount of warping.  If so, the physics of the decompactification becomes very complicated
(even in the simplest examples \cite{Etheredge:2023odp, Alvarez-Garcia:2023qqj}). Analyzing such cases systematically is
beyond the scope of our paper. However, in this subsection, we explain one example in which such warping occurs. We will see that (a) vertices of the frame simplex may vary as a function of the position in moduli space, resulting in a violation of our taxonomy rules in special, irregular limits of moduli space, but (b) generic limits of the moduli space are regular and satisfy our taxonomy rules. As a result, we can still use the polytopes satisfying the taxonomy rules for regular limits as building blocks to determine the global geometry of the tower vectors.

\begin{figure}[h]
\begin{center}
\includegraphics[width = 95mm]{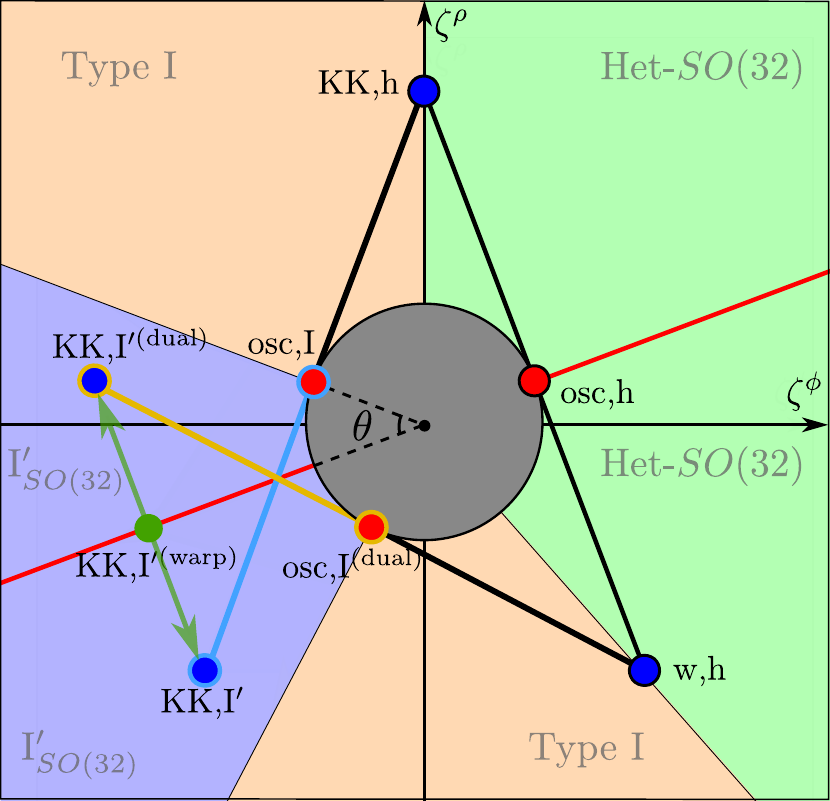}
\end{center}
\caption{Frame simplices in different asymptotic limits of the I$'_{SO(32)}$ regions in $SO(32)$ heterotic string theory compactified on $S^1$. In Type I$'$ infinite-distance limits parallel to the self-duality line (red), the taxonomy rules are violated, as the tower vector for the Type I$'$ KK modes takes values along a segment (green arrows) orthogonal to this line. Such limits are irregular due to the effects of warping. In all other Type I$'$ infinite-distance limits, however, the KK, I$'^{\rm(warp)}$ vector approaches either KK, I$'$ or KK, I$'^{\rm (dual)}$. Such limits are regular, as the effects of warping become negligible in the asymptotic limit, so the frame simplices of these limits satisfy the taxonomy rules. Note that
the heterotic towers (along the thin black segment) remain fixed in any limit.}
\label{f.slidingSO32}
\end{figure}

 With this, let us consider $SO(32)$ heterotic string theory compactified on $S^1$. As shown in \cite{Etheredge:2023odp}, this theory features a collection of principal towers whose tower vectors vary as a function of position in moduli space, as shown in Figure \ref{f.slidingSO32}. Within the upper Type I$'$ phase (i.e., when $\phi/\sqrt{7} <  \rho < - \phi/\sqrt{7} $), the vector labeled KK, I$'$ lies below the line $\zeta_{\rho} = \zeta_\phi / \sqrt{7}$. Conversely, upon crossing the self-duality line $\rho = \phi/\sqrt{7}$ into the lower Type I$'$ phase (i.e., when $\frac{5}{32 \sqrt{7}}\phi <  \rho <  \phi/\sqrt{7} $), this vector continuously slides above the line $\zeta_{\rho} = \zeta_\phi / \sqrt{7}$.
 
 As a result of this sliding, the taxonomy rules of Section \ref{s.rules} break down along an infinite-distance geodesic parallel to the self-duality line (shown in red). For example, for the infinite-distance limit on the self-duality line with $\rho / \phi = 1/\sqrt{7}$, $\rho \rightarrow - \infty$, the leading tower (labeled KK, I$'^{\text{(warp)}}$ in Figure \ref{f.slidingSO32}) has length $|\vec \zeta_{\text{KK, I'}^{\rm (warp)}}| = 5/\sqrt{28}$ and has an angle $\theta=\arccos\left(\frac{3}{4}\right)$ with $\vec{\zeta}_{\rm osc, I}$, which violates the taxonomy rules for the tower vectors \eqref{eqn:taxonomyRules}\footnote{Whether the pattern relating the tower and species vectors holds in this case is still an open question \cite{Castellano:2023jjt}. It could happen that the rule \eqref{rule1} for the light towers needs to be modified in this case while the pattern with the species scale \eqref{eq:pattern} still holds if sustained in a deeper quantum gravity constraint on the density of states.}. Such a violation occurs because the Type I$'$ decompactification limit does not lead to a 10-dimensional vacuum, but rather a running solution in which the string coupling varies, with the string oscillator tower not becoming light when moving in such direction. This violates the assumption of asymptotic emptiness, as the effects of warping cannot be neglected in this limit.
As a result, our derivation of the taxonomy rules is no longer valid for the family of infinite-distance limits parallel to the self-duality line.

Nonetheless, as discussed at length in \cite{Etheredge:2023odp}, the effects of warping vanish asymptotically far from the self-duality line, and a result the taxonomic rules are restored in any infinite-distance limit that is not parallel to this line. To be more precise, consider an infinite-distance limit in one of the Type I$'$ regions of moduli space, $\phi \rightarrow -\infty$ with fixed slope $\frac{5}{32\sqrt{7}}<\frac{ d  \rho}{ d  \phi}<-\frac{1}{\sqrt{7}}$. If this limit lies above the self-duality line, i.e., if $-\frac{1}{\sqrt{7}}<\frac{ d  \rho}{ d  \phi}<\frac{1}{\sqrt{7}}$, then the frame simplex is generated by the tower vectors labeled ``osc", I and KK, I$'$ in Figure \ref{f.slidingSO32}. These vectors \emph{do} satisfy our taxonomic rules in \eqref{eqn:taxonomyRules}. Similarly, below the self-duality line, i.e. for $\frac{5}{32\sqrt{7}}<\frac{ d  \rho}{ d  \phi}<\frac{1}{\sqrt{7}}$, the frame simplex is generated by the tower vectors labeled ``osc", I$^{\rm (dual)}$ and KK, I$'^{\rm (dual)}$ in Figure \ref{f.slidingSO32}: this frame simplex also satisfies the taxonomic rules. Likewise, the rules are satisfied in all infinite-distance limits in the Type I and heterotic phases.

In other words, whereas the tower vector of the Type I$'$ KK modes slides continuously as a function of position in the interior of moduli space, it jumps discontinuously in the asymptotic regime of moduli space as a function of the angle $\vartheta = \tan^{-1}(\rmd \rho / \rmd \phi )$, satisfying the taxonomic rules on either side of the self-duality line $\rho = \phi/\sqrt{7}$.\footnote{This discontinuous behavior in the asymptotic regime is indicative of an order-of-limits issue; continuous sliding gives way to discontinuous jumping in the same way that a family of continuous Gaussian functions may approach a discontinuous Dirac delta function in the asymptotic limit. While the exponential rate of the towers is defined for the infinite-distance points, the masses of the towers (and the associated scalar charge-to-mass ratio vectors) are local functions of the moduli space, their asymptotic expression needs not to be continuous in the space of infinite-distance limits.} The assumption of asymptotic emptiness is satisfied in these limits because, although there is nonzero warping of the decompactification manifold, the effects of this warping tend to zero in the asymptotic limit. 

We see, therefore, that the taxonomy rules may be violated when the regularity assumptions are violated. However, we also see that in the case at hand, generic infinite-distance limits are regular, so violations occurs only in special directions of moduli space. We also note that in this example, the tower vector slides along a line segment orthogonal to the family of irregular geodesics. In Appendix \ref{s.orthogonality}, we show that this phenomenon is much more general: tower vector sliding always occurs orthogonal to the direction of an irregular geodesic in the asymptotic limits of the moduli space. As a result, for the purposes of computing the scaling coefficient $\alpha$ of the leading tower, the sliding is irrelevant: the three KK, I$'$ tower vectors of Figure \ref{f.slidingSO32} all yield the same scaling coefficient $\alpha$ for the family of geodesics parallel to the self-duality line.

Finally, note that the directions $\hat{t}$ where leading towers become degenerate are
fixed by the geometry of the frame simplex, but this is not the case for
directions where non-asymptotically-empty decompactifications occur.
Predicting where the latter can happen is non-trivial and is
beyond the scope of our paper.

\subsection{Flat moduli spaces and axions\label{s:flat}}

In Section \ref{ss.towerpolytope}, we studied the conditions under which frame simplices can be glued together to form a tower polytope. We found that a sufficient condition for such gluing is the existence of an asymptotically flat slice of moduli space, such that a frame simplex of an infinite-distance limit in this slice admits a ``good'' projection onto the tangent space of the flat slice. For instance, in 10d Type IIA, heterotic, and Type I string theories, there is just one modulus, the dilaton. The entire moduli space is flat, and we may construct a tower polytope for this theory.

In 10d IIB string theory, there are two moduli: the axion and the dilaton. The moduli space is no longer flat, and the tower/species vectors of the $(p,q)$ string scales vary with position in moduli space. To avoid this complication, we consider a slice of the moduli space, such as $C_0=0$. This slice is a (flat) line, parameterized by the dilaton, and in the weak coupling limit the tower vector for the fundamental string lies tangent to the geodesic. Thus, our taxonomy rules can be applied globally along this slice to construct a tower polytope.

In general, in order to find a flat slice satisfying the assumptions of Section \ref{ss.towerpolytope}, a good rule of thumb to set all the axions to zero. For instance, compactifying M-theory on a $k$-torus $T^k$, axions arise both from the off-diagonal components of the $T^k$ metric as well as from $C_3$ reduced on various three cycles of the torus.\footnote{In $d\le 5$, there are also axions coming from the magnetically dual $C_6$ potential reduced on six cycles of the torus.} Setting all these axions to zero restricts us to a flat slice of moduli space consisting of M-theory compactified on a rectangular torus without Wilson lines, parameterized by $k$ radion moduli. One can check that the tower vectors of the KK modes of each of the 1-cycles of the torus admit a good projection onto the tangent space of this slice, as do the tower vectors for the other duality frames that appear as we move in this slice of the moduli space. As a result, our taxonomic rules can be applied throughout this slice of the moduli space.

  Typically, the existence of a flat slice of moduli space that satisfies the assumptions of Section \ref{ss.towerpolytope} is related to a discrete symmetry that is preserved along the slice and broken (partially or completely) away from it. This symmetry forces the tower and species vectors of each duality frame along the slice to lie tangent to it. For instance, in the case of M-theory on a rectangular $k$-torus, there are $k$ $\mathbb{Z}_2$ symmetries reflecting each of the $k$ circles (combined with spacetime parity in the external directions), under which the metric and $C_3$ axions for the corresponding directions are charged. There is also an 11d $\mathbb{Z}_2$ CP symmetry combining $C_3 \to -C_3$ with spacetime parity, under which all of the $C_3$ axions are charged. Only when all of the axions vanish are all of these discrete symmetries unbroken.

More generally, setting all of the axions to fixed, non-zero values typically also leads to flat (or, at least, asymptotically flat) slices. However, the tower and species vectors for the various dual descriptions will not always admit good projections onto the tangent space. For instance, in the case of M-theory on a torus, setting the metric axions to fixed, non-zero values generates a globally flat slice, and the frame simplex generated by the KK modes admits a good projection. In general, however, the tower and species vectors of the other duality frames in this slice do not admit good projections.\footnote{In fact, when the axions are fixed at irrational values, some of the ``asymptotic'' directions in such a slice will turn out to wind endlessly around the moduli space without traveling to infinite distance, leading to more drastic violations of the assumptions of Section \ref{s.rules}.}

However, when there is a discrete symmetry enhancement along the chosen axion slice, then we can once again safely fix the axions to these values. For instance, the $C_0 = 1/2$ slice of type IIB moduli space, fixed by $\bigl(\begin{smallmatrix}1 & -1 \\ 0 & -1 \end{smallmatrix}\bigr) \in GL(2,\mathbb{Z})$, is just as suitable for our taxonomy as the $C_0 = 0$ slice. For the same reason, we can also consider compactification on non-rectangular tori with suitable discrete isometries, such as $T^2$ with complex structure $\tau = e^{2\pi i/3}$.

Let us emphasize, however, that even when there is not an asymptotically flat slice of the moduli space, our rules can still be applied in a particular asymptotic region associated to a given set of light principal towers. In such cases, while we cannot construct the full tower polytope, we can still construct the frame simplex. This situation can occur, for instance, when compactifying on a manifold with a non-trivial fibration. In the limit where the base grows large in comparison to the fiber, the manifold looks locally like a product, and there is an approximately-flat radion-radion moduli space. However, the opposite limit in which the fiber grows relative to the base is more non-trivial, and it may even be obstructed.

Our work does not rule out the possibility of constructing and classifying tower polytopes in the absence of globally asymptotically-flat slices of the moduli space, though this may require some revision of the rules described above.
As a first step in this direction, in Appendix \ref{app:K3} we derive the convex hull of the leading principal towers for M-theory on K3, which corresponds to half of one of the tower polytopes of Section \ref{s.2dclassification}.\footnote{Since M-theory on K3 is dual to heterotic string theory on $T^3$, we expect that globally flat slices of moduli space do exist in this example, but they are not obvious from the M-theory perspective.}
Similar phenomena occur in Calabi-Yau threefold compactifications, see Appendix \ref{a.GeodesicallyIncomplete} and \cite{Castellano:2023jjt} for some examples.

\subsection{The direction vector and the principal
plane}\label{subsec:parallel}

As mentioned briefly in Section \ref{subsec:SDC}, the taxonomy rules
\eqref{eqn:taxonomyRules}, \eqref{eqn:dirrule} do not constrain the component
of the direction vector $\hat{t}^i = \frac{d \phi^i}{d s}$
perpendicular to the principal plane. This is not relevant for the derivation of the taxonomy rules, but it is 
essential if one wants to bound the rates at which the various towers become light asymptotically. Having $\hat{t}$ not lying within the principal plane would result in $\theta\neq 0$ in  \eqref{e.theta}, which could result in a violation of the Sharpened Distance Conjecture. In this section, we will explain in greater detail the difficulties involved with proving that $\hat{t}$ must lie parallel to the principal plane.

To begin, consider the decompactification along a manifold $X_n$ associated to
the leading KK tower in some regular infinite-distance limit.
At large volume, vacua in the moduli space are semiclassical backgrounds of
the $D = d + n$ dimensional theory with $d$-dimensional Lorentz invariance,
with the metric on moduli space determined by fluctuations about these
backgrounds. Because $X_n$ is asymptotically empty, the geometric (size and
shape) moduli of $X_n$ are independent of the other moduli, as are the
$D$-dimensional moduli. Thus, ignoring the axion and brane moduli, the moduli
space projects down to a direct product:
\begin{equation}
  \mathcal{M}_{\text{base}} =\mathbb{R}_{\text{rad}} \times
  \mathcal{M}_{\text{shape}} \times \mathcal{M}_{(D)} .
\end{equation}
For each point in $\mathcal{M}_{\text{base}}$, there is a moduli space
$\mathcal{M}_{\text{axion} / \text{brane}}$ of axion and brane moduli, where
the size and shape of this space depends on the choice of point in
$\mathcal{M}_{\text{base}}$. (For instance, the axions inherit an
$\mathcal{M}_{(D)}$-dependent metric from the $D$-dimensional gauge kinetic
terms).\footnote{Note that there is no clean separation between the axion
(Wilson lines) and brane moduli spaces. The branes can source fluxes that
alter the Wilson lines as they are moved around, and the Wilson lines can
create monodromies in the brane moduli space.} Thus, in the large volume limit
the $d$-dimensional moduli space $\mathcal{M}_{(d)}$ is a fibration:
\begin{equation}
  \mathcal{M}_{\text{axion} / \text{brane}} \hookrightarrow \mathcal{M}_{(d)}
  \rightarrow (\mathbb{R}_{\text{rad}} \times \mathcal{M}_{\text{shape}}
  \times \mathcal{M}_{(D)}) .
\end{equation}
Note that $\mathcal{M}_{\text{axion} / \text{brane}}$ refers only to the
axions and brane moduli \emph{associated to $X_n$}. Other ``axions,''
etc., with a different, higher-dimensional origin are included in
$\mathcal{M}_{(D)}$.

With this, the statement that the direction vector $\hat t$ lies within the principal plane is tantamount to the condition that its components in the axion/brane directions $\mathcal{M}_{\text{axion} / \text{brane}}$ and shape directions $\mathcal{M}_{\text{shape}}$ vanish, and its components in the higher-dimensional moduli space $\mathcal{M}_{(D)}$ are either radial directions of a further compactification, or else dilatonic directions associated with a fundamental string.

From our regularity assumptions in the previous section, we may safely conclude that the components of the direction vector in the shape directions $\hat{t}_{\rm shape}$ must vanish, since nonzero $\hat{t}_{\rm shape}$ would imply a splitting of the Kaluza-Klein scale into multiple parametrically different scales, violating the assumption of non-degeneracy. However, it is more difficult to show that the components of $\hat t$ vanish in the $\mathcal{M}_{\text{axion} / \text{brane}}$ directions of moduli space, particularly in cases where there is nontrivial mixing between the brane moduli and axions. Similarly, while an inductive argument could be used in principle to argue $\hat t$ points only in the radial directions of $\mathcal{M}_{(D)}$ for a Planckian phase, the case of a stringy phase is more challenging. We leave a thorough investigation of this issue for future work.

\section{Applications\label{s.Applications}}

In this section, we provide a classification of tower and species polytopes in various dimensions when the following conditions hold: (1) we have a globally flat slice of moduli space where the tower-vectors are constant functions of the moduli space,\footnote{Note that this is not a new assumption but rather a quick summary of Assumptions \ref{assump1} to \ref{assump7} from Section \ref{ss.assumpt}} (2) quantum gravity theories do not exist in more than eleven dimensions, and that (3) string theories do not exist in more than ten dimensions. We find that only a finite list of tower (or species) polytopes are consistent with our taxonomic rules. Many of the allowed polytopes are realized in maximal supergravity in higher dimensions, and they satisfy the recursive relations of Section \ref{s.inf dist str}. However, some of polytopes allowed by our classification have not yet been observed in the string landscape. This raises the question of whether additional constraints should be imposed to eliminate these possibilities, or instead whether these polytopes describe new, undiscovered regions of the quantum gravity landscape.

\subsection{Classification of polygons in diverse dimensions\label{s.2dclassification}}

We begin by considering 2d moduli spaces in various dimensions. These 2d moduli spaces may viewed as slices of higher-dimensional moduli spaces, and if these higher-dimensional moduli spaces are flat, then the polygons we find in the 2d moduli spaces are simultaneously slices and good projections of higher-rank polytopes. This will be illustrated in examples when we deal with higher-rank polytopes in Section \ref{s.fullpolytope}.

Following our taxonomy rules above, the vertices of the full tower polytope must come from KK modes, since string-oscillator modes have tower vectors of length $\frac{1}{\sqrt{d-2}}$, and thus lie in the interior of a facet of the tower polytope (see Figure \ref{f.sketch}). This convex hull must be generated by at least three KK tower vectors.

A tower polygon may be specified by a tuple of extended natural numbers
\begin{align}
	\vec n=(n_1,\dots,n_k),
\end{align}
where $n_i > 0$ represents the KK modes of $n_i$ compact dimensions, $n_i=\infty$ represents string oscillation modes, and $k\geq 3$ is the number of vertices. This $n_i=\infty$ notation for string oscillation modes follows from the fact that $|\vec{\zeta}_{\rm osc}|=\lim_{n\to\infty}|\vec{\zeta}_{{\rm KK},\;n}|$, see \eqref{vertexlengths}. Moreover, two string oscillator vertices cannot be connected to each other by an edge in the polytope, so $n_i$ and $n_{i+1}$ cannot both be infinite for any $i$.

Following \eqref{eqn:taxonomyRules}, the angle between consecutive vertices $\vec\zeta_i$ and $\vec\zeta_{i+1}$ is given by
\begin{align}\label{vertexangles again}
	\theta_{i} \equiv \arccos \sqrt{\frac{n_in_{i+1}}{(n_i+d-2)(n_{i+1}+d-2)}}.
\end{align}
With this, for a given $\vec n$-tuple, we can define the angle summation function $\Sigma_\theta(\vec n)$ as
\begin{align}
	\Sigma_\theta(\vec n)= \sum_{i=1}^n \theta_{i} = \sum_{i=1}^k\arccos \sqrt{\frac{n_in_{i+1}}{(n_i+d-2)(n_{i+1}+d-2)}},
\end{align}
where $n_{k+1} \equiv n_1$.
For $\vec n$ to give a consistent polygon, the interior angles must sum up to $2\pi$, so 
\begin{align}
	\Sigma_\theta(\vec n)=2\pi \,.
\end{align}

In the remainder of this subsection, we derive the allowed $\vec{n}$-tuples, up to permutation, for $d=9,\, 8,\,7,\, 6$, respectively. As stated above, we assume that we can decompactify our theory to no more than eleven dimensions, and we assume that any eleven-dimensional theory has no strings. The former assumption implies that neighboring KK-vertices with $n_i = p$, $n_{i+1} = q$ must satisfy $p+q\leq 11$. The latter requirement prohibits a string vertex from neighboring a vertex with $n_i = 11-d$. These assumptions are well motivated by the known string landscape, though it may prove interesting to explore the additional possibilities that arise when these assumptions are relaxed.

The results of our analysis--namely, the full list of 2d polygons $\Pi_{(d,\cdot)}$ allowed by our rules and assumptions in dimensions 9, 8, 7, and 6--are depicted in Table \ref{tab:BIG}. The associated species polygons $\Pi^\circ_{(d,\cdot)}$ are analogously depicted in Table \ref{tab:BIGsp}. We now present a derivation of the results shown in these tables, beginning in dimension 9.

\begin{sidewaystable}
\centering
\begin{tabular}{cccccccccccccccccccccccccccccc}\hline
 \multicolumn{6}{|c|}{$d=9$}&
 \multicolumn{12}{c|}{$d=8$}&
 \multicolumn{12}{c|}{$d=7$}\\\hline
 \multicolumn{6}{|c|}{
\includegraphics[scale=0.3]{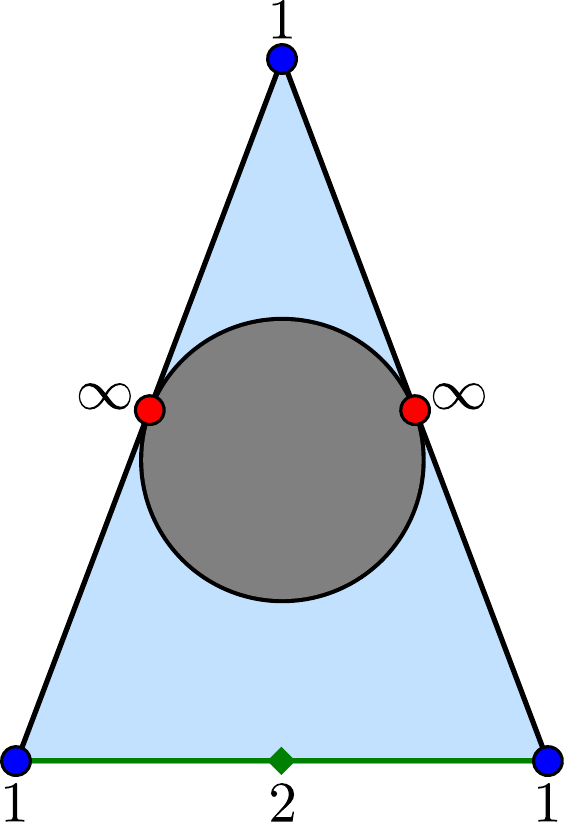}}&
 \multicolumn{6}{c}{\includegraphics[scale=0.35]{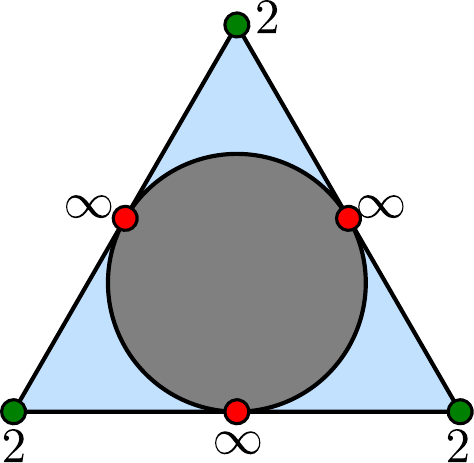}}&
 \multicolumn{6}{c|}{\includegraphics[scale=0.35]{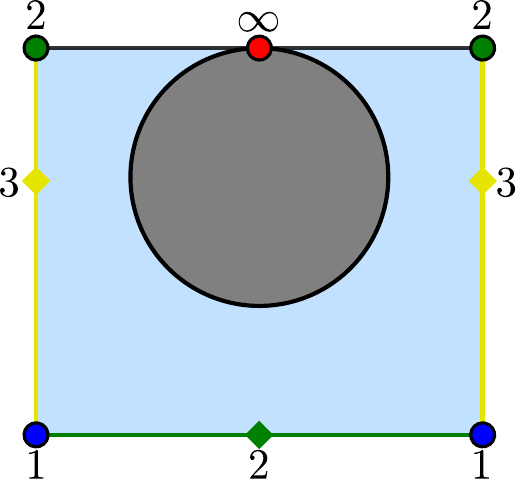}}&
 \multicolumn{6}{c}{\includegraphics[scale=0.35]{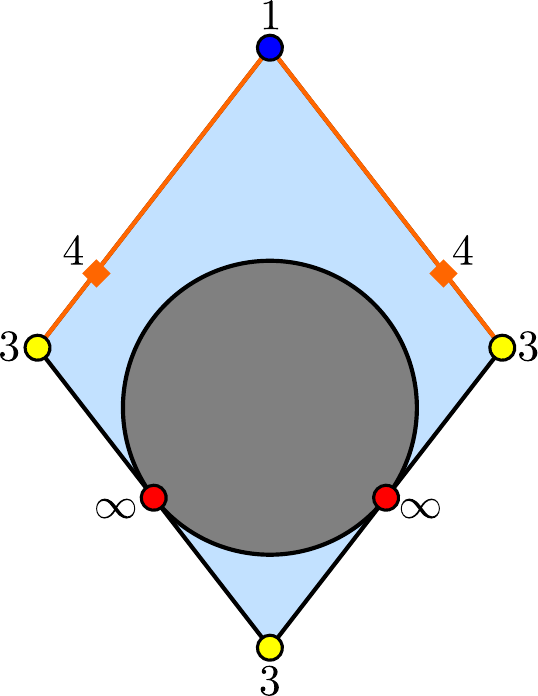}}&
 \multicolumn{6}{c|}{\includegraphics[scale=0.35]{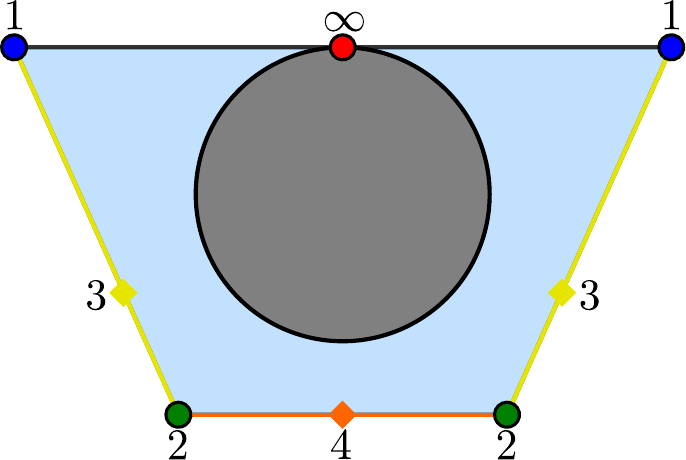}}\\
 \multicolumn{6}{|c|}{$\mathcal{P}_{(9)}=\Pi_{(9,I)}$}&
 \multicolumn{6}{c}{$\Pi_{\rm (8,I)}$}&
 \multicolumn{6}{c|}{$\Pi_{\rm (8,II)}$}&
 \multicolumn{6}{c}{$\Pi_{\rm (7,I)}$}&
 \multicolumn{6}{c|}{$\Pi_{\rm (7,II)}$}\\\hline
 \multicolumn{30}{|c|}{$d=6$}\\\hline
 \multicolumn{6}{|c}{\includegraphics[scale=0.35]{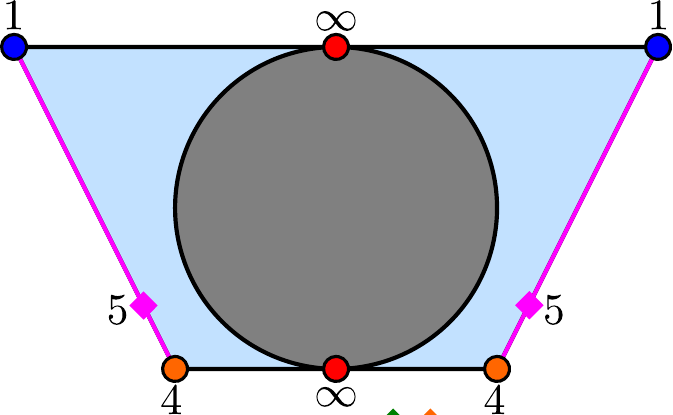}}&
 \multicolumn{6}{c}{\includegraphics[scale=0.35]{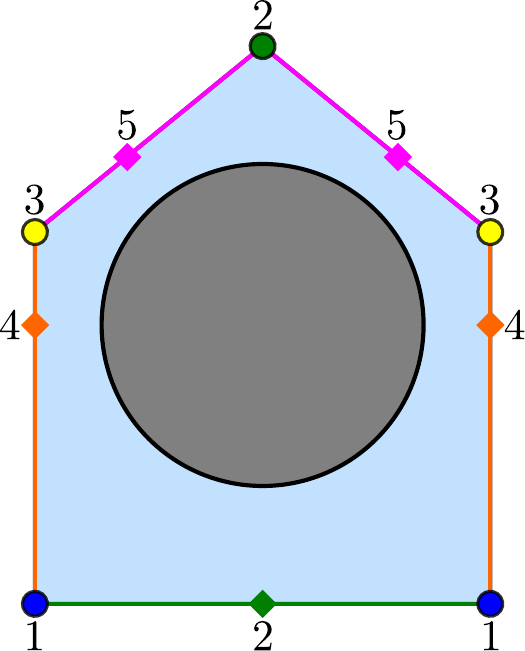}}&
 \multicolumn{6}{c}{\includegraphics[scale=0.35]{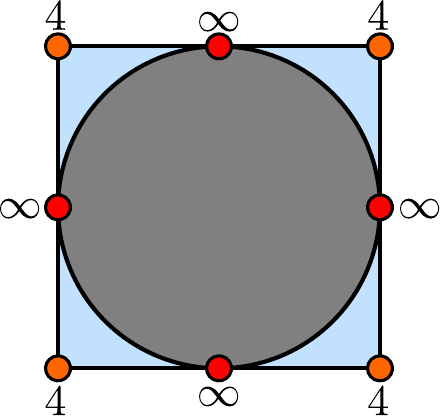}}&
 \multicolumn{6}{c}{\includegraphics[scale=0.35]{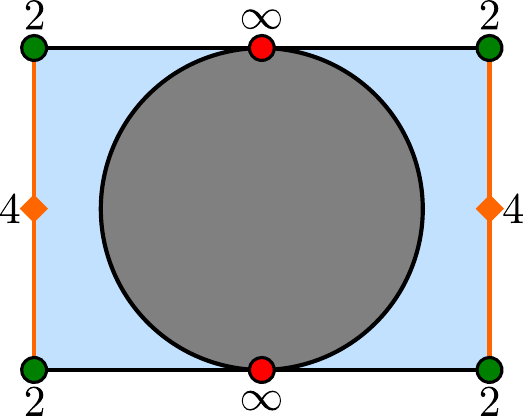}}&
 \multicolumn{6}{c|}{\includegraphics[scale=0.35]{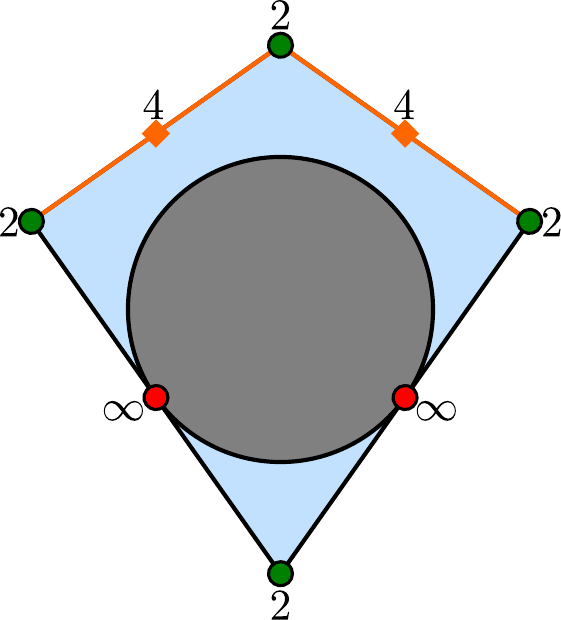}}\\
 \multicolumn{6}{|c}{$\Pi_{\rm (6,I)}$}&
 \multicolumn{6}{c}{$\Pi_{\rm (6,II)}$}&
 \multicolumn{6}{c}{$\Pi_{\rm (6,III)}$}&
 \multicolumn{6}{c}{$\Pi_{\rm (6,IV)}$}&
 \multicolumn{6}{c|}{$\color{red}\Pi_{\rm (6,V)}$}\\\hline
\multicolumn{30}{|c|}{$d=6$ (extra)}\\\hline
\multicolumn{5}{|c}{\includegraphics[scale=0.35]{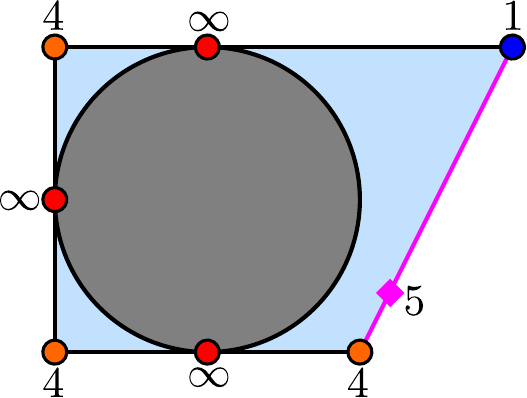}}&
 \multicolumn{5}{c}{\includegraphics[scale=0.35]{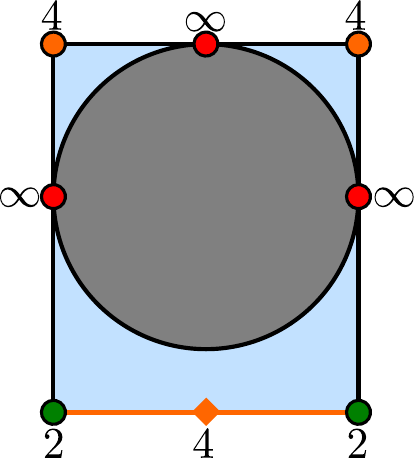}}&
 \multicolumn{5}{c}{\includegraphics[scale=0.35]{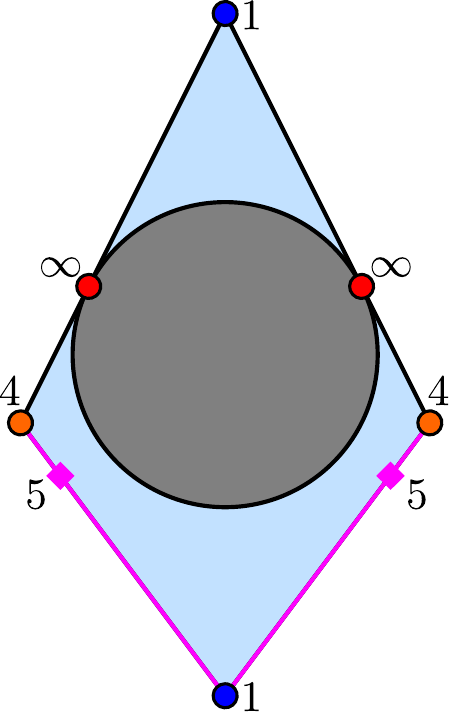}}&
 \multicolumn{5}{c}{\includegraphics[scale=0.35]{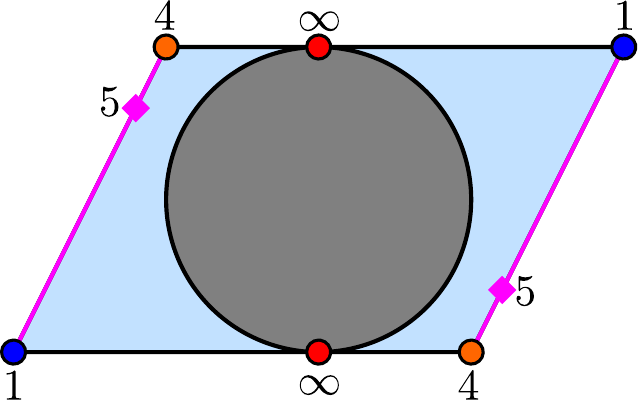}}&
 \multicolumn{5}{c}{\includegraphics[scale=0.35]{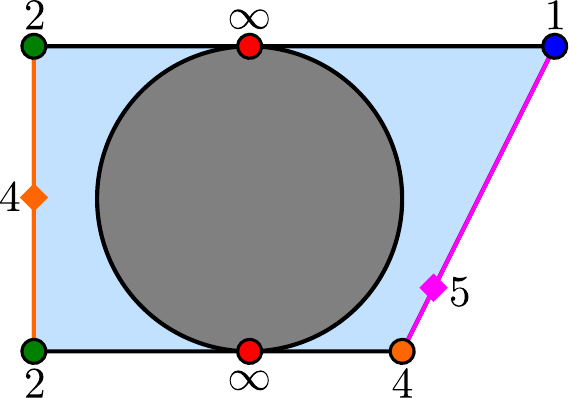}}&
 \multicolumn{5}{c|}{\includegraphics[scale=0.35]{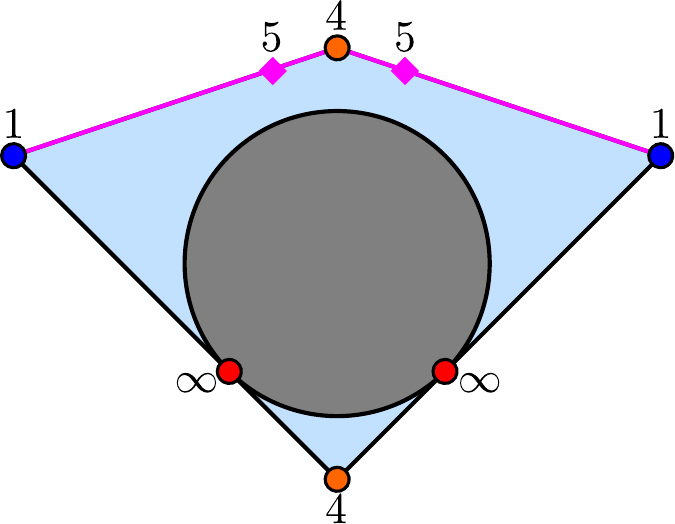}}\\
 \multicolumn{5}{|c}{$\color{red}\Pi_{\rm (6,VI)}$}&
 \multicolumn{5}{c}{$\color{red}\Pi_{\rm (6,VII)}$}&
 \multicolumn{5}{c}{$\color{red}\Pi_{\rm (6,VIII)}$}&
 \multicolumn{5}{c}{$\color{red}\Pi_{\rm (6,IX)}$}&
 \multicolumn{5}{c}{$\color{red}\Pi_{\rm (6,X)}$}&
 \multicolumn{5}{c|}{$\color{red}\Pi_{\rm (6,XI)}$}\\\hline
\end{tabular}
\caption{2d tower polygons for 9, 8, 7, and 6 dimensional theories, with 11d as the maximum decompactification dimension. String oscillators depicted in red \fcolorbox{black}{red}{\rule{0pt}{6pt}\rule{6pt}{0pt}}, while one, two, three, four and five dimensional KK modes appear in blue \fcolorbox{black}{blue}{\rule{0pt}{6pt}\rule{6pt}{0pt}}, green \fcolorbox{black}{dark-green}{\rule{0pt}{6pt}\rule{6pt}{0pt}}, yellow \fcolorbox{black}{yellow}{\rule{0pt}{6pt}\rule{6pt}{0pt}}, orange \fcolorbox{black}{orange}{\rule{0pt}{6pt}\rule{6pt}{0pt}} and pink \fcolorbox{black}{Magenta}{\rule{0pt}{6pt}\rule{6pt}{0pt}}, respectively. The point closest to the origin of edges associated to decompactification of several dimensions is highlighted. The disk of radius $\frac{1}{\sqrt{d-2}}$ is depicted in gray. The polygons of which no string embedding is known are highlighted in red.\label{tab:BIG}}
\end{sidewaystable}

\begin{sidewaystable}
\centering
\begin{tabular}{cccccccccccccccccccccccccccccc}\hline
 \multicolumn{6}{|c|}{$d=9$}&
 \multicolumn{12}{c|}{$d=8$}&
 \multicolumn{12}{c|}{$d=7$}\\\hline
 \multicolumn{6}{|c|}{
\includegraphics[scale=0.45]{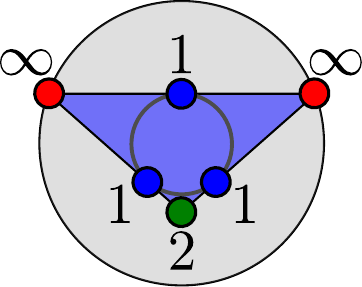}}&
 \multicolumn{6}{c}{\includegraphics[scale=0.5]{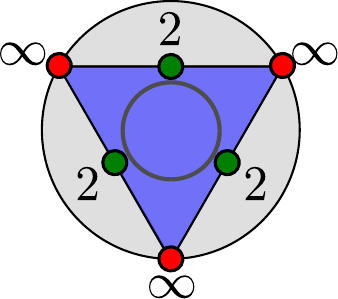}}&
 \multicolumn{6}{c|}{\includegraphics[scale=0.5]{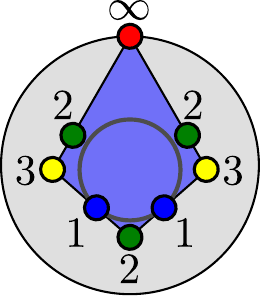}}&
 \multicolumn{6}{c}{\includegraphics[scale=0.5]{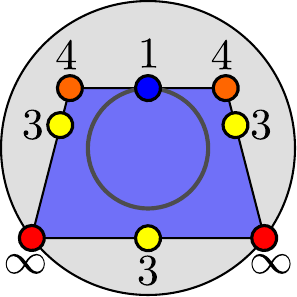}}&
 \multicolumn{6}{c|}{\includegraphics[scale=0.5]{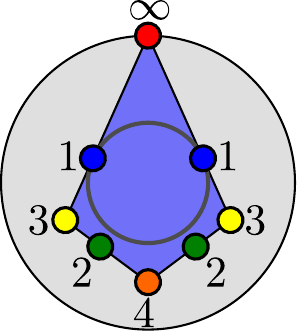}}\\
 \multicolumn{6}{|c|}{$\mathcal{P}^\circ_{(9)}=\Pi^\circ_{(9,I)}$}&
 \multicolumn{6}{c}{$\Pi^\circ_{\rm (8,I)}$}&
 \multicolumn{6}{c|}{$\Pi^\circ_{\rm (8,II)}$}&
 \multicolumn{6}{c}{$\Pi^\circ_{\rm (7,I)}$}&
 \multicolumn{6}{c|}{$\Pi^\circ_{\rm (7,II)}$}\\\hline
 \multicolumn{30}{|c|}{$d=6$}\\\hline
 \multicolumn{6}{|c}{\includegraphics[scale=0.5]{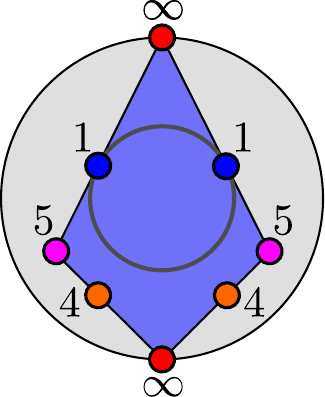}}&
 \multicolumn{6}{c}{\includegraphics[scale=0.5]{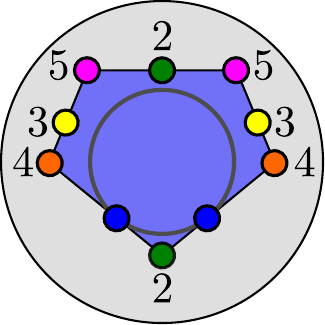}}&
 \multicolumn{6}{c}{\includegraphics[scale=0.5]{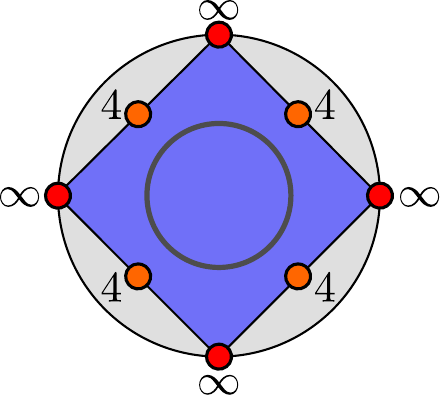}}&
 \multicolumn{6}{c}{\includegraphics[scale=0.5]{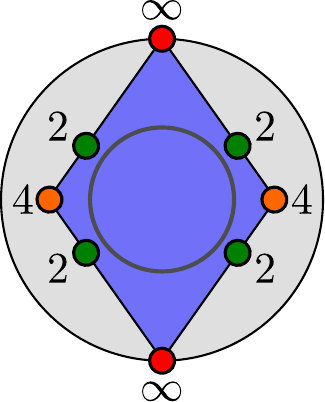}}&
 \multicolumn{6}{c|}{\includegraphics[scale=0.5]{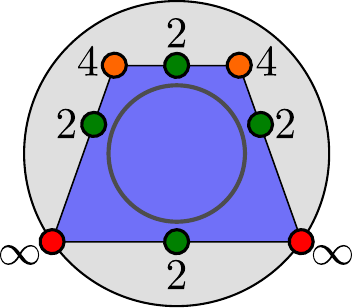}}\\
 \multicolumn{6}{|c}{$\Pi^\circ_{\rm (6,I)}$}&
 \multicolumn{6}{c}{$\Pi^\circ_{\rm (6,II)}$}&
 \multicolumn{6}{c}{$\Pi^\circ_{\rm (6,III)}$}&
 \multicolumn{6}{c}{$\Pi^\circ_{\rm (6,IV)}$}&
 \multicolumn{6}{c|}{$\color{red}\Pi^\circ_{\rm (6,V)}$}\\\hline
\multicolumn{30}{|c|}{$d=6$ (extra)}\\\hline
\multicolumn{5}{|c}{\includegraphics[scale=0.5]{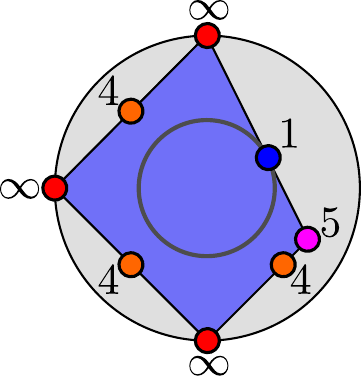}}&
 \multicolumn{5}{c}{\includegraphics[scale=0.5]{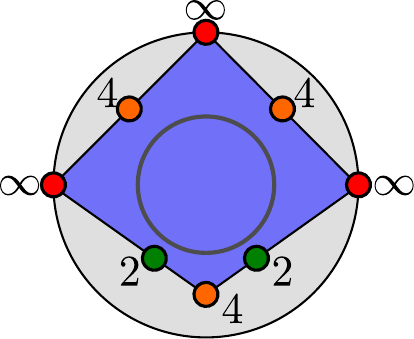}}&
 \multicolumn{5}{c}{\includegraphics[scale=0.5]{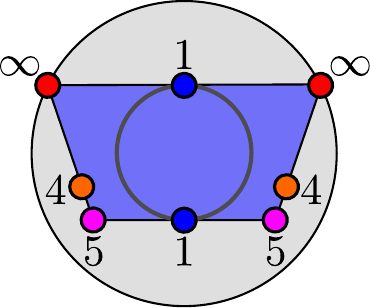}}&
 \multicolumn{5}{c}{\includegraphics[scale=0.5]{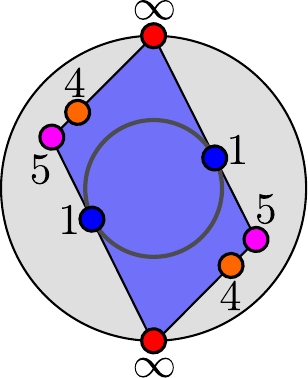}}&
 \multicolumn{5}{c}{\includegraphics[scale=0.5]{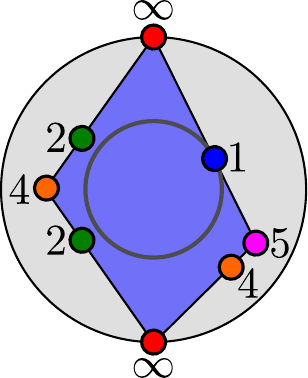}}&
 \multicolumn{5}{c|}{\includegraphics[scale=0.5]{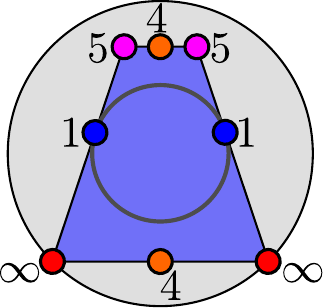}}\\
 \multicolumn{5}{|c}{$\color{red}\Pi^\circ_{\rm (6,VI)}$}&
 \multicolumn{5}{c}{$\color{red}\Pi^\circ_{\rm (6,VII)}$}&
 \multicolumn{5}{c}{$\color{red}\Pi^\circ_{\rm (6,VIII)}$}&
 \multicolumn{5}{c}{$\color{red}\Pi^\circ_{\rm (6,IX)}$}&
 \multicolumn{5}{c}{$\color{red}\Pi^\circ_{\rm (6,X)}$}&
 \multicolumn{5}{c|}{$\color{red}\Pi^\circ_{\rm (6,XI)}$}\\\hline
\end{tabular}
\caption{2d species polygons for 9, 8, 7, and 6 dimensional theories, with 11d as the maximum decompactification dimension. Species scales given by the string scale are depicted in red \fcolorbox{black}{red}{\rule{0pt}{6pt}\rule{6pt}{0pt}}, while Planck masses in 1, 2, 3, 4 and 5 dimensions more appear in blue \fcolorbox{black}{blue}{\rule{0pt}{6pt}\rule{6pt}{0pt}}, green \fcolorbox{black}{dark-green}{\rule{0pt}{6pt}\rule{6pt}{0pt}}, yellow \fcolorbox{black}{yellow}{\rule{0pt}{6pt}\rule{6pt}{0pt}}, orange \fcolorbox{black}{orange}{\rule{0pt}{6pt}\rule{6pt}{0pt}} and pink \fcolorbox{black}{Magenta}{\rule{0pt}{6pt}\rule{6pt}{0pt}}, respectively. The inner and outer disks have of radii $\frac{1}{\sqrt{d-2}}$ and $\frac{1}{\sqrt{(d-1)(d-2)}}$.  The polygons of which no string embedding is known are highlighted in red. \label{tab:BIGsp}}
\end{sidewaystable} 

\clearpage

\subsubsection*{9d}
Our approach to classifying the allowed tower polygons begins by finding the maximum number of vertices allowed for an $\vec n$-tuple. This is achieved by noting that the angles between the respective vertices cannot sum to greater than $2\pi$. For each dimension $d$, there exists a critical number of vertices $k_{\rm max}$ such that all $n$-tuples with length $k > k_{\rm max}$ have $\Sigma_{\theta}(\vec n) > 2\pi$. This leaves a finite list of polygons to check. We exhaust this list, and thus produce all the tower polygons allowed by our taxonomy.

For nine-dimensional theories, there cannot be an $\vec n$-tuple with six or more vertices. To see this, note that the six-component $\vec n$-tuple with the shortest angle summation function $\Sigma_\theta$ is\footnote{Recall that $n_i=\infty$ corresponds to a string oscillator vertex. Recall also that no more than two dimensions may be decompactified in any limit, since we assume that eleven is the maximum number of dimensions after decompactification.} $(2,\infty,2,\infty,2,\infty)$, but
\begin{align}
	\Sigma_\theta(2,\infty,2,\infty,2,\infty)\approx 1.03124\times 2\pi>2\pi,
\end{align}
Since the angle summation function for $\vec n$-tuples of seven or more components are strictly larger than this one, we conclude that the $\vec n$-tuple must have five or fewer vertices.

This leaves a finite list of possible $\vec n$-tuples to check, and there is only one $\vec n$-tuple (up to cyclic permutation) allowed, given by
\begin{align}
	\vec n=(1,1,\infty,1,\infty)\;,
\end{align}
corresponding to $\mathcal{P}_{(9)}=\Pi_{(9,I)}$ in Table \ref{tab:BIG}, which is precisely the example depicted in Figure \ref{f.alphas} from Type IIB string theory on a circle (or M-theory on a 2-torus)! As 
described in Section \ref{s:sliding} and depicted in Figure \ref{f.slidingSO32}, this is also the convex hull of $SO(32)$ heterotic string theory compactified on $S^1$ for asymptotic limits far away from the self-dual line.

The polygon $\mathcal{P}_{(9)}=\Pi_{(9,I)}$ describes multiple different decompactification limits, which lead to different theories. The limits that correspond to motion in the direction of any of the KK-vertices describe decompactification to type IIA or IIB string theories, whose polytopes are depicted in Figures \ref{fig:NOsdual10d} and \ref{fig:sdual10d}. The decompactification in the direction of the shortest facet corresponds to $T^2$ decompactification to 11d M-theory.

As explained in Section \ref{s.inf dist str}, the shape of the tower polygon uniquely fixes the shape of the species polygon $P^\circ$, which controls the asymptotic behavior of the species scale. The species polygon associated to the 9d tower polytope is given by  $\mathcal{P}_{(9)}^\circ=\Pi^\circ_{\rm (9,I)}$ in Table \ref{tab:BIGsp}.

\subsubsection*{8d}
We next consider the allowed tower polygons for an 8d theory. First, $\vec n$-tuples for this theory cannot have more than six components. To see this, note that the seven-component $\vec n$-tuple with the shortest angle summation function $\Sigma_\theta$ is given by $(3, \infty, 3,$ $\infty, 3, \infty,3)$, but
\begin{align}
\Sigma_\theta(3, \infty, 3, \infty, 3,\infty,3)\approx 1.10817\times 2\pi>2\pi.
\end{align}
This leaves a finite list of possible $\vec n$-tuples to check. Ultimately, we find only two options for $\vec n$-tuples up to cyclic permutation, which are given by
\begin{align}
	\vec n\in\{(2,\infty,2,\infty,2,\infty),\ (1,1,2,\infty,2)\}\;.
\end{align}
These correspond respectively to polygons $\Pi_{\rm(8,I)}$ and $\Pi_{\rm(8,II)}$ in Table \ref{tab:BIG}. Both of these polygons are realized as a planar slice of the tower polytope of maximal SUGRA in 8d, as depicted in Figure \ref{f.symFixed8d}, and indeed they correspond to the two inequivalent fixed planes of the symmetry group $\mathsf{G}_8=S_3\times S_2$. The associated species polytopes $\Pi^\circ_{\rm(8,I)}$ and $\Pi^\circ_{\rm(8,II)}$ appear in Table \ref{tab:BIGsp}.

\subsubsection*{7d}

In 7d, there are no $\vec n$-tuples allowed with more than seven components. The eight-component $\vec n$-tuple with the shortest angle summation function $\Sigma_\theta$ is $(4,\infty,4,\infty,4,\infty,$ $4,\infty)$, but
\begin{align}
	\Sigma_\theta(4,\infty,4,\infty,4,\infty,4,\infty)\approx 1.07088\times 2\pi>2\pi.
\end{align}
 This means that the $\vec n$-tuple must have at most seven components, which leaves a finite list of possible $\vec n$-tuples to investigate. In the end, there are only two possible $\vec n$-tuples up to cyclic permutation, which are given by
\begin{align}
	\vec n&\in\{(1,3,\infty,3,\infty,3),(1,2,2,1,\infty)\}\;.
\end{align}
 These correspond respectively to $\Pi_{\rm(7,I)}$ and $\Pi_{\rm(7,II)}$ in Table \ref{tab:BIG}. The associated species polytopes can be found in Table \ref{tab:BIGsp}.

As discussed in more depth in Appendix \ref{s.symmetry}, $\Pi_{\rm(7,I)}$ and $\Pi_{\rm(7,II)}$ can be recovered from the 4d polytope associated with M-theory on $T^4$ (introduced in Section \ref{s.fullpolytope}, see Figure \ref{fig: 7from8}), representing the 2-dimensional loci that are invariant under the $S_5$ symmetry group. Furthermore, $\Pi_{\rm(7,II)}$ also appears in a two-dimensional slice of the moduli space of M-theory on K3, as shown in Appendix \ref{app:K3}.

\subsubsection*{6d}
We finally turn our attention to tower polygons in $d=6$ dimensions. In this case, by an analogous argument to the ones above, $\vec n$-tuples can have at most eight components. This leaves a finite list of possible $\vec n$-tuples to check.

This time, we find eleven options of $\vec n$-tuples, up to cyclic permutation. These are given by
\begin{align}\label{eq.sols 6d}
	\vec n\in\{&
	(1,\infty,1,4,\infty,4),
	(2,3,1,1,3),
	(4,\infty,4,\infty,4,\infty,4,\infty),
	(2,\infty,2,2,\infty,2),
	\nonumber\\&
	(2,2,\infty,2,\infty,2),
	(4,\infty,1,4,\infty,4,\infty),
	(4,\infty,4,\infty,2,2,\infty),
	(1,\infty,4,1,4,\infty),
	\nonumber\\	&
	(4,\infty,1,4,\infty,1)
	(2,\infty,1,4,\infty,2),
	(4,1,\infty,4,\infty,1)
	\}.
\end{align}
These options correspond respectively to polygons $\Pi_{\rm(6,I)}$ through $\Pi_{\rm(6,XI)}$ in Table \ref{tab:BIG}.
Of these possibilities, only the first four (polygons $\Pi_{\rm(6,I)}$ through $\Pi_{\rm(6,IV)}$) can be recovered from slices of the polytope describing M-theory on $T^5$ with all axions set to zero.

The list of polygons in 6d features several novelties that were not observed in the higher-dimensional cases above.
To begin, whereas all of the 7d-9d polygons feature string oscillator vertices, in 6d it is possible to have consistent $\vec n$-tuples with no string oscillator vertices (namely, polygon $\Pi_{\rm(6,II)}$).

Furthermore, whereas any string oscillator vertex in 7d-9d lies on an edge connecting KK modes of the same number of decompactifying dimensions, in 6d six of the possibilities (namely, $\Pi_{\rm(6,VI)}$ to $\Pi_{\rm(6,XI)}$) allow string oscillator vertices to be located along edges spanned by KK vertices of different $n$. None of these six polygons are realized as slices in the M-theory compactifications studied in this work or other polygons found in the literature. On the other hand, our current rules do not exclude them. It may be possible to rule out these polygons, perhaps using sigma-model/worldsheet CFT methods. We leave a more thorough investigation of this possibility to future work.

This leaves $\Pi_{\rm(6,V)}$ as the most mysterious possibility remaining. It does not appear as a slice of the polytope for M-theory on $T^5$ with axions turned off, but it also does not feature two KK vertices with different $n$ separated by a string oscillator vertex. One possible reason for excluding this polytope is the fact that it does not reduce nicely to a 3-dimensional polytope in 5d, but rather to one in 4d that only allows for decompactification of an even number of dimensions, as we explain further in Section \ref{app:dim red}. A more optimistic scenario is that $\Pi_{\rm(6,V)}$ may arise in a more exotic context, such that the classification results here do not apply to the resulting 5d theory after dimensional reduction. More work is needed to determine which of these possibilities is the correct one.
 
\subsubsection*{Lower dimensions}

While we do not classify the possible polytopes in $d<6$ in this work, we note in passing the following interesting fact: our rules for angles of the polytope \eqref{vertexangles again} are invariant under a uniform rescaling $(n, d-2) \rightarrow  (\lambda n, \lambda (d-2))$ for each vertex. This means that, up to an overall rescaling of the polytope by a factor of $\sqrt{2}$, the polytopes in $d$ dimensions whose KK vertices are all labeled by even $n$ will reappear in $\frac{d-2}{2}+2$ dimensions, with $n$ replaced by $n/2$. For example, since every KK vertex of $\Pi_{\rm{(8,I)}}$ is labeled by an even number $n$, this polytope will reoccur in 5d, rescaled by an overall factor of $\sqrt{2}$. Since every vertex of $\Pi_{\rm{(6,III)}}$, $\Pi_{\rm{(6,IV)}}$, $\Pi_{\rm{(6,V)}}$ and $\Pi_{\rm{(6,VII)}}$ is labeled by an even number $n$, these polytopes will reoccur in 4d, rescaled by a factor of $\sqrt{2}$. Furthermore, since every KK vertex of $\Pi_{\rm{(6,III)}}$ has $n=4$, we can rescale $(n, d-2)$ by yet another factor of $\lambda=1/2$, concluding that the same polytope will also appear in 3d, rescaled by an overall factor of 2.

 Of course, not all polygons in $d <6$ dimensions descend from polygons in higher dimensions in this fashion. We defer further study of these lower-dimensional cases to future work.

 \subsection{Classification of maximal tower and species polytopes \label{s.fullpolytope} }
 
If eleven is the maximum number of dimensions allowed after decompactification, then a theory in $d$ dimensions can have at most $11-d$ radion moduli. This means that the tower and species polytopes associated with the slice of moduli space parametrized by these radion moduli can be at most $(11-d)$-dimensional. We will thus use the term \emph{maximal} to describe rank-$(11-d)$ polytopes that obey our taxonomy rules in $(11-d)$-dimensional, flat, geodesically complete slices in $d$ spacetime dimensions.

In this section, we once again assume that eleven is the maximum number of dimensions, there are no strings in the eleven-dimensional theory, and the frame simplices from the various limits of the moduli space can be combined globally into a single tower polytope, where our taxonomy rules apply. We then fully classify the allowed tower and species polytopes of rank $11-d$ for dimensions $d\in\{10,9,8\}$, and we show examples satisfying the taxonomy rules for $d=7$ and 6.  In dimension 10, our classification reproduces the Type IIA and Type IIB polytopes, as well as a polytope not yet known to exist in the string landscape. In dimensions 8 and 9, our classification reproduces all of the tower and species polytopes for the radion moduli of M-theory on orthogonal tori.

The radion-tower vectors from M-theory on orthogonal tori are computed in Appendix \ref{a.MTheory}. In this section, we perform the computation in dimensions 10 through 6.

\subsubsection*{10d\label{s.10d}}

\begin{figure}[htp]
\begin{subfigure}{\textwidth}
\centering
\includegraphics[width=0.6\textwidth]{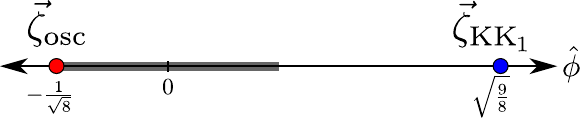}
\caption{$\mathcal{P}_{(10,A)}$: IIA\label{fig:NOsdual10d}}
\end{subfigure}
\bigskip
\begin{subfigure}{\textwidth}
\centering
\includegraphics[width=0.6\textwidth]{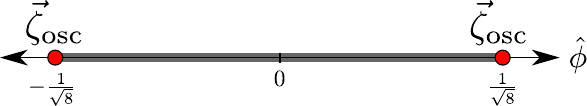}
\caption{$\mathcal{P}_{(10,B)}$: IIB, I, heterotic.\label{fig:sdual10d}}
\end{subfigure}
\bigskip
\begin{subfigure}{\textwidth}
\centering
\includegraphics[width=0.6\textwidth]{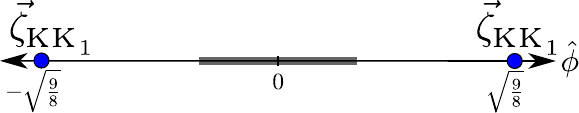}
\caption{$\mathcal{P}_{(10,C)}$: No known example.\label{fig:KKKK10d}}
\end{subfigure}
\caption{Allowed configurations of scalar charge-to-mass ratio vectors $\vec{\zeta}$ for leading towers in $d=10$. In gray the radius $\frac{1}{\sqrt{d-2}}=\frac{1}{\sqrt{8}}$ 1-ball is depicted.\label{fig:10figs}}
\end{figure}

We begin in ten dimensions. Here, we have only a one dimensional moduli space, and the tower polytope is merely a line segment, with edges corresponding to either KK modes or string oscillator modes. There are three possible tower polytopes shown in Figure \ref{fig:10figs}:
\begin{align}
	\mathcal{P}_{(10)}= \begin{cases}
	\mathcal{P}_{(10,A)}=\left\{-\frac{1}{\sqrt{d-2}},\sqrt{\frac{d-1}{d-2}}\right\},&\begin{matrix}\text{one emergent string and one}\vspace{-0.2cm}\\\text{ decompactification limit,}\end{matrix}\\
		\mathcal{P}_{(10,B)}=\left\{\pm\frac{1}{\sqrt{d-2}}\right\},&\text{two emergent string limits,}\\
		\mathcal{P}_{(10,C)}=\left\{\pm \sqrt{\frac{d-1}{d-2}}\right\},&\text{two decompactification limits.}
	\end{cases}
\end{align}

The case where there is one emergent string limit and one decompactification limit, $\mathcal{P}_{(10,A)}$, occurs in Type IIA string theory. There, the emergent string limit is the weak-coupling limit of the theory, and the decompactification limit in the decompactification into M-theory.

\begin{figure}[htp]
\begin{subfigure}{\textwidth}
\centering
\includegraphics[width=0.6\textwidth]{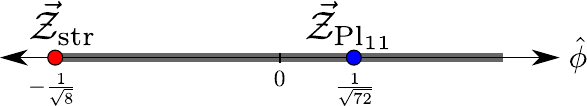}
\caption{$\mathcal{P}_{(10,A)}^\circ$: IIA\label{fig:NOsdual10dsp}}
\end{subfigure}
\bigskip
\begin{subfigure}{\textwidth}
\centering
\includegraphics[width=0.6\textwidth]{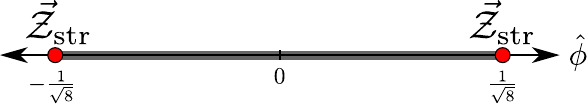}
\caption{$\mathcal{P}_{(10,B)}^\circ$: IIB, I, heterotic.\label{fig:sdual10dsp} }
\end{subfigure}
\bigskip
\begin{subfigure}{\textwidth}
\centering
\includegraphics[width=0.6\textwidth]{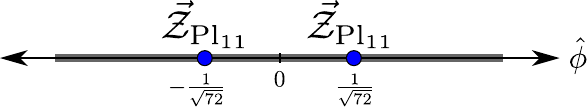}
\caption{$\mathcal{P}_{(10,C)}^\circ$: No known example.\label{fig:KKKK10dsp}}
\end{subfigure}
\caption{Different allowed configurations for scalar charge-to-mass ratio vectors $\vec{\mathcal{Z}}$ for the species scales in $d=10$. In gray the $\frac{1}{\sqrt{d-2}}=\frac{1}{\sqrt{8}}$ 1-disk is depicted.\label{fig:10figssp}}
\end{figure}

Our classification also fixes the species polytope as the dual of the tower polytope. These are given respectively by
\begin{align}
	\mathcal{P}_{(10)}^\circ=\begin{cases}
	\mathcal{P}_{(10,A)}^\circ=\left\{-\frac{1}{\sqrt{d-2}},\frac{1}{\sqrt{(d-1)(d-2)}}\right\},&\begin{array}{c}\text{one emergent string and one} \vspace{-0.2cm}\\\text{decompactification limit,}\end{array}\\
		\mathcal{P}_{(10,B)}^\circ=\left\{\pm\frac{1}{\sqrt{d-2}}\right\},&\text{two emergent string limits,}\\
		\mathcal{P}_{(10,C)}^\circ=\left\{\pm \frac{1}{\sqrt{(d-1)(d-2)}}\right\},&\text{two decompactification limits.}
	\end{cases}
\end{align}
These three species polytopes are depicted in Figure \ref{fig:10figssp}.

The case of two emergent string limits, $\mathcal{P}_{(10,B)}$ occurs in Type IIB string theory (where the two strings are the fundamental strings and D1-branes), and also in Type I and heterotic string theories. In these cases, the slice of moduli space considered here is parametrized by the dilaton.

The remaining case, $\mathcal{P}_{(10,C)}$, features two decompactification limits. No theory with this polytope is known in the landscape.\footnote{Of course, it would be extremely exciting if such a theory does exist, but we leave this possibility to future research.} As will be shown in Section \ref{app:dim red}, this polytope can be recovered from the decompactification of the polygon $\Pi_{\rm (6,XI)}$ in $d=6$ from Table \ref{tab:BIG}, which itself also has no identified occurrence in the string landscape.

\subsubsection*{9d}

\begin{figure}[h]
\begin{center}
\includegraphics[width = 80mm]{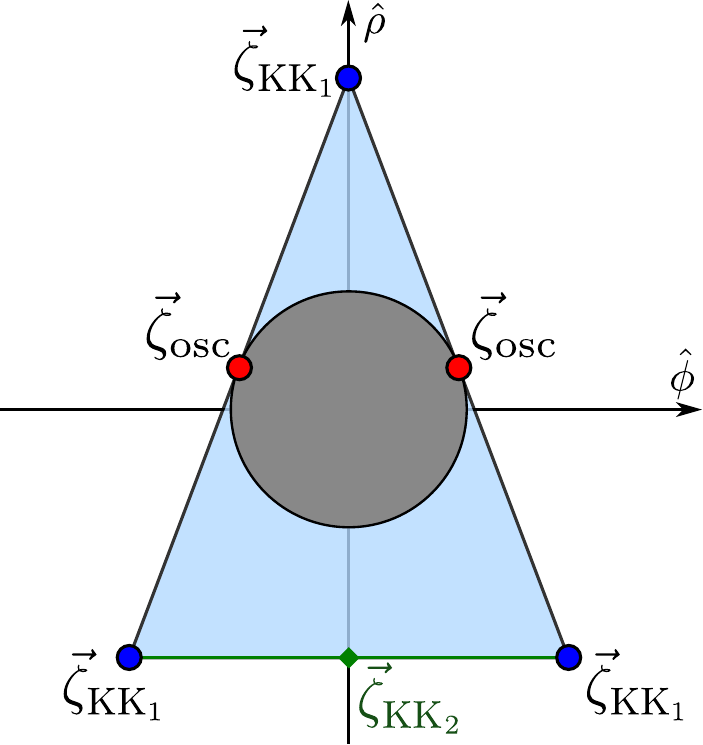}
      \caption{Maximal tower polytope for 9d theories. Red and blue points  respectively correspond to string oscillators and KK towers associated with decompactifying 1 dimension. The edge colored in green, with its closest point highlighted, is associated to decompactification of 2 dimensions. \label{f.9dmaximal}}
\end{center}
\end{figure}

In 9d, the relevant slice of moduli space is now two-dimensional. Thus, our above classification of tower polygons in 9d has already described this case. As depicted in Figure \ref{f.9dmaximal}, the tower polytope is now generated by the following tower vectors,
\begin{align}
	\mathcal{P}_{(9)}=\left\{\left(0,\sqrt{\frac{8}{7}}\right),\left(\pm\frac{1}{ \sqrt{2}},-\frac{3}{\sqrt{14}}\right)\right\}
\end{align}
This tower polytope describes the radion-radion components of M-theory on an orthogonal 2-torus, or Type II string theory on a circle. From an M-theory perspective, the three KK vertices correspond to $\frac{1}{2}$ BPS states, with the top vertex from M2 branes wrapped on $T^2$ and the other two from KK modes of either 1-cycle of the 2-torus. 

Unlike in the 10d case, both the tower and species polytopes are unique. The latter is given by
\begin{align}
	\mathcal{P}_{(9)}^\circ=\left\{\left(0,\frac{1}{\sqrt{56}}\right),\left(\pm\frac{1}{ 4},-\frac{3}{4\sqrt{7}}\right)\right\}\,,
\end{align}
as previously depicted in Figure \ref{f.lambdas} (also $P^\circ_{(9)}$ in Table \ref{tab:BIGsp}). This agrees with previous results in the literature, \cite{Calderon-Infante:2023ler}.

\subsubsection*{8d} 

\begin{figure}[h]
\begin{center}
\includegraphics[width=0.6\textwidth]{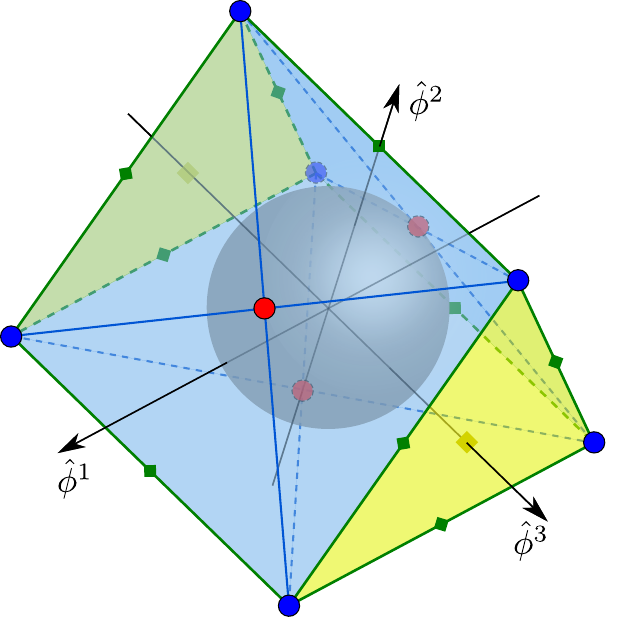}
\caption{Maximal tower polytope of the 8d theory, $\mathcal{P}_{(8)}$. The string and $\vec{\zeta}_{\rm KK_1}$ towers are depicted in red \fcolorbox{black}{red}{\rule{0pt}{6pt}\rule{6pt}{0pt}} and blue blue \fcolorbox{black}{blue}{\rule{0pt}{6pt}\rule{6pt}{0pt}}. The edges and facets associated to decompactification of two and three internal dimensions are depicted in green \fcolorbox{black}{dark-green}{\rule{0pt}{6pt}\rule{6pt}{0pt}} and yellow \fcolorbox{black}{yellow}{\rule{0pt}{6pt}\rule{6pt}{0pt}}, with their closest point to the origin highlighted. The ball of radius $\frac{1}{\sqrt{d-2}}=\frac{1}{\sqrt{6}}$ is presented in gray and the triangulation of the polytope into frame simplices is depicted with blue lines.\label{fig:ch8d}}
\end{center}
\end{figure}

In 8d, the only 3-dimensional tower polytope allowed is generated by six vertices:
\begin{align}
	\mathcal{P}_{(8)}=\left\{\left(0,\pm\frac{1}{\sqrt{2}},\sqrt{\frac{2}{3}}\right),\left(\pm\frac{1}{\sqrt{2}},\pm'\frac{1}{\sqrt{2}},-\frac{1}{\sqrt{6}}\right)\right\}\,,
\end{align}
where the $\pm$ signs are uncorrelated. This is depicted in Figure \ref{fig:ch8d}, and it matches previous results in the literature \cite{Etheredge:2022opl, Castellano:2023jjt}.

The species polytope is also generated by six vertices:
\begin{align}
	\mathcal{P}_{(8)}^\circ=\left\{\left(0,\pm\frac{1}{\sqrt{14}},\sqrt{\frac{2}{21}}\right),\left(\pm\frac{1}{\sqrt{14}},\pm'\frac{1}{\sqrt{14}},-\frac{1}{\sqrt{42}}\right)\right\}\,,
\end{align}
where again the $\pm$ signs are uncorrelated. This is depicted in Figure \ref{fig:ch8d}, and also agree with previous results \cite{Calderon-Infante:2023ler}.

\begin{figure}[htp]
\begin{center}
\includegraphics[width=0.55\textwidth]{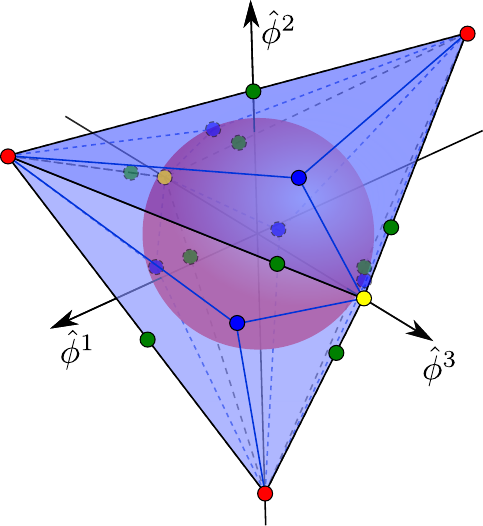}
\caption{Maximal species polytope for 8d theories. Points associated to the string scale \fcolorbox{black}{red}{\rule{0pt}{6pt}\rule{6pt}{0pt}}, while 9, 10 and 11 dimensional Planck mass appear in blue \fcolorbox{black}{blue}{\rule{0pt}{6pt}\rule{6pt}{0pt}}, green \fcolorbox{black}{dark-green}{\rule{0pt}{6pt}\rule{6pt}{0pt}} and yellow \fcolorbox{black}{yellow}{\rule{0pt}{6pt}\rule{6pt}{0pt}}, respectively. The sphere of radius $\frac{1}{\sqrt{(d-1)(d-2)}}=\frac{1}{\sqrt{42}}$ is depicted in red and the triangulation of the species polytope in blue lines.
\label{fig:ch8dsp}}
\end{center}
\end{figure}

\subsection{Illustration of recursion relations\label{s.applications.recursionillustration}}

In this subsection, we show how the tower and species polytopes of higher-dimensional theories are encoded in the tower and species polytopes of lower-dimensional theories.

\begin{figure}
\begin{center}
\includegraphics[width=0.425\textwidth]{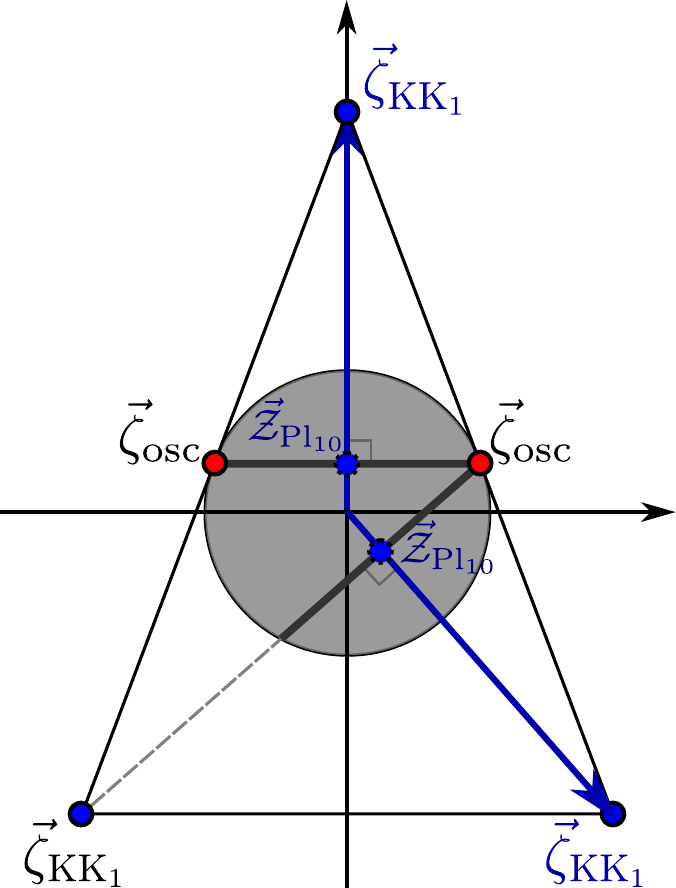}
\caption{Shaded in black, the $\mathcal{P}_{(10,A)}$ (lower) and $\mathcal{P}_{(10,B)}$ (upper) tower polytopes are obtained from $\mathcal{P}_{(9)}$ after decompactifying along the two inequivalent $\vec{\zeta}_{\rm KK_1}$ vertices. The disk of radius $\frac{1}{\sqrt{(d-2)}}=\frac{1}{\sqrt{7}}$ is depicted in gray. 
\label{fig: 10from9}}
\end{center}
\end{figure}

We begin with the maximal tower polytope $\mathcal{P}_{(9)}$ for the 9d theory. This is depicted in Figure \ref{fig: 10from9}. Consider first the two tower vectors that are adjacent to the $\vec \zeta_{\text{KK}_1}$ in the bottom right of the figure; these are labeled by $\vec \zeta_{\text{KK}_1}$ (lower left) and $\vec \zeta_\text{osc}$ (right). As shown in that figure, the components of these two adjacent tower vectors that are perpendicular to $\vec \zeta_{\text{KK}_1}$ form the tower polytope $\mathcal{P}_\text{10,A}$. Alternatively, consider the two tower vectors that are adjacent to the $\vec \zeta_{\text{KK}_1}$ at the top of the triangle; these are each labeled by $\vec \zeta_\text{osc}$. The components of these two adjacent tower vectors that are perpendicular to $\vec \zeta_{\text{KK}_1}$ at the top of the triangle form the tower polytope $\mathcal{P}_\text{10,B}$.

\begin{figure}[h]
\begin{center}
\begin{subfigure}{0.505\textwidth}
\center
\includegraphics[width=\textwidth]{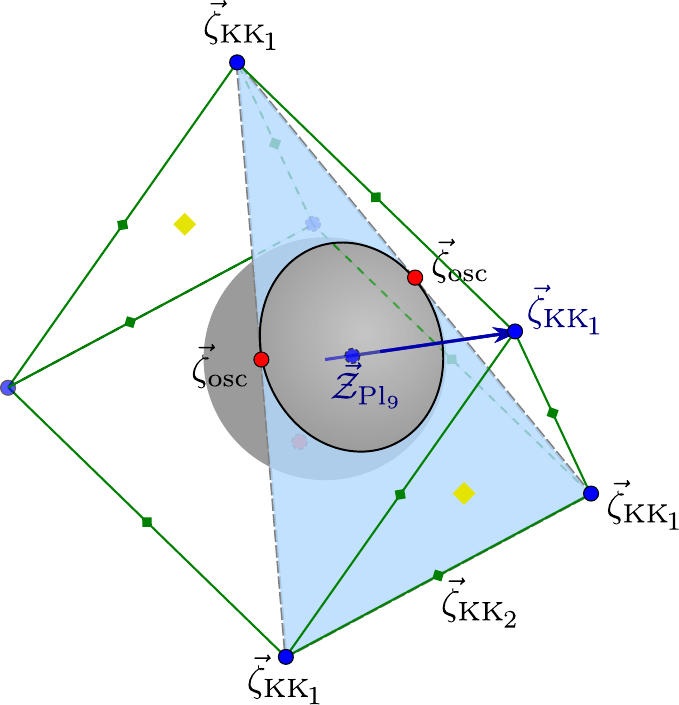}
\caption{$\mathcal{P}_{(9)}$ embedded in $\mathcal{P}_{(8)}$} \label{fig:9from8}
\end{subfigure}
\begin{subfigure}{0.475\textwidth}
\center
\includegraphics[width=\textwidth]{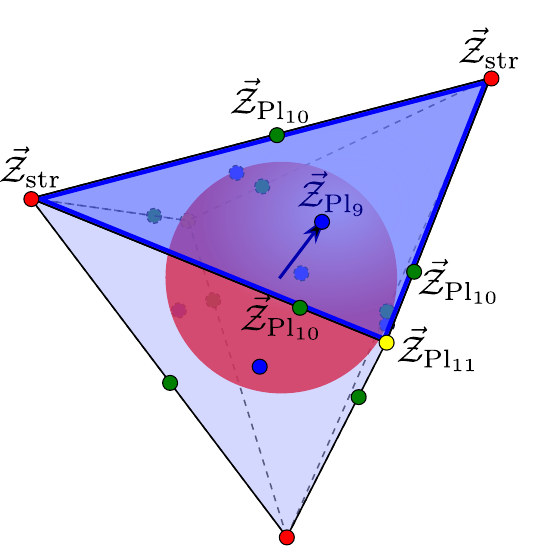}
\caption{$\mathcal{P}_{(9)}^\circ$ embedded in $\mathcal{P}_{(8)}^\circ$} \label{fig:9from8sp}
\end{subfigure}
\caption{In \subref{fig:9from8} the $\mathcal{P}_{(9)}$ tower polytope is obtained from $\mathcal{P}_{(8)}$ after decompactifying along any $\vec{\zeta}_{\rm KK_1}$ vertex, as all of them are equivalent. The sphere of radius $\frac{1}{\sqrt{(d-2)}}=\frac{1}{\sqrt{6}}$ is depicted in gray, while the $\frac{1}{\sqrt{d-2+1}}=\frac{1}{\sqrt{7}}$ circle in the 9-dimensional theory is in a darker shade. Analogously, in \subref{fig:9from8sp} the species polytope $\mathcal{P}_{(9)}^\circ$, associated to the species scale, is recovered as a facet from $\mathcal{P}_{(8)}^\circ$, perpendicular to the direction we are moving. 
\label{fig: 9from8}}
\end{center}
\end{figure}

Let us next consider the tower polytope $\mathcal{P}_{(8)}$, depicted in Figure \ref{fig: 9from8}. Consider any of the KK vertices $\vec \zeta_{\text{KK}_1}$. Consider the tower vectors that are adjacent to this vertex. The components of these adjacent tower vectors that are perpendicular to $\vec \zeta_{\text{KK}_1}$ form the triangular tower polytope $\mathcal{P}_{(9)}$ of the 9d theory.

\begin{figure}[h]
\begin{center}
\begin{subfigure}{0.515\textwidth}
\center
\includegraphics[width=\textwidth]{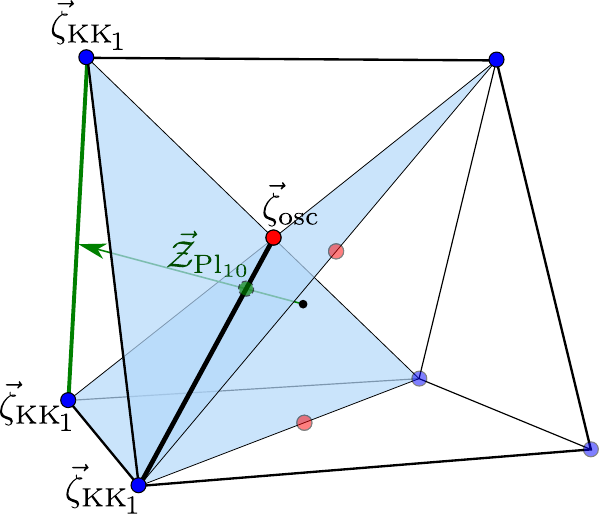}
\caption{$\mathcal{P}_{(10,A)}$ embedded in $\mathcal{P}_{(8)}$} \label{fig: 10from8A}
\end{subfigure}
\begin{subfigure}{0.475\textwidth}
\center
\includegraphics[width=\textwidth]{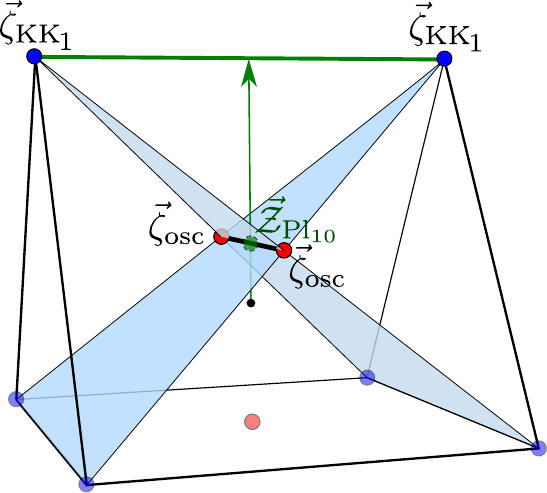}
\caption{$\mathcal{P}_{(10,B)}$ embedded in $\mathcal{P}_{(8)}$} \label{fig: 10from8B}
\end{subfigure}
\caption{$\mathcal{P}_{(10,A)}$ and $\mathcal{P}_{(10,B)}$ tower polytopes (depicted in thick black lines) obtained from $\mathcal{P}_{(8)}$ after homogeneous decompactification of two dimensions, in a direction given by the pericenter of the edge (thick green line) spanned by two $\vec{\zeta}_{\rm KK_1}$ vectors. Not all these edges/limits are equivalent, and the two different possibilities, depicted in \subref{fig: 10from8A} and \subref{fig: 10from8B}, result in the aforementioned 10d polytopes. Note that (1) these polytopes are given by the intersection of the 9d polytopes $\mathcal{P}_{(9)}$ (in blue) that would be obtained from decompactifying o¡in one of the two $\vec{\zeta}_{\rm KK_1}$ directions and (2) lay at the height of the resulting species scale, given by $\vec{\mathcal{Z}}_{\rm Pl_{10}}$ (in green).
\label{fig: 10from8}}
\end{center}
\end{figure}

Note that, because the 9d polytope can be recovered by the 8d polytope, and the 10d polytopes $\mathcal{P}_{(10,A)}$ and $\mathcal{P}_{(10,B)}$ can be recovered from the 9d polytope, these can also be obtained from the 8d polytopes (see Figures \ref{fig: 10from9} and \ref{fig: 10from8}).

One can check that analogous procedures can be carried on also for $\mathcal{P}_{(7)}$ and $\mathcal{P}_{(6)}$, as depicted in Figures \ref{fig: 7from8} and \ref{fig: 6from7} (see Appendix \ref{app: polytope coord} for the generating vertices of $\mathcal{P}_{(7)}$ and $\mathcal{P}_{(6)}$).

We can also apply this approach to study the recursive relations to species polytopes. For instance, in Figure \ref{fig:9from8sp}, the faces of the maximal species polytope in the 8d theory is the species polytope of the 9d theory.

\begin{figure}[h]
\begin{center}
\begin{subfigure}{0.44\textwidth}
\center
\includegraphics[width=60mm]{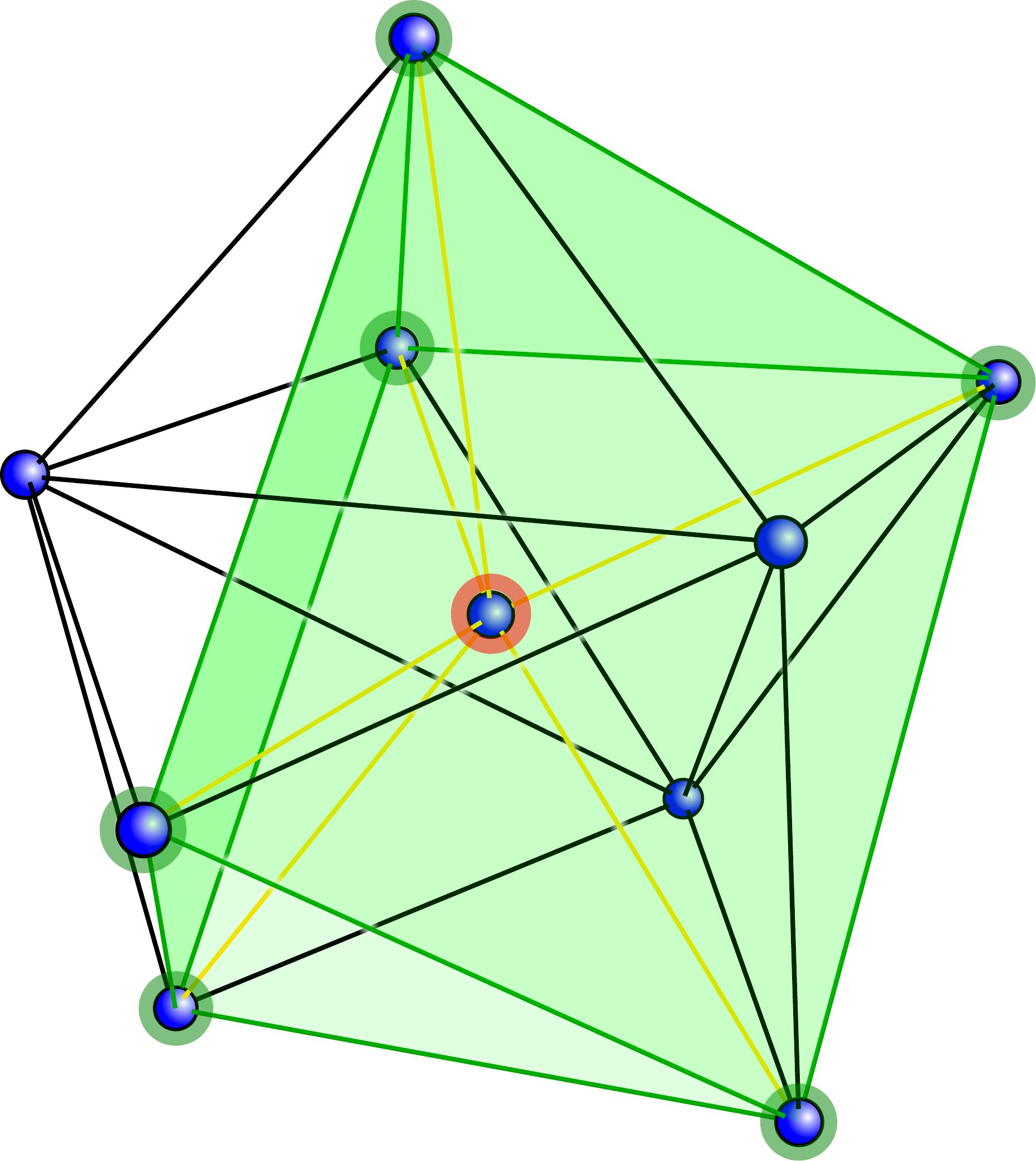}
\caption{$\mathcal{P}_{(8)}$ embedded in $\mathcal{P}_{(7)}$} \label{fig: 7from8}
\end{subfigure}
\begin{subfigure}{0.55\textwidth}
\center
\includegraphics[width=82mm]{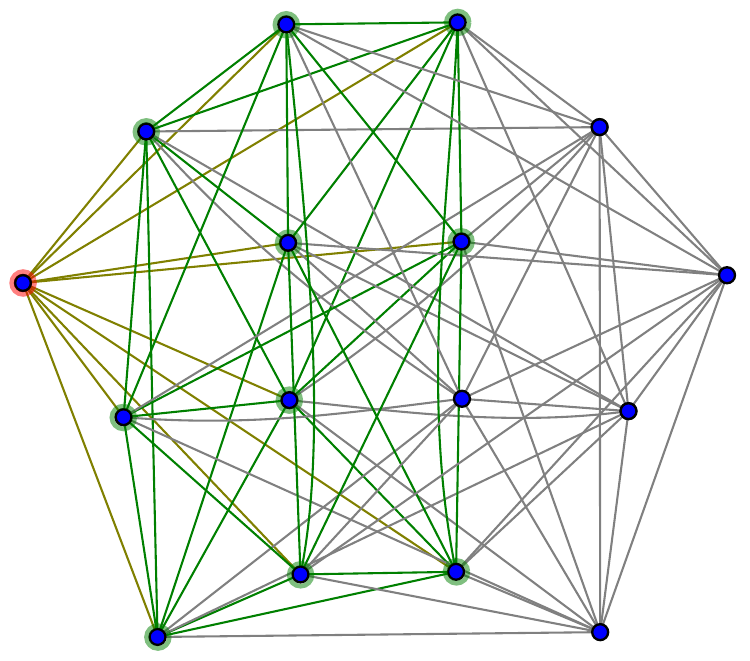}
\caption{$\mathcal{P}_{(7)}$ embedded in $\mathcal{P}_{(6)}$} \label{fig: 6from7}
\end{subfigure}
\caption{An illustration depicting the $\vec{\zeta}_{\text{KK}_1}$ vectors (highlighted in green) corresponding to the higher dimensional polytope resulting from decompactifying in a given $\vec{\zeta}_{\text{KK}_1}$ direction (highlighted in red) for $\mathcal{P}_{(7)}$ \subref{fig: 7from8} and $\mathcal{P}_{(6)}$ \subref{fig: 6from7}. Notice that the tower sub-polytope generated by these vertices corresponds to $\mathcal{P}_{(8)}$ and $\mathcal{P}_{(7)}$, respectively. The resulting tower polytope is located on a plane perpendicular to the red vector $\vec{\zeta}_{\rm KK_ 1}$ at a distance of $\vec{\mathcal{Z}}_{\text{Pl}_{d+1}}$ from the origin.
\label{f.MT3radions}}
\end{center}
\end{figure}

\subsection{Mysterious polytopes and dimensional reduction\label{app:dim red}}

Many of the polytopes we have encountered in the above subsections have known realizations in the string landscape, but others do not. It remains an open problem whether these new possibilities are part of some uncharted region of the landscape, or if instead they violate some presently unknown swampland constraint.

For instance, our classification allows for a 10d theory with a moduli space of $\mathbb R$ featuring two decompactification limits, which is not currently known to exist in the string landscape. Additionally, in our classification in 6d, we found several more polytopes that do not occur in maximal supergravity compactifications on tori (see those labeled in red Tables \ref{tab:BIG} and \ref{tab:BIGsp}). In most of these examples, namely $\Pi_{\rm (6,VI)}$ to $\Pi_{\rm (6,XI)}$, there exist KK vectors on either side of string oscillator vectors that are no symmetric (i.e., they come from decompactifications of different numbers of dimensions). It is possible that some worldsheet argument forbids asymmetric arrangement of KK vertices around oscillator ones, and thus all these polytopes belong in the swampland, but this remains an open problem for future work. 

Even if one could develop an argument against such an asymmetric arrangement, however, this would not address every indeterminate example in Table \ref{tab:BIG}. In particular, the polytope $\Pi_{\rm (6,V)}$ is allowed by our taxonomic rules and features symmetric towers on each side of every string oscillator vector, yet it is not realized as a slice of the tower polytope of M-theory on $T^5$.

In what follows, we consider what happens to this example under dimensional reduction. We will see that the theory cannot be viewed as the decompactification of a five-dimensional theory that satisfies our assumptions (including the assumption of symmetry about string oscillator vectors), but it \emph{can} be viewed as the decompactification of a four-dimensional theory that satisfies our assumptions. 

\begin{figure}[htp]

\begin{subfigure}{0.24\textwidth}
\centering
\includegraphics[scale=0.25]{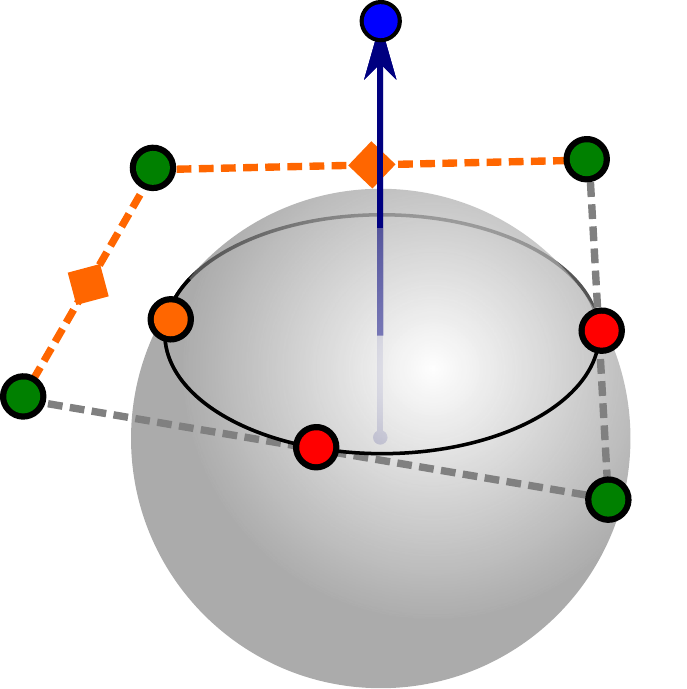}
\caption{\label{sf.8e_from_5d-1}}
\end{subfigure}
\bigskip
\begin{subfigure}{0.24\textwidth}
\centering
\includegraphics[scale=0.25]{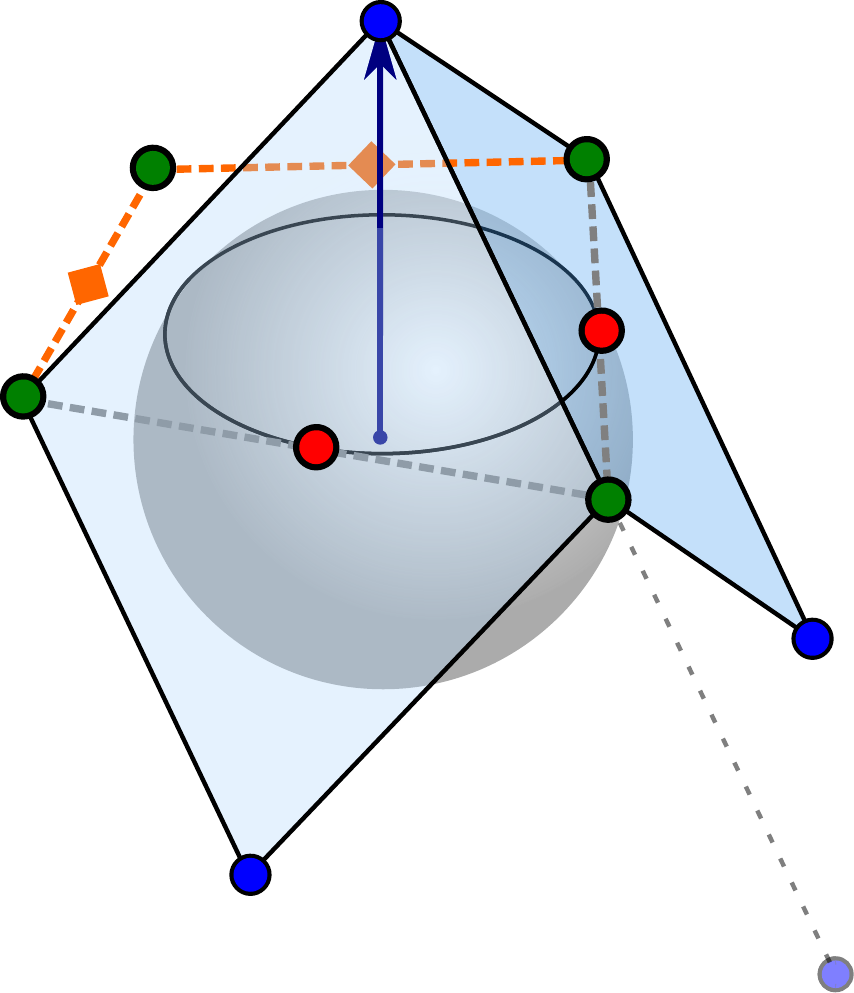}
\caption{\label{sf.8e_from_5d-2}}
\end{subfigure}
\bigskip
\begin{subfigure}{0.24\textwidth}
\centering
\includegraphics[scale=0.25]{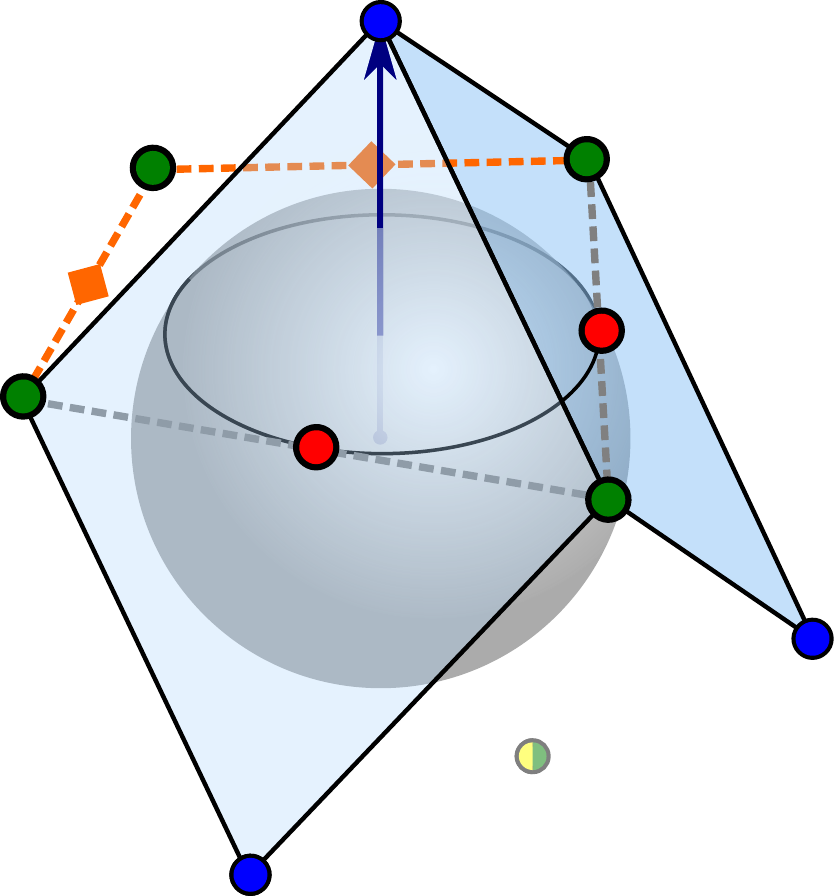}
\caption{\label{sf.8e_from_5d-3}}
\end{subfigure}
\bigskip
\centering
\begin{subfigure}{0.24\textwidth}
\centering
\includegraphics[scale=0.25]{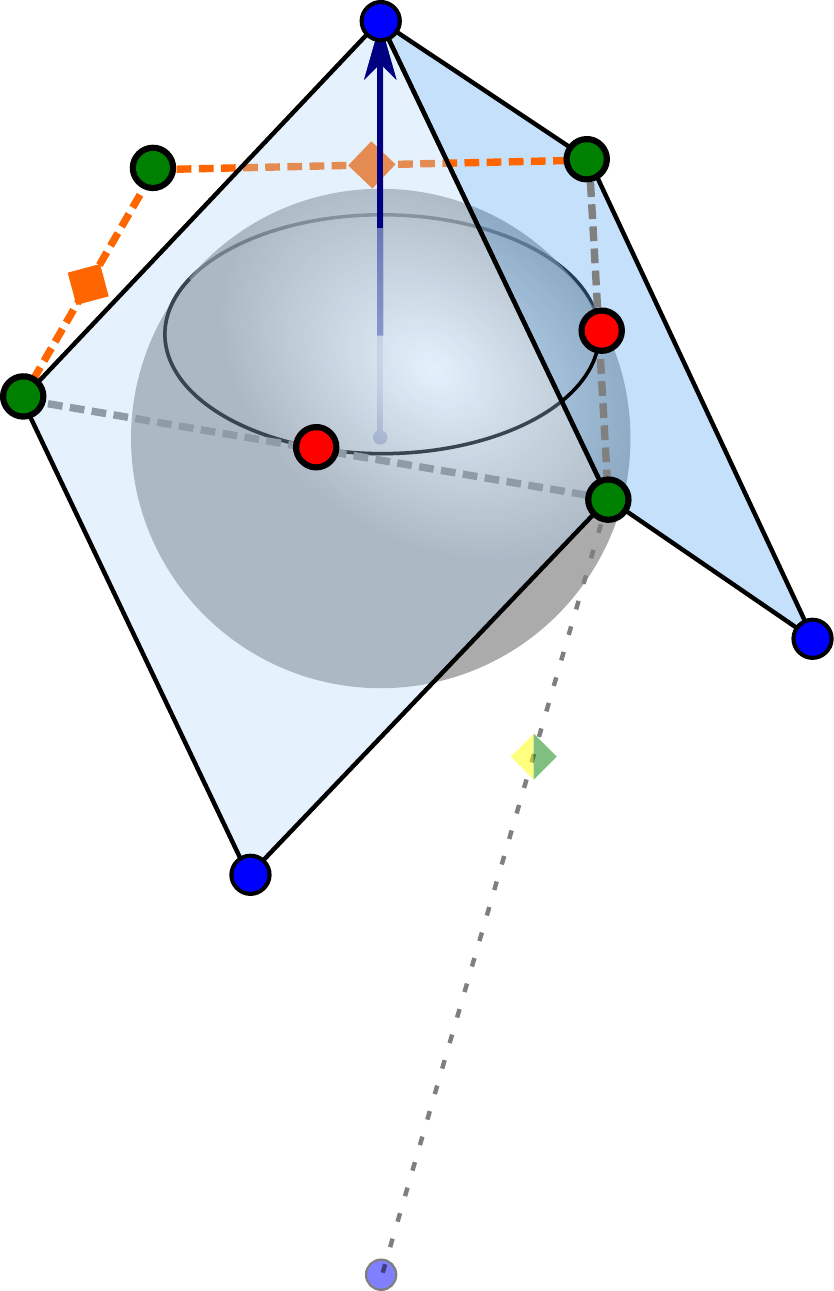}
\caption{\label{sf.8e_from_5d-4}}
\end{subfigure}
\caption{Tentative steps in trying to consistently compactify the polygon $\Pi_{\rm (6,V)}$ from Table \ref{tab:BIG} (in dashed gray lines) from $d=6$ to 5 dimensions. The depicted sphere has radius $\frac{1}{\sqrt{5-2}}$, while the black circumference has radius $\frac{1}{\sqrt{6-2}}$. The string towers are depicted in \fcolorbox{black}{red}{\rule{0pt}{6pt}\rule{6pt}{0pt}}, whilst KK towers associated to decompactification of one, two, three and four dimensions are in colors \fcolorbox{black}{blue}{\rule{0pt}{6pt}\rule{6pt}{0pt}}, \fcolorbox{black}{dark-green}{\rule{0pt}{6pt}\rule{6pt}{0pt}}, \fcolorbox{black}{yellow}{\rule{0pt}{6pt}\rule{6pt}{0pt}} and \fcolorbox{black}{orange}{\rule{0pt}{6pt}\rule{6pt}{0pt}}, respectively. \label{f.8e_from_5d}
}
\end{figure}

We begin by embedding the tower polytope $\Pi_{\rm (6,V)}$ in $\mathbb{R}^3$ at a plane located at a distance of $\frac{1}{2\sqrt{3}}$ from the origin (i.e., the length of $\vec{\mathcal{Z}}_{\rm Pl_6}$, which gives the species scale in that limit). Along this plane, the circle of radius $\frac{1}{\sqrt{6-2}}$ appears as a section of the 2-sphere of radius $\frac{1}{\sqrt{5-2}}$. From this, one can apply the taxonomy rules to obtain the KK vector $\vec{\zeta}_{\rm KK_1}$ corresponding to the decompactification limit of the 5d theory to 6d that would result in the 2d slice of interest, as depicted in Figure \ref{sf.8e_from_5d-1}. Assuming a reflection symmetry about each string oscillator, there are two additional points $\vec{\zeta}_{\rm KK_1}$ obtained from reflecting the $\vec{\zeta}_{\rm KK_1}$ vertex towards which we are decompactifying. If one tries to add an extra $\vec{\zeta}_{\rm KK_1}$ vertex by reflecting the vertex associated with the decompactification across the $\vec{\zeta}_{\rm KK_2}$ vertex between the two $\vec{\zeta}_{\rm osc}$, its norm would be larger than the required length $|\vec{\zeta}_{\rm KK_1}|=\sqrt{\frac{5-1}{5-2}}$, so this possibility is excluded. As depicted in Figure \ref{sf.8e_from_5d-2}, two facets of the prospective tower polytope are thus determined. As can be seen in Figure \ref{sf.8e_from_5d-3}, the next natural step would be to attempt to join these additional
$\vec{\zeta}_{\rm KK_1}$ vectors through an edge. However, this edge would not have the proper length, as can be seen by noting that the midpoint of the edge (marked in that figure) has a length of $\sqrt{\frac{5+3-2}{3(5-2)}}=|\vec{\zeta}_{\rm KK_3}|$, which is inconsistent with the construction. This  marked point could represent the closest point of a face generated by three $ \vec{\zeta}_{\rm KK_1}$ vectors, but this would require an additional $ \vec{\zeta}_{\rm KK_1}$ vertex, as shown in Figure \ref{sf.8e_from_5d-4}. However, fulfilling \eqref{e:zeta min} implies that the norm of this vertex is too large. Having exhausted all the possibilities, we conclude that one cannot obtain $\Pi_{\rm(6,V)}$ from a three-dimensional tower polytope in $d=5$ that satisfies our taxonomy rules.

\begin{figure}[h]
\begin{center}
\includegraphics[width = 0.425\textwidth]{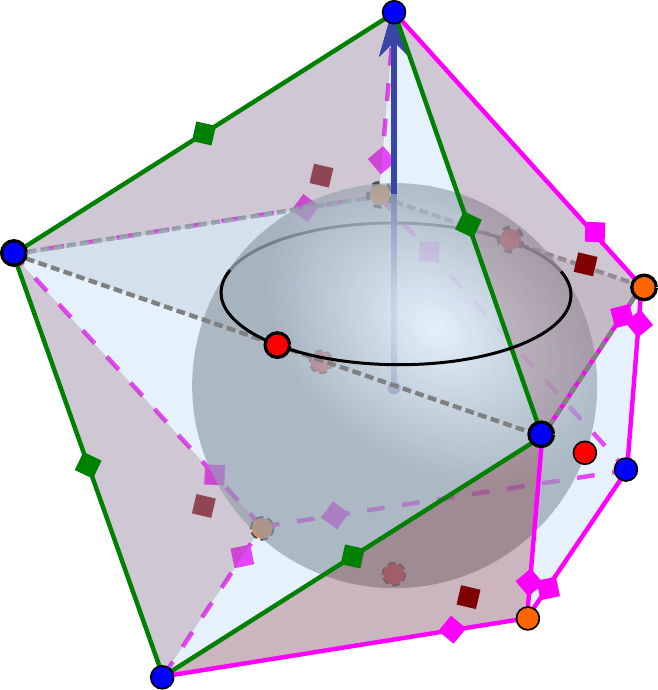}
\end{center}
\caption{Consistent 3d convex hull in $d=5$ from which the $d=6$ polygon $\Pi_{\rm (6,I)}$ depicted in Table \ref{tab:BIG} (in dashed gray lines) can be recovered from decompactifying one dimension. The depicted sphere has radius $\frac{1}{\sqrt{5-2}}$, while the black circumference has radius $\frac{1}{\sqrt{6-2}}$. The string towers are depicted in red \fcolorbox{black}{red}{\rule{0pt}{6pt}\rule{6pt}{0pt}}, while KK towers and edges/facets (in these cases with the points closest to the origin highlighted) associated to decompactification of one, two, three, four, five and six dimensions appear in blue \fcolorbox{black}{blue}{\rule{0pt}{6pt}\rule{6pt}{0pt}}, green \fcolorbox{black}{dark-green}{\rule{0pt}{6pt}\rule{6pt}{0pt}}, yellow \fcolorbox{black}{yellow}{\rule{0pt}{6pt}\rule{6pt}{0pt}}, orange \fcolorbox{black}{orange}{\rule{0pt}{6pt}\rule{6pt}{0pt}}, pink \fcolorbox{black}{Magenta}{\rule{0pt}{6pt}\rule{6pt}{0pt}} and  brown \fcolorbox{black}{Brown}{\rule{0pt}{6pt}\rule{6pt}{0pt}}, respectively.}
\label{f.8a_from_5d}
\end{figure}

This is not the case for the first four 2d slices in $d=6$, $\Pi_{\rm (6, I)}$-$\Pi_{\rm (6, IV)}$, which preserve our assumptions under dimensional reduction on a circle and thus can be obtained from decompactification of higher rank and lower $d$ polytopes. For each 6d case, there is a 5d case following our assumptions that decompactifies into that 6d case. For example, one possibility for the polygon $\Pi_{ \rm (6,I)}$ is depicted in Figure \ref{f.8a_from_5d}.

\begin{figure}[h]
\begin{center}
\includegraphics[width = 0.35\textwidth]{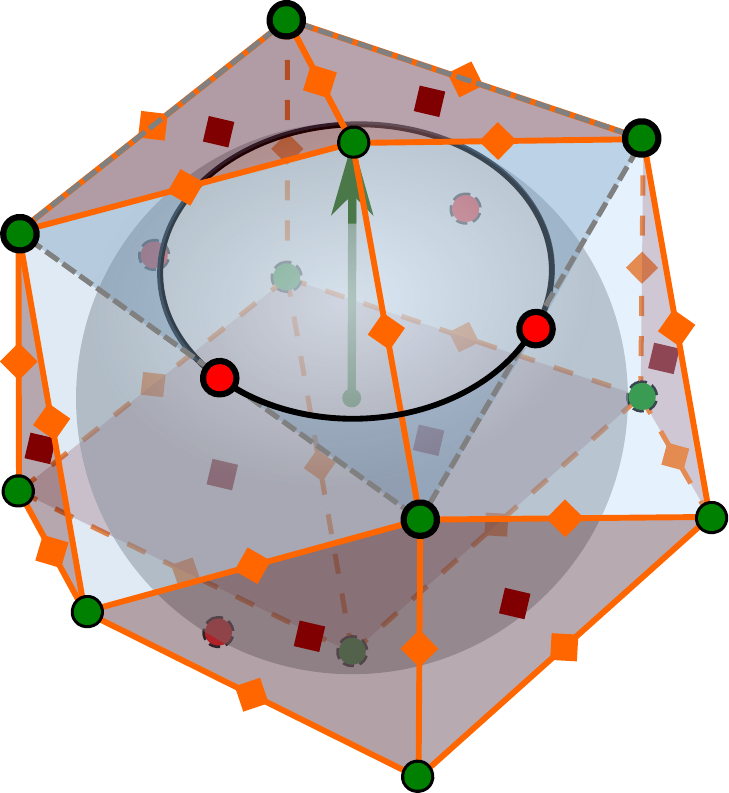}
\end{center}
\caption{Consistent 3d convex hull in $d=4$ from which the $d=6$ polygon $\Pi_{\rm (6,V)}$ depicted in Table \ref{tab:BIG} (in dashed gray lines) can be recovered from decompactifying \emph{two} dimensions. The depicted sphere has radius $\frac{1}{\sqrt{4-2}}$, while the black circumference has radius $\frac{1}{\sqrt{6-2}}$. The string towers are depicted in red \fcolorbox{black}{red}{\rule{0pt}{6pt}\rule{6pt}{0pt}}, whilst KK towers and edges/facets (for these the closest point to the origin is highlighted) associated to decompactification of two, four and six dimensions appear in green \fcolorbox{black}{dark-green}{\rule{0pt}{6pt}\rule{6pt}{0pt}}, orange \fcolorbox{black}{orange}{\rule{0pt}{6pt}\rule{6pt}{0pt}} and  brown \fcolorbox{black}{Brown}{\rule{0pt}{6pt}\rule{6pt}{0pt}}, respectively. Note that in this slice of the moduli space there would not be decompacitification limits to an odd number of dimensions.}
\label{f.8e_from_4d}
\end{figure}

This does not necessarily imply that $\Pi_{\rm (6, V)}$ cannot be observed in a consistent theory of quantum gravity. Indeed, as illustrated in Figure \ref{f.8e_from_4d}, one can find a 4d theory that follows our assumptions and decompactifies to this 6d theory. Note that, for the example depicted in Figure \ref{f.8e_from_4d}, all the decompactification limits in this latter polytope lead to an even number of dimensions--6, 8 or 10. One can check that, depending on which of the two inequivalent $\vec{\zeta}_{\rm KK_2}$ vertices we decompactify, either $\Pi_{\rm (6,IV)}$ or $\Pi_{\rm (6,V)}$ are recovered in $d=6$.

Another, maybe simpler possibility,  is that perhaps there is nothing fundamentally wrong with having $\vec{\zeta}_{\rm KK}$ vertices decompactifying a different number of dimensions at opposite sides of $\vec{\zeta}_{\rm osc}$. The argument above relied in assuming only symmetric vertices, so it is possible that relaxing this allows for reduction to $d=5$. For the sake of brevity we will not engage in the construction of such rank 3 polytope.

\begin{figure}[h]
\begin{center}
\includegraphics[width = 0.45\textwidth]{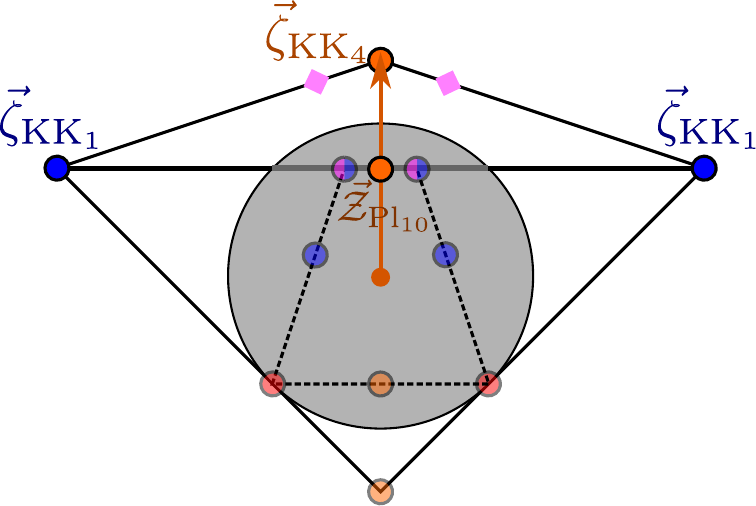}
      \caption{Recovery of the tower and species polytopes $\mathcal{P}_{(10,C)}$ and $\mathcal{P}_{(10,C)}^\circ$ of the 10d theory from decompactification of a 4d theory with tower and species polygons $\Pi_{\rm (6,XI)}$ (in solid lines) and $\Pi^\circ_{\rm (6,XI)}$ (dashed), see Tables \ref{tab:BIG} and \ref{tab:BIGsp}. These are located at the same height as the species scale $\vec{\mathcal{Z}}_{\rm Pl_{10}}$. \label{fig:p10C}}
\end{center}
\end{figure}

As a final comment, we note that a similar story applies to the mysterious 1-dimensional polytope $\mathcal{P}_{(10,C)}$, which involves two decompactification limits from 10d to 11d and no emergent string limits (see Figure \ref{fig:KKKK10d}).
Figure \ref{fig:p10C} shows how this polytope can be recovered by decompactifying four internal dimensions in the $d=6$ polygon $\Pi_{\rm (6,XI)}$ from Table \ref{tab:BIG}. While this does not by itself shed light on the existence of the exotic, undiscovered polytopes $\mathcal{P}_{(10,C)}$ or $\Pi_{\rm (6,XI)}$, it does at least suggest that the two are related. It would be very interesting if these examples are ultimately realized in some undiscovered corner of the string landscape.

\section{Conclusions\label{s.conc}}

Using the Emergent String Conjecture, we have developed constraints on the behavior of the species scale and the towers of light states in generic infinite-distance limits of moduli space. These constraints are encoded in a set of taxonomic rules for the  scalar products of tower vectors, which describe the couplings of light towers of particles to the moduli of the theory. From this, we have reproduced several well-known constraints on the light towers and the species scale.

Under certain additional assumptions, we have applied these rules globally to produce a finite list of tower polytopes satisfying our taxonomic rules. This list includes many polytopes known to exist in the quantum gravity landscape, as well as some polytopes that have not yet been discovered in the quantum gravity landscape. We have also seen that the resulting tower and species polytopes satisfy recursive relations, whereby polytopes of higher-dimensional theories can be used to construct polytopes of lower-dimensional theories by KK reduction. Additionally, polytopes of lower-dimensional theories can be used to recover polytopes of higher-dimensional theories after decompactification.

Our work provides a new avenue to uncover the structure of dualities of the moduli space by defining the frame simplex generated by the light towers of states in each duality frame and exploring the ways to glue them together across different duality frames. These frame simplices can therefore be viewed as geometric building blocks for the tower polytope, and they provide a first crucial step towards a taxonomy of infinite-distance limits of the moduli space. 

However, our work is based on certain assumptions, which must be either justified or relaxed in order to attain a complete classification. First of all, we have focused on regular limits, as defined in Section \ref{sec:rules}. Although we have argued that irregular limits are of measure zero in the space of infinite-distance limits, the rules for gluing individual frame simplices can change when crossing an irregular limit. It would be interesting to explore these cases in more detail.

Secondly, the particular classification of polytopes in Section \ref{s.Applications} applies to asymptotically flat slices of moduli spaces. It would be interesting to investigate to what extent a classification of tower polytopes is possible in spaces that are not asymptotically flat. In particular, it would be interesting to study the effects of axions, which typically lead to curvature of the moduli space.\footnote{Natural objects of study for non-flat moduli spaces are geodesics and towers of scalar charge-to-mass ratios that align with these geodesics (see \cite{Etheredge:2023usk, Etheredge:2023zjk}).}

Moreover, our classification of polytopes in Section \ref{s.Applications} deals with polytopes of rank 1, 2, and $11-d$. It would be natural to extend this classification to polytopes of more general rank. It would be also interesting to relax some of the assumptions that went into these classifications, such as the limit on the maximal number of dimensions of spacetime or the assumption that decompactification limits necessarily lead to higher-dimensional vacua.

On the other hand, it may be possible to justify further assumptions, leading to tighter constraints on our classification of tower polytopes. For instance, in Section \ref{s.2dclassification}, our classification of 2d polytopes uncovered examples with KK-mode vertices associated with the decompactification of \emph{different} numbers of dimensions on either side of a string oscillator vertex. It would be worthwhile to either find an argument that rules out these polytopes or else find an example in string theory in which these polytopes arise.

The present work has focused on towers of light particles. Using dimensional reduction and the duality web, however, these light particles are often related to branes.
As a result, the constraints on particles discussed here also lead to constraints on the scalar charge-to-tension ratio vectors ($-\vec\nabla \log \mathcal{T}$) for extended objects \cite{Palti:2017elp,Font:2019cxq, Lanza:2020qmt, Herraez:2020tih, Alvarez-Garcia:2021pxo, Etheredge:2023zjk}. These constraints will be further explored in a forthcoming work \cite{Etheredge:BraneDC, Etheredge:BraneTaxonomy}.

We saw in Section \ref{subsec:SDC} that the Sharpened Distance Conjecture only follows from our taxonomic rules if we further assume that the tangent vector of the infinite-distance geodesic lies in the subspace of the tangent space generated by the light tower vectors (referred to as the principal plane above). We further saw in Section \ref{subsec:parallel} that this assumption is difficult to prove in full generality. It would be worthwhile to either prove this assumption or else find an example in which it is violated.

Our work has relied crucially on the Emergent String Conjecture. At present, the evidence for this conjecture comes primarily from known examples in string/M-theory. A bottom-up argument for the Emergent String Conjecture would be most desirable, and it would significantly strengthen the foundation on which our work rests.

Our understanding of infinite-distance limits in quantum gravity has grown immensely since the pioneering work of Ooguri and Vafa \cite{Ooguri:2006in}. One of main insights of the present work\footnote{See also \cite{Bedroya:2023xue, vandeHeisteeg:2024lsa}.} is that continuous families of infinite-distance limits can be sorted into a discrete set of duality frames that share a common perturbative limit, and this process in turn produces additional constraints on the individual limits and how they can fit together in the moduli space. These constraints are very powerful, yet they still leave some room for new theories outside the known landscape. Time will tell whether  this gap can be narrowed further, or if yet undiscovered theories may inhabit the uncharted territory between landscape and swampland.

\section*{Acknowledgements}
We are grateful for conversations with Alek Bedroya, Alberto Castellano, Albrecht Klemm, Matthew Reece, and Cumrun Vafa, and we thank Jacob McNamara for collaboration in the early stages of this work. We thank the organizers of Swamplandia, as well as the Simons Summer Workshop, where part of the research on this project was carried out. M.E. and B.H. received support from NSF grant PHY-2112800. The work of T.R. was supported in part by STFC through grant ST/T000708/1. ME was supported in part by the Heising-Simons Foundation, the Simons Foundation, and grant no. PHY-2309135 to the Kavli Institute for Theoretical Physics (KITP). I.R. wishes to acknowledge the hospitality of the Department of Theoretical Physics at CERN and the Department of Physics of Harvard University during the late stages of this work. I.R. and I.V. acknowledge the support of the Spanish Agencia Estatal de Investigaci\'on through the grant ``IFT Centro de Excelencia Severo Ochoa'' CEX2020-001007-S and the grant PID2021-123017NB-I00, funded by MCIN/AEI/10.13039/ 501100011033 and by ERDF A way of making Europe. The work of I.R. is also supported by the Spanish FPI grant No. PRE2020-094163. The work of I.V. is also partly supported by the grant RYC2019-028512-I from the MCI (Spain) and the ERC Starting Grant QGuide-101042568 - StG 2021.

\appendix

\section{M-theory on tori and K3\label{a.MTheory}}

In this appendix, we calculate the radion components of tower vectors for non-sliding theories on orthogonal tori. We consider the tower vectors of KK modes and also wrapped branes. Many of these calculations have not been performed elsewhere in the literature. 
The resulting polytopes can be compared with our taxonomy. We discuss symmetries of the polytopes, and we also consider a slice of the polytope from M-theory on K3.

\subsection{Moduli, tensions, and masses}

\subsubsection{Diagonal tori moduli}
Let us compactify a $D$-dimensional theory on a diagonal $n$-torus with the following ansatz, 
\begin{align}
	{\rm d}s_D^2=|g_{ij}|^{-\frac 1{d-2}} {\rm d}  s^2_d+g_{ij}{\rm d}\theta^i {\rm d}\theta^j,\qquad
	g_{ij}=\delta_{ij}\exp(-2\rho^i).
\end{align}

With this, the $D$-dimensional Einstein-Hilbert action,
\begin{align}
	S_D=\frac{1}{2\kappa_D^2}\int {\rm d}^Dx\sqrt{-g} R_D,
\end{align}
reduces to
\begin{align}
	S_d
	&=\frac{1}{2\kappa_d^2}\int {\rm d} ^dx\sqrt{-g}\left(R-\sum_i(\nabla \rho^i)^2-\frac{1}{d-2}\sum_{ij}\nabla \rho^i\nabla \rho^j\right).
\end{align}
From this, we can read off the metric and inverse metric for the radions $\rho^i$,
\begin{align}
	\mathsf{G}_{ij}=\delta_{ij}+\frac1{d-2}1_i1_j.
\end{align}
The inverse metric is given by
\begin{align}
	\mathsf{G}^{ij}=\delta^{ij}-\frac{1}{D-2}1^i1^j,
\end{align}
since this satisfies
\begin{align}
	\mathsf{G}_{ij}\mathsf{G}^{jk}=\left(\delta_{ij}+\frac1{d-2}1_i1_j\right)\left(\delta^{jk}-\frac{1}{D-2}1^j1^k\right)=\delta_i^k.
\end{align}
Here $1_i$ refers to the vector where each entry is 1.

\subsubsection{Brane tensions and KK mode masses}
Consider a $(P-1)$ brane in the $D$-dimensional theory with tension $T_P^{(D)}$. In toroidally reducing to the $d$-dimensional theory, this brane can wrap multiple cycles of the tori. Suppose this brane wraps the $i$th cycle $k_i$ times, where $k_i$ is either 0 or 1. Then the tension of the resulting $(p-1)$-brane in the $d$-dimensional theory is
\begin{align}
T_p^{(d)}\sim \exp\left\{\left(\frac p{d-2}1_i-k_i\right)\rho^i\right\}T_P^{(D)},
\end{align}
where $P=p+k=p+\sum_i k_i$.

For example, when $n=1$ and $k=0$, $\mathsf{G}_{\rho \rho}=1+\frac{1}{d-2}=\frac{d-1}{d-2}$
\begin{align}
T_p^{(d)}\sim \exp\left\{\frac p{d-2}\rho\right\}T_P^{(D)}=\exp\left\{-\frac{p}{\sqrt{(d-1)(d-2)}}\hat\rho\right\}T_P^{(D)},
\end{align}
where $\hat \rho$ is canonically normalized.
Another example is when $n=1$ and $k=1$, in which case
\begin{align}
T_p^{(d)}\sim \exp\left\{\frac {p-d+2}{d-2}\rho\right\}T_P^{(D)}=\exp\left\{-\frac{-d+p+2}{\sqrt{(d-1)(d-2)}}\hat\rho\right\}T_P^{(D)}.
\end{align}

Meanwhile, KK modes have mass in $d$-dimensional Planck units
\begin{align}
	m^2=|g_{ij}|^{-\frac{1}{d-2}}g^{ij}n_in_j=\sum_i\exp\left (2\rho^i+\frac{2}{d-2}\sum_j \rho^j\right)(n^i)^2.
\end{align}

\subsubsection{Canonically normalized moduli}

Currently, the radions $\rho^i$ have a complicated metric:
\begin{align}
	\mathsf{G}_{ij}\rho^i\rho^j&=\rho^2+\frac{1}{d-2}(\rho\cdot 1)(\rho\cdot 1)
\end{align}
(where dots refer to product with respect to $\delta_{ij}$)

The metric on moduli space for the $\rho^i$ is not the identity matrix, and thus the moduli $\rho^i$ are not canonically normalized. To canonically normalize, we define the new moduli $R^i$ in terms of the radions $\rho^i$:
\begin{align}
	R^i&\equiv-\rho^i- \frac{1}{d-2\pm\sqrt{(d-2)(D-2)}}(\rho \cdot 1)1^i
\end{align}
With these new radions $R^i$, the moduli space's metric is the identity matrix,
\begin{align}
	S_d=\frac{1}{\kappa_d^2}\int d^dx\sqrt{-g}\left(R-\delta_{ij}(\nabla R^i)\cdot (\nabla R^j)\right).
\end{align}
There is a $\pm$ in the relation between $R^i$ and $\rho^i$. Let us choose the $+$ choice (the $-$ choice also works just as fine, but the details are different). Thus,
\begin{align}
	R^i&\equiv-\rho^i- \frac{1}{d-2+\sqrt{(d-2)(D-2)}}(\rho \cdot 1)1^i.
\end{align}
We can express $\rho^i$ in terms of $R^i$,
\begin{align}
	\rho^i&=-\left(R^i+\frac{R\cdot 1}n\left(-1+\sqrt{\frac{d-2}{D-2}}\right)1^i\right).
\end{align}

So, we can express the tension in terms of these canonically-normalized diagonalized radions $R^i$,
\begin{align}
T_p^{(d)}
&\sim \exp\left\{-R^i \left[\left(\frac{k }{D-d}\left(1-\sqrt{\frac{d-2}{D-2}}\right)+\frac{p }{\sqrt{(d-2) (D-2)}}\right)1_i-k_i\right]\right\}T_P^{(D)}.
\end{align}

Also, KK modes have masses in $d$-dimensional Planck units of
\begin{align}
	m^2&=\sum_i\exp\bigg(-2\left(R^i+\frac{R\cdot 1}n\left(-1+\sqrt{\frac{d-2}{D-2}}\right)1^i\right)\nonumber\\
	&-\frac{2}{d-2}\sum_j \left(R^j+\frac{R\cdot 1}n\left(-1+\sqrt{\frac{d-2}{D-2}}\right)1^j\right)\bigg ](n^i)^2.
\end{align}

\subsection{Polytopes from M-theory on $T^1$ through $T^5$\label{app: polytope coord}}

We are interested in computing the tower polytopes from M-theory on a $k$ torus. Suppose that $k\leq5$, so that the only relevant towers are KK modes, fully-wrapped branes, and string oscillators. Dealing with $k\geq6$ requires an analysis with KK-monopoles, and we postpone that case to future work.

The only way to get particles in these theories is from KK modes, wrapped branes, and string oscillators. However, with the exception of 10d, the string oscillators are unimportant in obtaining the full tower polytopes. Thus, for $d\leq 9$ we need to consider only fully wrapped branes and KK modes.

The KK modes have a general formula, and all of these can be classified by considering all of the different ways in which they can vibrate through the compact dimensions. Meanwhile, for a $(p-1)$-brane to produce a non-oscillation particle, we must have it wrap $p-1$ of the  1-cycles.

In general, there are finitely many different ways for $(p-1)$-branes to wrap $p-1$ of the 1-cycles on the $k$-torus, and there can be many different cycles of the torus for the KK modes to have momentum in. The algorithm we take is to collect all of the different $(p-1)$-branes can wrap $p-1$ of the cycles on the $k$-torus, and all of the different KK modes with momentum in precisely one, to $k$, different cycles of the torus. This way, we find a sufficient collection of tower vectors that generate the tower polytope $\mathcal{P}_{(d)}$.

In this section we use the above algorithm to collect the tower vectors for the maximal supergravity tower polytopes $\mathcal{P}_{(d)}$ with $d\in\{10,9,8,7,6\}$.

For dimension 10, 9, and 8, we state all of the tower vectors for $\mathcal{P}_{(d)}$. For dimensions 7 and 6, we write only the tower vectors for decompactifications of 1-dimension, as all of the tower vectors follow from these, and there are too many to write in this paper. 

\subsubsection*{10d}
As explained in section \ref{s.10d}, there are two possibilities for the 10d maximal SUGRA tower polytope. The 10d IIA tower polytope has a string oscillator and single KK mode, $\vec{\zeta}_{\rm osc}=-\frac{1}{\sqrt{8}}$ and $\vec{\zeta}_{\rm KK_1}=\frac{3}{\sqrt{8}}$. The  IIB tower polytope has two oscillator modes, $\vec{\zeta}_{\rm osc}=\pm\frac{1}{\sqrt{8}}$.

\subsubsection*{9d}
The tower polytope $\mathcal{P}_{(9)}$ is spanned by three $\vec{\zeta}_{\rm KK_1}$ vectors,
$$
\left(0,\sqrt{\frac{8}{7}}\right),\;\left(\frac{1}{ \sqrt{2}},-\frac{3}{\sqrt{14}}\right),\;\left(-\frac{1}{ \sqrt{2}},-\frac{3}{\sqrt{14}}\right)\,.
$$
There are also two string oscillators towers, $\vec{\zeta}_{\rm osc}=\left(\pm\frac{1}{2 \sqrt{2}},\frac{1}{2 \sqrt{14}}\right)$, and a KK vector corresponding to decompactification to 11d dimensions, $\vec{\zeta}_{\rm KK_1}=\left(0,-\frac{3}{\sqrt{14}}\right)$. See Figure \ref{f.9dmaximal}.

\subsubsection*{8d}

In $d=8$, $\mathcal{P}_{(8)}$ is generated by six $\vec{\zeta}_{\rm KK_1}$ vectors, 
\begin{equation*}
\begin{array}{rrr}
\left(0,\frac{1}{\sqrt{2}},\sqrt{\frac{2}{3}}\right),&\left(0,-\frac{1}{\sqrt{2}},\sqrt{\frac{2}{3}}\right),&\left(\frac{1}{\sqrt{2}},\frac{1}{\sqrt{2}},-\frac{1}{\sqrt{6}}\right),\\
	\left(\frac{1}{\sqrt{2}},-\frac{1}{\sqrt{2}},-\frac{1}{\sqrt{6}}\right),&\left(-\frac{1}{\sqrt{2}},\frac{1}{\sqrt{2}},-\frac{1}{\sqrt{6}}\right),&\left(-\frac{1}{\sqrt{2}},-\frac{1}{\sqrt{2}},-\frac{1}{\sqrt{6}}\right)\,,
\end{array}	
\end{equation*}

This tower polytope also contains nine $\vec{\zeta}_{\rm KK_2}$ vectors,
\begin{equation*}
\begin{array}{rrrrr}
\left(0,0,\sqrt{\frac{2}{3}}\right),\left(\frac{1}{\sqrt{2}},0,-\frac{1}{\sqrt{6}}\right),\left(-\frac{1}{\sqrt{2}},0,-\frac{1}{\sqrt{6}}\right),\left(\frac{1}{2\sqrt{2}},\frac{1}{\sqrt{2}},\frac{1}{2\sqrt{6}}\right),\left(\frac{1}{2\sqrt{2}},-\frac{1}{\sqrt{2}},\frac{1}{2\sqrt{6}}\right),\\
\left(-\frac{1}{2\sqrt{2}},\frac{1}{\sqrt{2}},\frac{1}{2\sqrt{6}}\right),\left(-\frac{1}{2\sqrt{2}},-\frac{1}{\sqrt{2}},\frac{1}{2\sqrt{6}}\right),\left(0,\frac{1}{\sqrt{2}},-\frac{1}{\sqrt{6}}\right),\left(0,-\frac{1}{\sqrt{2}},-\frac{1}{\sqrt{6}}\right),
\end{array}
\end{equation*}
two $\vec{\zeta}_{\rm KK_3}$,
$$
\left(0,\frac{1}{\sqrt{2}},0\right),\;\left(0,-\frac{1}{\sqrt{2}},0\right),
$$
as well as three string oscillator modes $\vec{\zeta}_{\rm osc}$
$$
\left(0,0,-\frac{1}{\sqrt{6}}\right),\left(\frac{1}{2\sqrt{2}},0,\frac{1}{2\sqrt{6}}\right),\left(-\frac{1}{2\sqrt{2}},0,\frac{1}{2\sqrt{6}}\right)\;,
$$
see Figure \ref{fig:ch8d}.

\subsubsection*{7d}
Here $\mathcal{P}_{(7)}$ is generated by ten $\vec{\zeta}_{\rm KK_1}$:
\begin{equation*}
\begin{array}{rrr}
\left(0,\frac{2 \sqrt{10}-5}{15} ,\frac{2 \sqrt{10}-5}{15} ,\frac{2\sqrt{10}+10}{15}	\right),&\left(	0,\frac{2 \sqrt{10}-5}{15},\frac{\sqrt{10}+10}{15} ,\frac{2 \sqrt{10}-5}{15}\right),\\
\left(0,\frac{2\sqrt{10}+10}{15},\frac{2 \sqrt{10}-5}{15},\frac{2 \sqrt{10}-5}{15}\right),&\left(\frac{3}{2 \sqrt{2}},\frac{1}{2 \sqrt{10}},\frac{1}{2 \sqrt{10}},\frac{1}{2 \sqrt{10}}	\right),\\
\left(-\frac{1}{\sqrt{2}},\frac{1}{3}-\frac{1}{3 \sqrt{10}},\frac{1}{3}-\frac{1}{3 \sqrt{10}},-\frac{2}{3}-\frac{1}{3 \sqrt{10}}	\right),&\left(	-\frac{1}{\sqrt{2}},\frac{1}{3}-\frac{1}{3 \sqrt{10}},-\frac{2}{3}-\frac{1}{3 \sqrt{10}},\frac{1}{3}-\frac{1}{3 \sqrt{10}}\right),\\
\left(-\frac{1}{\sqrt{2}},-\frac{2}{3}-\frac{1}{3 \sqrt{10}},\frac{1}{3}-\frac{1}{3 \sqrt{10}},\frac{1}{3}-\frac{1}{3 \sqrt{10}}\right),&\left(\frac{1}{2 \sqrt{2}},\frac{2}{3}-\frac{7}{6 \sqrt{10}},-\frac{1}{3}-\frac{7}{6 \sqrt{10}},-\frac{1}{3}-\frac{7}{6 \sqrt{10}}	\right),\\
\left(\frac{1}{2 \sqrt{2}},-\frac{1}{3}-\frac{7}{6 \sqrt{10}},\frac{2}{3}-\frac{7}{6 \sqrt{10}},-\frac{1}{3}-\frac{7}{6 \sqrt{10}}\right),&\left(\frac{1}{2 \sqrt{2}},-\frac{1}{3}-\frac{7}{6 \sqrt{10}},-\frac{1}{3}-\frac{7}{6 \sqrt{10}},\frac{2}{3}-\frac{7}{6 \sqrt{10}}\right),\\
\end{array}
\end{equation*}
From these, we have thirty $\vec{\zeta}_{\rm KK_2}$'s,
thirty $\vec{\zeta}_{\rm KK_3}$'s, 
five $\vec{\zeta}_{\rm KK_4}$'s, and five string oscillators $\vec{\zeta}_{\rm osc}$ vectors. For this case, as well as the 6d case, we do not state the 2d and higher KK modes, as well as string oscillator modes, because there are many of them and they also follow from the positions of the 1d KK modes. Because a four-dimensional polytope cannot be properly embedded in $\mathbb{R}^2$ or even in $\mathbb{R}^3$, in Figure \ref{fig: 7from8} we depict the adjacency relations of the $\vec{\zeta}_{\rm KK_1}$ vertices, with the other $\vec{\zeta}_{{\rm KK}, n}$ and $\vec{\zeta}_{\rm osc}$ being obtained as described in \S \ref{s.inf dist str}. It can also be shown that the ball of radius $\frac{1}{\sqrt{d-2}}=\frac{1}{\sqrt{5}}$ is contained inside $\mathcal{P}_{(7)}$.

\subsubsection*{6d}
$\mathcal{P}_{(6)}$ is generated by sixteen $\vec{\zeta}_{\rm KK_1}$'s:
\begin{equation*}
\begin{array}{rr}
\left(0,\frac{\sqrt{2}-1}{4} ,\frac{\sqrt{2}-1}{4} ,\frac{\sqrt{2}-1}{4} ,\frac{\sqrt{2}+3}{4}	\right),&\left(0,\frac{\sqrt{2}-1}{4} ,\frac{\sqrt{2}-1}{4} ,\frac{\sqrt{2}+3}{4} ,\frac{\sqrt{2}-1}{4} 	\right),\\
\left(0,\frac{\sqrt{2}-1}{4} ,\frac{\sqrt{2}+3}{4} ,\frac{\sqrt{2}-1}{4} ,\frac{\sqrt{2}-1}{4} \right),&\left(0,\frac{\sqrt{2}+3}{4} ,\frac{\sqrt{2}-1}{4} ,\frac{\sqrt{2}-1}{4} ,\frac{\sqrt{2}-1}{4} 	\right),\\
\left(\frac{3}{2 \sqrt{2}},\frac{1}{4 \sqrt{2}},\frac{1}{4 \sqrt{2}},\frac{1}{4 \sqrt{2}},\frac{1}{4 \sqrt{2}}	\right),&\left(-\frac{1}{\sqrt{2}},\frac{1}{4},\frac{1}{4},\frac{1}{4},-\frac{3}{4}	\right),\\
\left(	-\frac{1}{\sqrt{2}},\frac{1}{4},\frac{1}{4},-\frac{3}{4},\frac{1}{4}\right),&\left(-\frac{1}{\sqrt{2}},\frac{1}{4},-\frac{3}{4},\frac{1}{4},\frac{1}{4}	\right),\\
\left(-\frac{1}{\sqrt{2}},-\frac{3}{4},\frac{1}{4},\frac{1}{4},\frac{1}{4}	\right),&\left(-\frac{1}{2 \sqrt{2}},-\frac{3}{4 \sqrt{2}},-\frac{3}{4 \sqrt{2}},-\frac{3}{4 \sqrt{2}},-\frac{3}{4 \sqrt{2}}	\right),\\
\end{array}
\end{equation*}
\begin{equation*}
\begin{array}{rr}
\left(\frac{1}{2 \sqrt{2}},\frac{4-\sqrt{2}}{8} ,\frac{4-\sqrt{2}}{8},-\frac{4+\sqrt{2}}{8},-\frac{4+\sqrt{2}}{8}\right),&\left(\frac{1}{2 \sqrt{2}},\frac{4-\sqrt{2}}{8} ,-\frac{4+\sqrt{2}}{8},\frac{4-\sqrt{2}}{8},-\frac{4+\sqrt{2}}{8}\right),\\
\left(\frac{1}{2 \sqrt{2}},-\frac{4+\sqrt{2}}{8},\frac{4-\sqrt{2}}{8} ,\frac{4-\sqrt{2}}{8},-\frac{4+\sqrt{2}}{8}\right),&\left(\frac{1}{2 \sqrt{2}},-\frac{4+\sqrt{2}}{8},\frac{4-\sqrt{2}}{8},-\frac{4+\sqrt{2}}{8} ,\frac{4-\sqrt{2}}{8}\right),\\
\left(\frac{1}{2 \sqrt{2}},-\frac{4+\sqrt{2}}{8},-\frac{4+\sqrt{2}}{8} ,\frac{4-\sqrt{2}}{8},\frac{4-\sqrt{2}}{8}\right),&\left(\frac{1}{2 \sqrt{2}},\frac{4-\sqrt{2}}{8} ,-\frac{4+\sqrt{2}}{8},-\frac{4+\sqrt{2}}{8}	,\frac{4-\sqrt{2}}{8}\right),
\end{array}
\end{equation*}
There are also 80 $\vec{\zeta}_{\rm KK_2}$'s, 160 $\vec{\zeta}_{\rm KK_3}$'s, 107 $\vec{\zeta}_{\rm KK_4}$'s, 16 $\vec{\zeta}_{\rm KK_5}$'s,
and 10 string oscillators $\vec{\zeta}_{\rm osc}$. For this case we do not sate the 2d and higher KK modes, as well as string oscillator modes, because there are many of them and they also follow from the positions of the 1d KK modes.

The edges between these $\vec{\zeta}_{\rm KK_1}$ vertices are depicted in Figure \ref{fig: 6from7}. The radius $\frac{1}{\sqrt{d-2}}=\frac{1}{2}$ ball fits inside $\mathcal{P}_{(6)}$.

\subsection{Symmetries \label{s.symmetry}}

The different maximal supergravity $\mathcal{P}_{(d)}$  tower polytopes described in section \ref{s.fullpolytope} are invariant under the action of specific finite symmetry groups $\mathsf{G}_d$. As it will be shown in this appendix, these symmetries are associated with U-duality groups of lower dimensional gauged supergravity obtained from toroidal compactifications of M-theory.

Consider for this some general EFT in $d$ dimensions such that its $k$-dimensional moduli space $\mathcal{M}$ is parameterized by some moduli $\{\varphi^i\}_{i=1}^{k}$. Its U-duality group $G$ maps these moduli into each other. Some of these transformations act only on the non-compact directions of $\mathcal{M}$ (parameterized by \emph{radions}), while others do so on the compact ones (\emph{axions}). 
In analogy with the \emph{maximal torus} defined for \emph{compact} Lie groups, for our \emph{non-compact} $G$ we consider $T_G^{\rm (r)}\simeq \mathbb{R}^n$ as the subgroup of diagonal matrices, acting only on the radions, by rescaling them. In general this will not affect the structure of the infinite-distance limits, so the polytopes obtained by using taxonomy rules should not be affected by these transformations.

In general $T_G^{\rm (r)}$ is not a normal subgroup of $G$, so one cannot simply quotient out these transformations, for this operation is generically ill-defined. In order to properly quotient $G$ by $T_G^{\rm (r)}$, we must introduce the \emph{normalizer} $N_GT_G^{\rm (r)}=\{g\in G:gT_G^{\rm (r)}=T_G^{\rm (r)}g\}$, i.e., the largest subgroup of $G$ such that $T_G^{\rm (r)}$ is a normal subgroup. The \emph{Weyl group} of $G$ is then defined as $W(G):=N_GT_G^{\rm (r)}/T_G^{\rm (r)}$,\footnote{Analogously as for the compact case, the Weyl group is unique up to isomorphism.} and now corresponds with the symmetries of the tower polytope, $\mathsf{G}_d\simeq W(G)$, as exchanging vertices is equivalent to mapping asymptotic directions among themselves. The Weyl group is finite (there are only finitely many distinct ways of exchanging vertices of a tower polytope) and a subgroup of $GL(\mathbb{R}^{n})$, where $n=k-a$ is the number of unbounded moduli/radions.

While $\mathsf{G}_d\simeq W(G)$ is not the full U-duality group (it lacks information about the symmetries of compact scalars and rescaling of the radions), it greatly constrains $G$, as $W(G)$ is the group generated by the reflections over hyperplanes perpendicular to the roots of the Lie algebra $\mathfrak{g}$ of $G$.
\begin{figure}
\begin{center}
\includegraphics[width=0.45\textwidth]{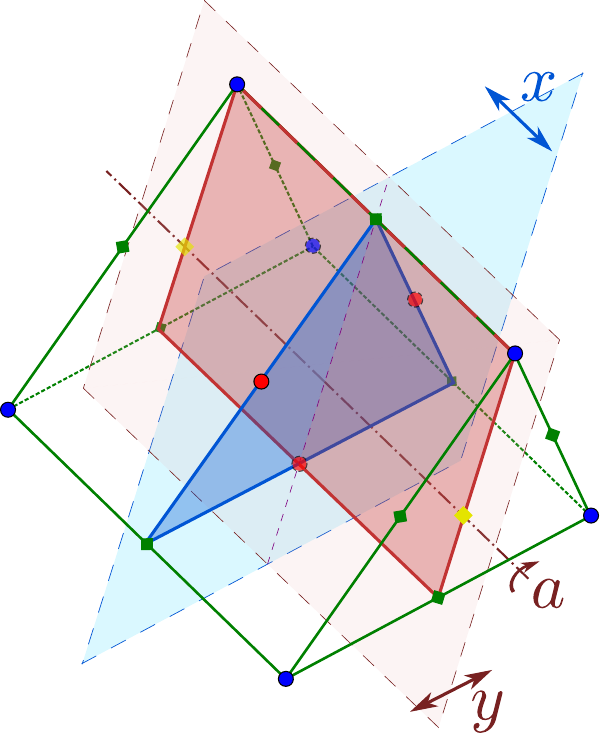}
\caption{Tower polytope $\mathcal{P}_{(8)}$ and the action on it by the group $\mathsf{G}_{8}=S_3\times S_2=\left \langle y,\,a: y^2=a^3=e,\, yay=a^{-1} \right\rangle\times\left\langle x: x^2=e\right\rangle$. The fixed loci correspond to those invariant under $x$ ($\Pi_{\rm (8,I)}$ in Table {\ref{tab:BIG}}) and $y$ ($\Pi_{\rm (8,II)}$).\label{f.symFixed8d}}
\end{center}
\end{figure}
In the case of toroidal compactifications of M-theory to $d=11-n$ dimensions the moduli space is given by the homogeneous space $E_{n(n)}/K_n$, where $E_{n(n)}$ are the split real forms of the compact exceptional Lie groups $E_n$, and $K_n$ is the maximal compact subgroup of $E_{n(n)}$. This way $E_{n(n)}$ acts a global symmetry on the moduli. For the UV completion of maximal supergravity by string theory compactifications, this symmetry is broken to a discrete subgroup $E_{n(n)}(\mathbb{Z})$, though the Weyl group stays the same \cite{Brcker1985RepresentationsOC}. This way, the symmetry group of the maximal supergravity tower polytopes $\mathcal{P}_{(d)}$ can be computed, as in Table \ref{tab.sym Pd}.
\begin{table}
\centering
\begin{tabular}{|c|c|c|c|c|c|}
\hline 
$d$ & $E_{n(n)}$ & $K_n$ & $\mathsf{G}_{d}\simeq W(E_{n(n)})$&$|\mathsf{G}_{d}|$&$N_d$ \\ 
\hline 
10A & $\mathbb{R}^+$ & $1$ & $1$&1	&1$^\ast$	 \\ 
10B & $SL(2,\mathbb{R})$ & $SO(2)$ & $\mathbb{Z}_2$ &2	&2	 \\ 
9 & $\!SL(2,\mathbb{R})\!\times\!SO(1,1)\!$ & $SO(2)$ & $S_2\simeq \mathbb{Z}_2$&2	&3	 \\  
8 & $\!SL(3,\mathbb{R})\!\times\!SL(2,\mathbb{R})\!$ & $SO(3)\times SO(2)$ & $S_3\times S_2$&12&6	 \\  
7 & $SL(5,\mathbb{R})$ & $SO(5)$ & $S_5$&120	&	10 \\ 
6 & ${\rm Spin}(5,5)$ & $({\rm Spin(5)}\!\times\!{\rm Spin(5)})/\mathbb{Z}_2\!$ & $SG(5)$&$1\,920$&16	 \\ 
\color{gray} 5 &\color{gray} $E_{6(6)}$ &\color{gray} $USp(8)/\mathbb{Z}_2$ & \color{gray}$W(E_{6(6)})$&\color{gray}$51840$&\color{gray}27	 \\ 
\color{gray} 4 &\color{gray} $E_{7(7)}$ &\color{gray} $SU(8)/\mathbb{Z}_2$ & \color{gray} $W(E_{7(7)})$ & \color{gray}$\!2903040\!$& \color{gray}56	 \\ 
\hline 
\end{tabular}
\caption{Global symmetry groups $E_{n(n)}$ of the $d$-dimensional supergravities and associated maximal compact subgroups and Weyl groups, for $10\geq d \geq 4$ (for completeness we also depict $d=5$ and 4, even if their associated $\mathcal{P}_{(d)}$ have not been computed). Notice the two possibilities for $d=10$, corresponding to Type IIA and Type IIB 10d supergravities. Following \cite{Brcker1985RepresentationsOC},  $W(SL(n))\simeq S_n$ and $W({\rm Spin}(n,n))\simeq SG(n)$, the group of even permutations $\sigma$ of $\{-n,...,-1,1,...,n\}$ such that $\sigma(-x)=-\sigma(x)$. The order of these groups and the number $N_d$ of $\vec{\zeta}_{\rm KK_1}$ vertices generating $\mathcal{P}_{(d)}$ is also given. Note that $N_{\rm 10A}=1$ as $\mathcal{P}_{(\rm 10A)}$ is generated by a $\vec{\zeta}_{\rm KK_1}$ and a $\vec{\zeta}_{\rm osc}$ vertices.\label{tab.sym Pd}}
\end{table}

To obtain the symmetry group directly from $\mathcal{P}_{(d)}$, we use the fact that $\mathsf{G}_{d}\leq GL(\mathbb{R},{n})$ (with $n=11-d$ for toroidal compactifications), and each element $g\in \mathsf{G}_d$ is determined by its action on $n$ linearly independent vertices out of the $N_d$ different $\vec{\zeta}_{\rm KK_1}$ vertices that generate $\mathcal{P}_{(d)}$. There are $\frac{N_d!}{n!}$ possibilities, represented by $M_g\in GL(\mathbb{R},{n})$ matrices, which can be used to check if the tower polytope is left invariant, and thus $g\in \mathsf{G}_d$. This determines $\mathsf{G}_d$, as the matrices $M_g$ give a representation of its elements. By this method one recovers the Weyl groups from Table \ref{tab.sym Pd}, at least for $d\in\{10,9,8,7,6\}$.

\begin{figure}
\begin{center}
\includegraphics[width=0.4\textwidth]{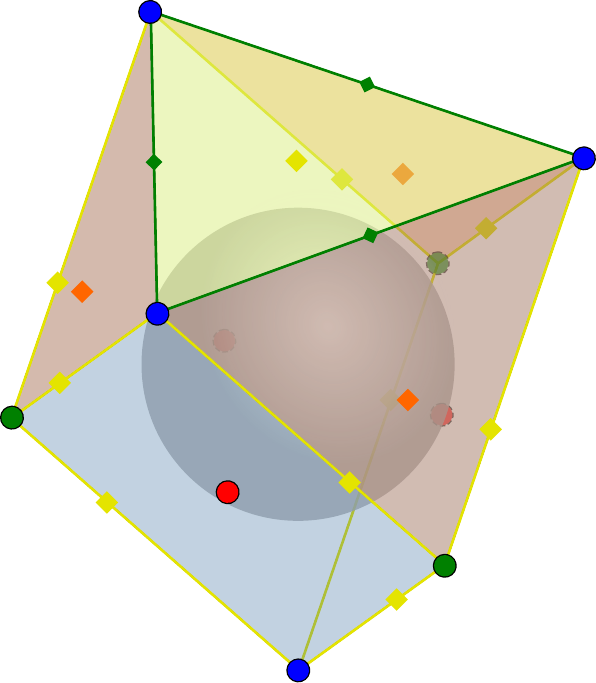}
\caption{Unique three-dimensional polytope obtained by applying the taxonomy rules for $d=7$. The string, $\vec{\zeta}_{\rm KK_1}$ and $\vec{\zeta}_{\rm KK_2}$ towers are depicted in red \fcolorbox{black}{red}{\rule{0pt}{6pt}\rule{6pt}{0pt}}, blue \fcolorbox{black}{blue}{\rule{0pt}{6pt}\rule{6pt}{0pt}} and green \fcolorbox{black}{dark-green}{\rule{0pt}{6pt}\rule{6pt}{0pt}}. Edges and facets associated to decompactifications of two, three and four dimensions appear in green \fcolorbox{black}{dark-green}{\rule{0pt}{6pt}\rule{6pt}{0pt}}, yellow \fcolorbox{black}{yellow}{\rule{0pt}{6pt}\rule{6pt}{0pt}} and orange \fcolorbox{black}{orange}{\rule{0pt}{6pt}\rule{6pt}{0pt}}, with their closest point to origin highlighted. Note the sphere with radius $\frac{1}{\sqrt{d-2}}=\frac{1}{\sqrt{5}}$ is contained inside it.\label{f.7d3}}
\end{center}
\end{figure}

\begin{figure}[htp]
\begin{subfigure}{0.36\textwidth}
\centering
\includegraphics[width=.85\textwidth]{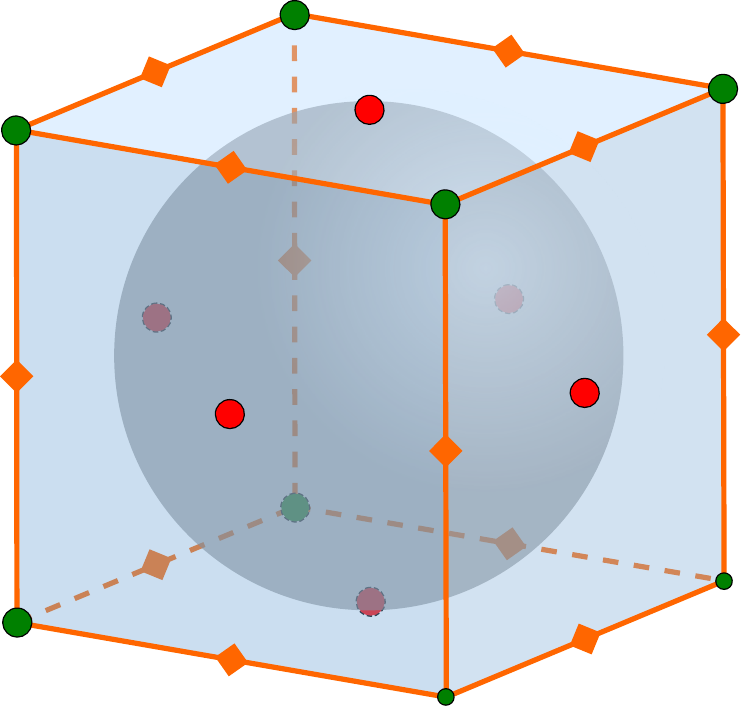}
\caption{\label{sf.3d-6d1}}
\end{subfigure}
\begin{subfigure}{0.24\textwidth}
\centering
\includegraphics[width=\textwidth]{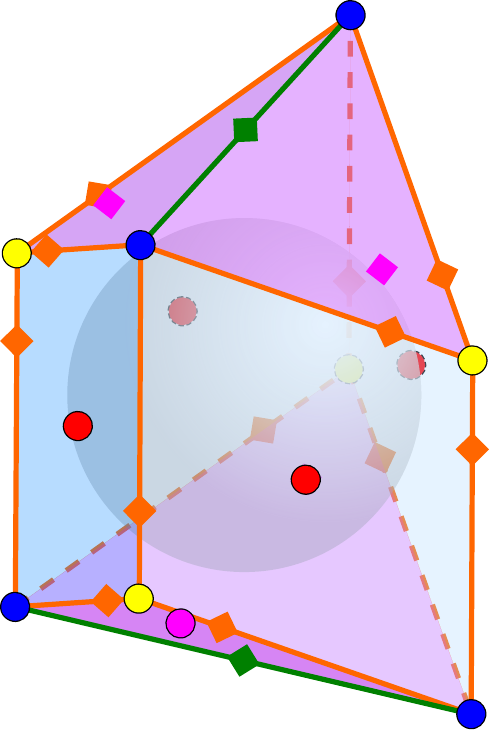}
\caption{\label{sf.3d-6d2}}
\end{subfigure}
\begin{subfigure}{0.36\textwidth}
\centering
\includegraphics[width=\textwidth]{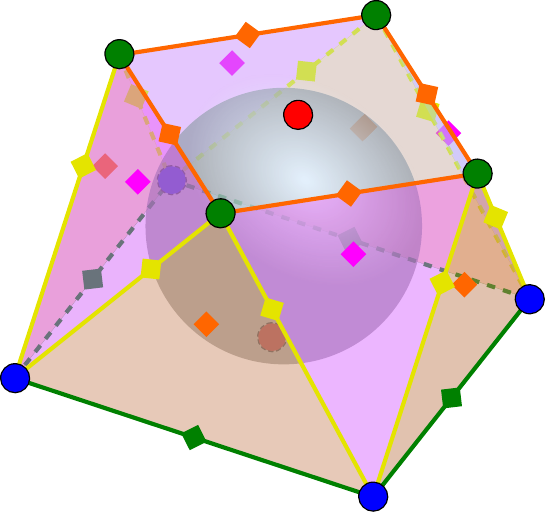}
\caption{\label{sf.3d-6d3}}
\end{subfigure}
\caption{Representation of the three rank 3 tower polytopes in $d=6$ obtained as slices of the maximal $\mathcal{P}_{(6)}$ tower polytope. The string, $\vec{\zeta}_{\rm KK_1}$, $\vec{\zeta}_{\rm KK_2}$ and $\vec{\zeta}_{\rm KK_3}$ towers are depicted in red \fcolorbox{black}{red}{\rule{0pt}{6pt}\rule{6pt}{0pt}}, blue \fcolorbox{black}{blue}{\rule{0pt}{6pt}\rule{6pt}{0pt}}, green \fcolorbox{black}{dark-green}{\rule{0pt}{6pt}\rule{6pt}{0pt}} and yellow \fcolorbox{black}{yellow}{\rule{0pt}{6pt}\rule{6pt}{0pt}}. Edges and facets associated to decompactifications of two, three, four and five dimensions appear in green \fcolorbox{black}{dark-green}{\rule{0pt}{6pt}\rule{6pt}{0pt}}, yellow \fcolorbox{black}{yellow}{\rule{0pt}{6pt}\rule{6pt}{0pt}}, orange \fcolorbox{black}{orange}{\rule{0pt}{6pt}\rule{6pt}{0pt}} and pink \fcolorbox{black}{Magenta}{\rule{0pt}{6pt}\rule{6pt}{0pt}}, with their closest point to origin highlighted. In gray the sphere with radius $\frac{1}{\sqrt{d-2}}=\frac{1}{2}$ is depicted. Note that the cube depicted in Figure \ref{sf.3d-6d1} is nothing but that from Figure \ref{f.cube4dTOW} under the rescaling described before Section \ref{s.fullpolytope}.\label{f.6d3}}
\end{figure}

\begin{figure}[h]
\begin{center}
\centering
\includegraphics[width=0.35\textwidth]{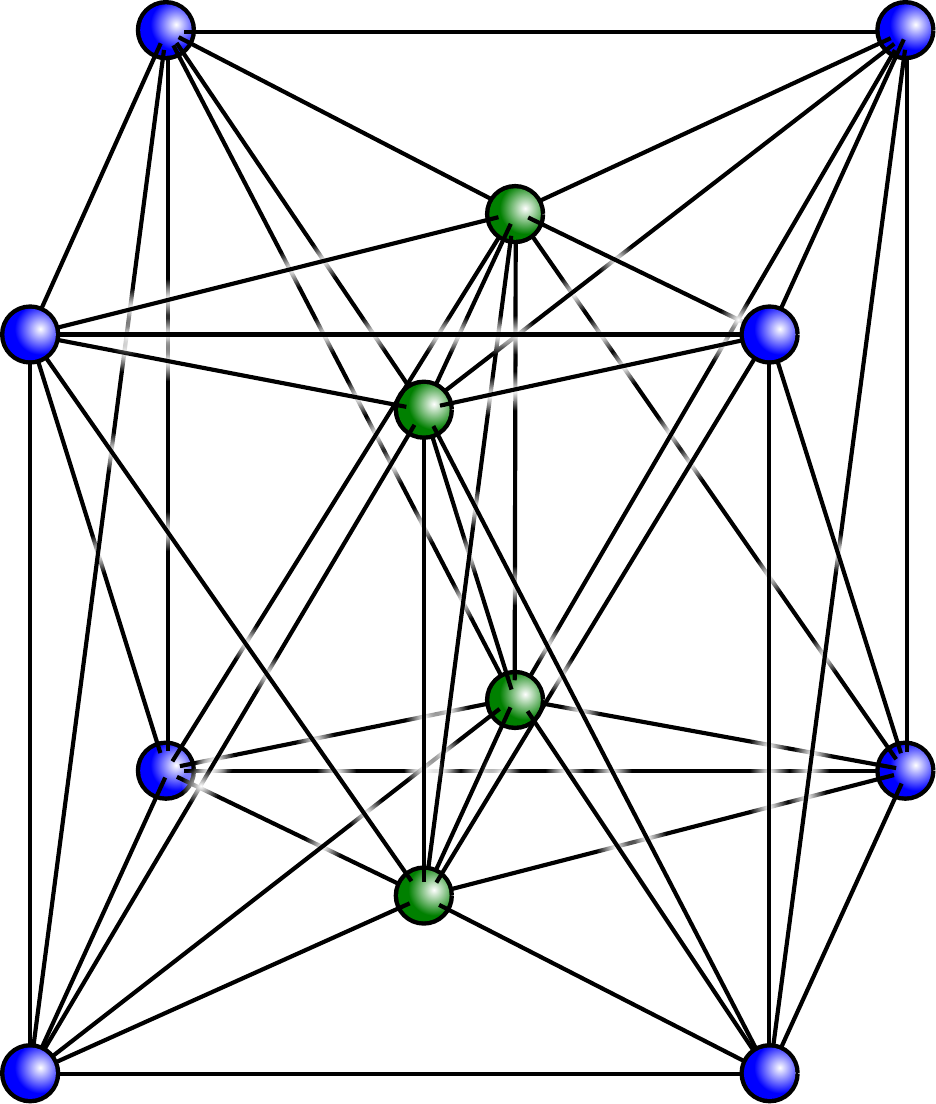}
\caption{An illustration of the unique four-dimensional tower polytope in $d=6$ obtained as a slice of the maximal tower polytope $\mathcal{P}_{(6)}$ of M-theory compactified on $T^5$ (note that other rank 4 tower polytopes following the taxonomy rules could exist, associated to non-toroidal compactifications). For simplicity, the only depicted towers are those corresponding to vertices generating the polytope, associated to decompactifications of either one or two dimensions, respectively in blue \fcolorbox{black}{blue}{\rule{0pt}{6pt}\rule{6pt}{0pt}} or green \fcolorbox{black}{dark-green}{\rule{0pt}{6pt}\rule{6pt}{0pt}}.\label{f.4d-6d1}}
\end{center}
\end{figure}

One can see whether an $n$-dimensional tower polytope with $n<11-d$ built using the $d$-dimensional taxonomy rules is embedded in $\mathcal{P}_{(d)}$ as part of an $n$-hyperplane containing the origin. One way to obtain such a polytope is by studying the invariant loci of $\mathcal{P}_{(d)}$ under the different elements of $\mathsf{G}_{(d)}$. Here, we will consider tower polytopes of rank $n\geq 2$.
\begin{itemize}
	\item We begin with $d=8$, with $\mathcal{P}_{(8)}$ being 3-dimensional and $\mathsf{G}_{8}=S_3\times S_2$ having $|S_3\times S_2|=3!\cdot 2!=12$ elements. As depicted in Figure \ref{f.symFixed8d}, there are two classes of 2-dimensional fixed loci, respectively being the fixed points of the order 2 elements of $S_2$ and $S_3$. These are the polygons depicted as $\Pi_{\rm (8, I)}$ and  $\Pi_{\rm (8, II)}$ in Table \ref{tab:BIG}, precisely the two only rank 2 polytopes given by taxonomy in $d=8$.
	\item For $d=7$, the four-dimensional polytope $\mathcal{P}_{(7)}$ is spanned by ten $\vec{\zeta}_{\rm KK_1}$, and $\mathsf{G}_{7}=S_5$ has $5!=120$ elements. In this case we obtain two possible rank 2 polytopes as fixed loci, precisely the unique two polygons $\Pi_{\rm (7, I)}$ and  $\Pi_{\rm (7, II)}$ depicted in Table \ref{tab:BIG}, as well as a single three-dimensional polytope, depicted in Figure \ref{f.7d3}. 
	\item For $d=6$, the five-dimensional polytope $\mathcal{P}_{(6)}$ (with $\vec{\zeta}_{\rm KK_1}$ depicted in Figure \ref{fig: 6from7}) has symmetry group $\mathsf{G}_6=W(\rm{Spin(5,5)})$, of order 1920. When obtaining the fixed loci, those of dimension 2 correspond to the 2-polygons $\Pi_{\rm (6, I)}$, $\Pi_{\rm (6, II)}$, $\Pi_{\rm (6, III)}$ and $\Pi_{\rm (6, IV)}$ from Table \ref{tab:BIG}.\footnote{Note that $\Pi_{\rm (6, V)}$ from Table \ref{tab:BIG} does not appear as an invariant locus of $\mathcal{P}_{(6)}$, though seemingly it is allowed by the rules. However, as shown in Section \ref{app:dim red}, the polygon is consistent, simply not appearing for toroidal compactifications of maximal SUGRA, but rather some other (unidentified) compactification without cycles of odd dimensionality.} There are no pairs of $\vec{\zeta}_{\rm osc}$ and $\vec{\zeta}_{\rm KK_4}$ vectors with an $\arccos\frac{1}{3}$ angle between them (preventing the $\Pi_{\rm (6, VIII)}$, $\Pi_{\rm (6, X)}$ and $\Pi_{\rm (6, XI)}$), no pairs of $\vec{\zeta}_{\rm KK_4}$  and $\vec{\zeta}_{\rm KK_5}$ vectors separated a distance $\sqrt{\frac{37}{20}}$ (polygons $\Pi_{\rm (6, VI)}$, $\Pi_{\rm (6, IX)}$ and $\Pi_{\rm (6, XI)}$) and no pairs of $\vec{\zeta}_{\rm KK_2}$  and $\vec{\zeta}_{\rm KK_4}$ vectors separated a distance $\frac{1}{2}+\frac{1}{\sqrt{2}}$ ($\Pi_{\rm (6, VII)}$ and $\Pi_{\rm (6, X)}$). As for the 3-dimensional fixed loci, three of them are found, depicted in Figure \ref{f.6d3}. Finally, a single type of fixed locus of dimension 4, depicted in Figure \ref{f.4d-6d1}, is recovered.
\end{itemize}
On principle one could consider the possibility that the $\Pi_{(d,\cdot)}$ polygons in question do not belong to an invariant plane, but still can be embedded on $\mathcal{P}_{(d)}$. A careful study (computing the distances between the different vertices and trying to connect them with the lengths in $\Pi_{(d,\cdot)}$) shows that this is not the case, i.e. all the $\Pi_{(d,\cdot)}$ polygons embedded in $\mathcal{P}_{(d)}$ appear in fixed planes.

From the recursive behavior of $\mathcal{P}_{(d)}$ when decompactifying, the following relation is recovered. As shown earlier in this appendix and in \cite{Etheredge:2022opl}, for $d<9$ all $N_d$ different $\vec{\zeta}_{\rm KK_1}$ vertices generating $\mathcal{P}_{(d)}$ are KK modes of one dimension reduction, allowing recovery of $\mathcal{P}_{(d+1)}$. As these tower polytopes have symmetry groups $\mathsf{G}_d$ and $\mathsf{G}_{d+1}$, from the orbit-stabilizer group it is straightforward that
\begin{equation}
	|\mathsf{G}_d|=N_d |\mathsf{G}_{d+1}|\qquad\text{for }d\leq 8\;,
\end{equation}
as noted in Table \ref{tab.sym Pd}.

\begin{figure}
\begin{center}
\includegraphics[width=0.45\textwidth]{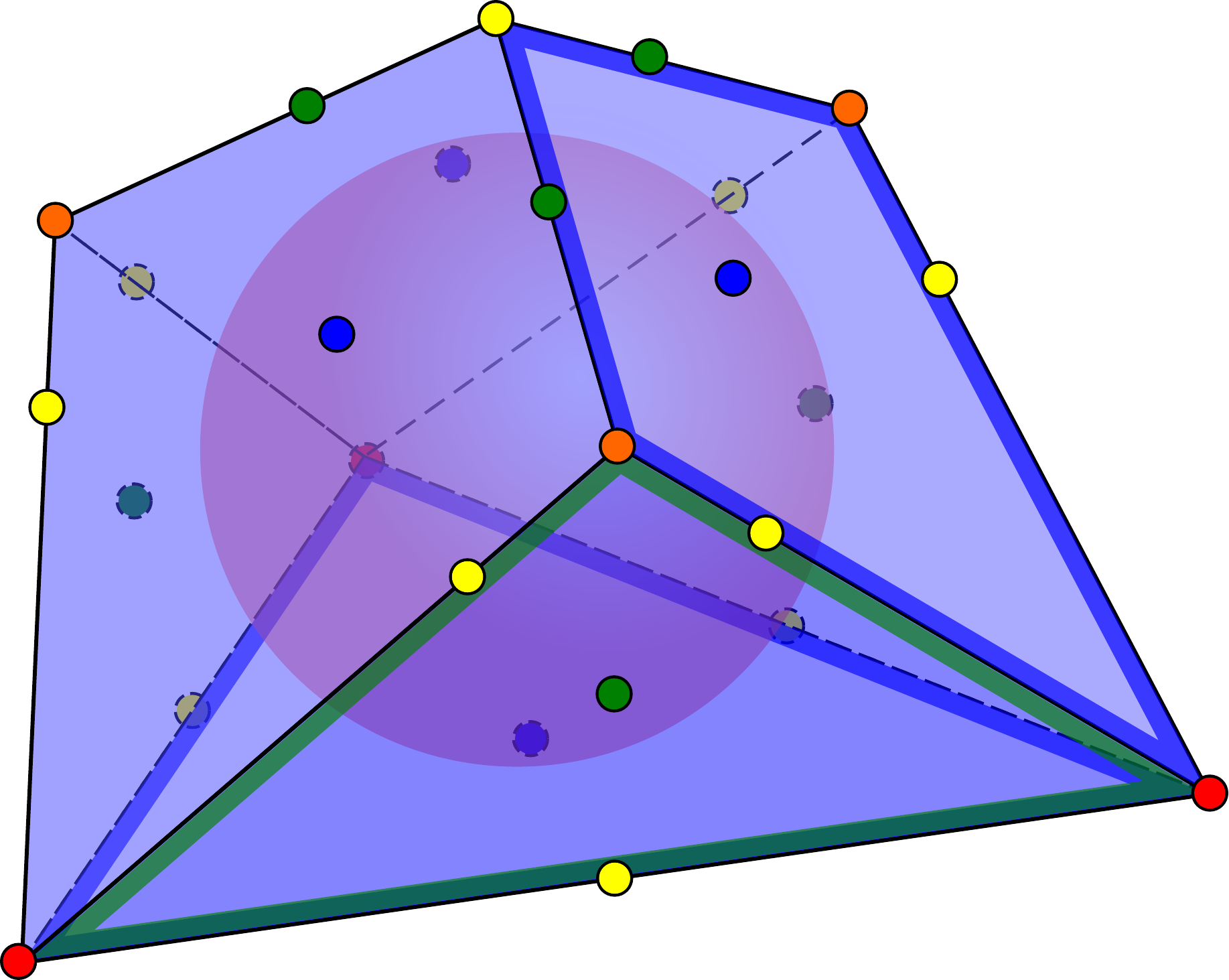}
\caption{Species polytope in $d=7$ dual to that pictured in Figure \ref{f.7d3}. The $\mathcal{Z}$-vectors associated to the string scale are depicted in red \fcolorbox{black}{red}{\rule{0pt}{6pt}\rule{6pt}{0pt}}, while those to the 8, 9, 10 and 11-dimensional Planck mass appear in blue \fcolorbox{black}{blue}{\rule{0pt}{6pt}\rule{6pt}{0pt}}, green \fcolorbox{black}{dark-green}{\rule{0pt}{6pt}\rule{6pt}{0pt}}, yellow \fcolorbox{black}{yellow}{\rule{0pt}{6pt}\rule{6pt}{0pt}} and orange \fcolorbox{black}{orange}{\rule{0pt}{6pt}\rule{6pt}{0pt}}, respectively. The sphere with radius $\frac{1}{\sqrt{(d-2)(d-1)}}=\frac{1}{\sqrt{30}}$ is depicted. The facets recovered from rank 2 species polytopes for $(D=7+n)$-dimensional theories are outlined in blue ($n=1$, corresponding to $\Pi^\circ_{\rm (8,I)}$ and $\Pi^\circ_{\rm (8,II)}$ in Table \ref{tab:BIGsp}) and green ($n=2$, $P^\circ_{(9)}=\Pi^\circ_{\rm (9,I)}$). The facets not outlined are congruent to those that are. The $n$ associated to each facet must be subtracted from each vertex to recover the appropriate one in the $D$-dimensional theory. \label{f.sp7d3}}
\end{center}
\end{figure}

\subsection{M-theory on K3\label{app:K3}}

Our rules not only apply to toroidal compactifications of M-theory, but also to more involved internal manifolds, such as a K3-surface. In this subsection, we consider a particular 2d slice of M-theory compactified on a K3 surface to produce a seven-dimensional theory with 32 supercharges. We show that this example is captured by our polygon classification for 7d theories with 11d as the maximum decompactification dimension.\footnote{For an in-depth discussion of the physics behind this example, see \cite{Lee:2019xtm,Castellano:2023jjt, Castellano:2023stg}.}

\begin{figure}[h]
\begin{center}
\includegraphics[width=100mm]{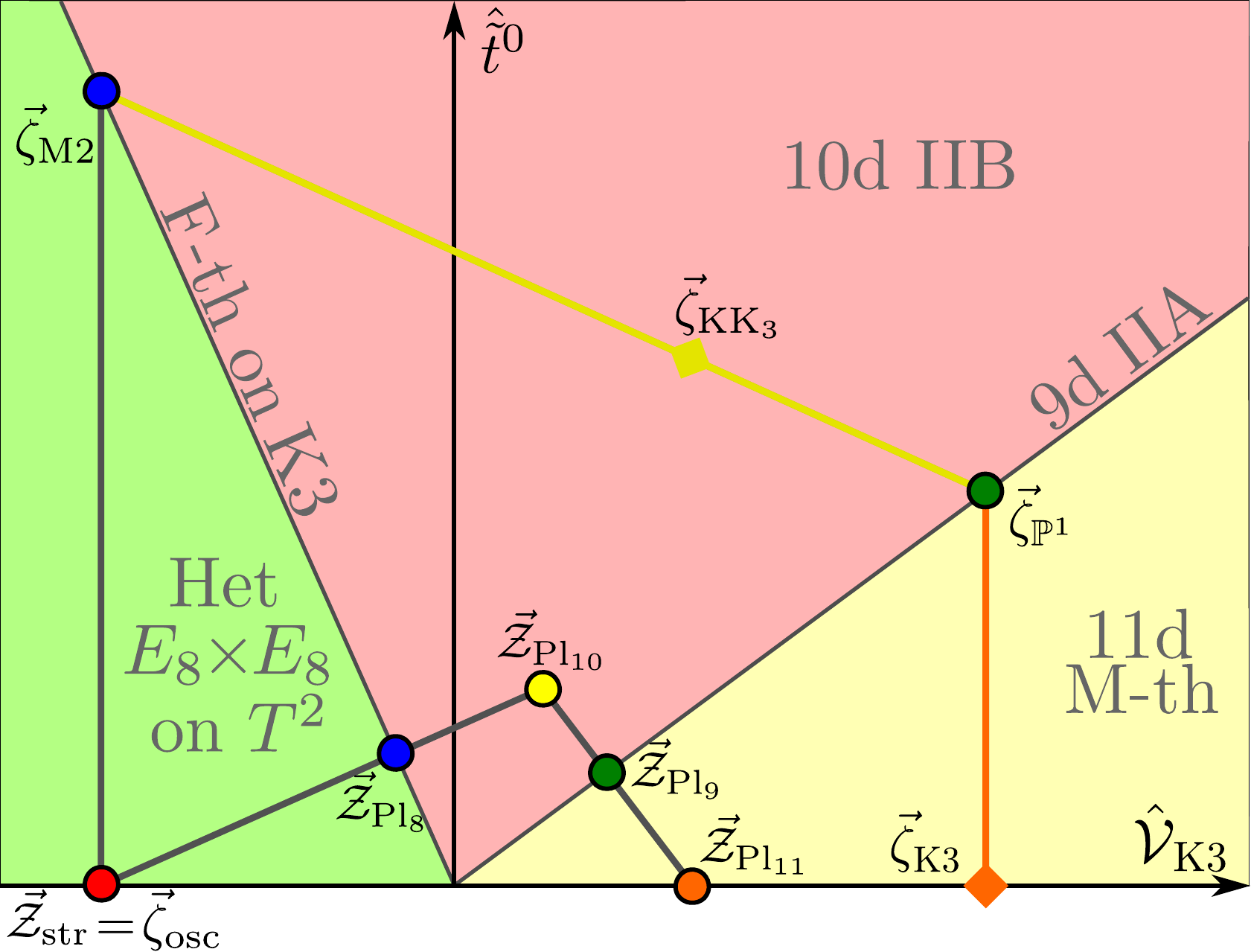}
      \caption{Scalar charge-to-mass ratio vectors for the different towers (outer polygon) and cut-offs (inner polygon), for M-theory compactified on ${\rm K3}$, in the moduli space slice spanned by the canonically normalized $\hat{\mathcal{V}}_{{\rm K3}}\in\mathbb{R}$ and $\hat{\tilde{t}}_0>0$, \cite{Castellano:2023jjt, Castellano:2023stg}. Note that they correspond to half of the polygon $\Pi_{\rm(7,II)}$ depicted in Table \ref{tab:BIG} and its dual polygon ($\Pi^\circ_{\rm(7,II)}$ in Table \ref{tab:BIGsp}). The different theories resulting from each limit are also depicted.\label{fig:Mth K3}}
\end{center}
\end{figure}

For simplicity we consider \emph{attractive K3s} (i.e. with maximal rank $20=h^{1,1}({\rm K3})$ for the Picard group ${\rm Pic}( {\rm K3})=H^{1,1}({\rm K3})\cap H^2({\rm K3},\mathbb{Z})$), such that the complex structure is completely fixed \cite{Moore:1998zu}, and the 7d action and the masses depend on only the K\"ahler moduli $\{t^a\}_{a=1}^{20}$. The relevant part of the 7d action is
\begin{equation}
	S_{\rm 7d}\supset \frac{1}{2\kappa_7^2}\int {\rm d}^7 x\sqrt{-g}\left[\mathcal{R}-\frac{9}{20}(\partial\log\mathcal{V}_{ {\rm K3}})^2-\mathsf{G}_{ab}\partial\tilde{t}^a\cdot\partial\tilde{t}^b\right]\;,
\end{equation} 
where $\mathcal{V}_{{\rm K3}}=\frac{1}{2}\eta_{ab}t^a t^b$ is the ${\rm K3}$ volume in Planck units, $\eta_{ab}=\omega_a\cdot\omega_b$ are the intersection numbers in a basis $\{\omega_a\}$ of $H^{1,1}({\rm K3},\mathbb{Z})$, such that  the K\"ahler form is expressed as $J=t^a\omega_a$, and $\{\tilde{t}^a=t^a/\mathcal{V}_{{\rm K3}}^{1/2}\}_{a=1}^{20}$ are rescaled moduli with $\frac{1}{2}\eta_{ab}\tilde{t}^a \tilde{t}^b\equiv 1$. The (classically exact) moduli space is given by $\mathcal{M}_{{\rm K3}}=O(3,19;\mathbb{Z})\backslash O(3,19;\mathbb{R})/O(19)\times \mathbb{R}_+$. The coset piece, parameterized by $\{\tilde{t}^a\}_{a=1}^{20}$, admits the natural metric
\begin{equation}
	\mathsf{G}_{ab}=\int_{{\rm K3}}\omega_a\wedge\star\omega_b=\tilde{t}^a\tilde{t}^b-\eta_{ab}\;,
\end{equation}
while the $\mathbb{R}_+$ is parameterized by the global volume, with metric $\mathsf{G}_{\mathcal{V}_{{\rm K3}}\mathcal{V}_{{\rm K3}}}=\frac{9}{20\mathcal{V}_{{\rm K3}}^2}$. As shown in \cite{Lee:2019xtm} the attractive ${\rm K3}$ admits an elliptic fibration $\mathcal{C}_0$ over a $\mathbb{P}^1$-base, such that the associated 2-form $\omega_0$ has intersection $\eta_{00}=\omega_0\cdot\omega_0=0$. One can then work in the (flat) 2d-slice of $\mathcal{M}_{{\rm K3}}$ spanned by the canonically normalized moduli $\{\hat{\mathcal{V}}_{K_3}=\frac{3}{2\sqrt{5}}\log\mathcal{V}_{{\rm K3}}\in\mathbb{R},\hat{\tilde{t}}_0=\log \tilde{t}_0>0\}$.\footnote{Do not confuse the canonically normalized modulus $\hat{\tilde{t}}_0$ with the unit tangent vector $\hat{t}$ given some trajectory.} There are five possible leading towers becoming light in the different infinite-distance limits (see Section 4.2 from \cite{Castellano:2023jjt} and references therein for more details):
\begin{itemize}
	\item In the large volume limit, $\mathcal{V}_{{\rm K3}}\to  \infty$, with the K\"ahler saxions fixed, we decompactify to 11d M-theory, with
	\begin{equation}
		\frac{m_{{\rm KK},\; {\rm K3}}}{M_{\rm Pl,7}}\sim \mathcal{V}_{{\rm K3}}^{-9/20}\;\Longrightarrow\;\vec{\zeta}_{{\rm K3}}=\left(\frac{3}{2\sqrt{5}},0\right)\;.
	\end{equation}
	\item In the small volume limit, $\mathcal{V}_{{\rm K3}}\to 0$, an emergent, heterotic-like, string appears as a result of M5-branes wrapped over the whole K3 surface becoming light, with
	\begin{equation}
		\frac{m_{\rm M5}}{M_{\rm Pl,7}}\sim \mathcal{V}_{{\rm K3}}^{3/10}\;\Longrightarrow\;\vec{\zeta}_{\rm osc}=\left(-\frac{1}{\sqrt{5}},0\right)\;.
	\end{equation}
	\item Associated to the $\mathbb{P}^1$-base, in the $\tilde{t}^0\to\infty$ limit we have that its volume grows asymptotically, resulting in a KK tower
	\begin{equation}
		\frac{m_{\mathbb{P}^1}}{M_{\rm Pl,7}}\sim (\tilde{t}^0)^{-1/2}\mathcal{V}_{{\rm K3}}^{9/20} \;\Longrightarrow\;\vec{\zeta}_{\mathbb{P}^1}=\left(\frac{3}{2\sqrt{5}},\frac{1}{2}\right)\;,
	\end{equation}
	which decompactifies to 9d Type IIA string theory.
	\item Additionally there exits $\frac{1}{2}$-BPS states obtained from wrapped M2-branes on the genus-one fibre $\mathcal{C}_0$ shrinking in that limit, with 
	\begin{equation}
	\frac{m_{\rm M2}}{M_{\rm Pl,7}}\sim (\tilde{t}^0)^{-1}\mathcal{V}_{{\rm K3}}^{3/10} \;\Longrightarrow\;\vec{\zeta}_{\rm M2}=\left(-\frac{1}{\sqrt{5}},1\right)\;,
	\end{equation}
	amounting to decompactification to an 8d theory given by F-theory on ${\rm K3}$.
	\item There is an extra tower charged both under the $\mathbb{P}^1$ KK and the M2 winding modes, equivalent to decompactifying three internal dimensions to 10d Type IIB string theory,
	\begin{equation}
		\frac{m_{\rm KK_3}}{M_{\rm Pl,7}}\sim \mathcal{V}_{{\rm K3}}^{-1/5}(\tilde{t}^0)^{-2/3}\;\Longrightarrow\;\vec{\zeta}_{\rm KK_3}=\left(\frac{2}{3\sqrt{5}},\frac{2}{3}\right)\;.
	\end{equation}
\end{itemize}
The species scales for each of the different limits can be computed as the Planck masses of the theories we are decompactifying to, or the string scale in the case of the wrapped M5 limit. In Figure \ref{fig:Mth K3} the different vectors are plotted, both for the towers and species scales. The resulting convex hull corresponds to half of the polygon $\Pi_{\rm(7,II)}$ represented in Table \ref{tab:BIG}, divided by the symmetry axis given by $\vec{\zeta}_{\rm KK_4}$ and $\vec{\zeta}_{\rm osc}$. Even if the asymptotic region of the moduli space (in our flat frame) subtends only an angle of $\pi$ (rather than $2\pi$), one can still apply the taxonomy rules to fill the regions between the above limiting vectors, which are known to be there beforehand from the $\hat{\mathcal{V}}_{{\rm K3}}\to\pm\infty$ limits. 

This example demonstrates that our taxonomy rules apply to examples beyond toroidal compactifications, but also to more complicated ones--even cases in which the moduli spaces are not geodesically complete (for more on this see Appendix \ref{a.GeodesicallyIncomplete}). On principle, one can try to find a 3-dimensional flat slice on $\mathcal{M}_{{\rm K3}}$, obtaining additional towers, in a such a way that part the tower polytope depicted in Figure \ref{f.7d3} is recovered.

\section{Orthogonality of sliding \label{s.orthogonality}}

In this section, we discuss orthogonality of sliding for flat moduli spaces. 

Consider first a flat 2d moduli space with coordinates $\{x,y\}$, and let $m=m(x,y)$ be the characteristic mass of an infinite tower as a function of the moduli. Defining $f(x,y)=-\log m(x,y)$, we have
\begin{align}
	\vec \zeta=\vec\nabla f.
\end{align}
Geodesics are straight lines in the plane, with all of them going to infinite-distance limits. Let us assume that asymptotically far out along any geodesic, $\vec \zeta$ goes to a constant, finite vector. This implies, for instance, that if the geodesic points in the $\hat{t}=(\hat{t}^x,\hat{t}^y)=(1,0)$ direction (with any fixed value of $y$), then we must have
\begin{align}
	\partial_x \zeta_x \rightarrow 0\qquad \text{and}\qquad \partial_x \zeta_y \rightarrow 0\qquad \text{as}\qquad x\rightarrow \infty, \quad \text{fixed }y.
\end{align}
But then, since $\zeta_x=\partial_x f$, $\zeta_y=\partial_y f$, and assuming partial derivatives commute, this implies that
\begin{align}
	\partial_y\zeta_x\rightarrow 0\qquad\text{as}\qquad x\rightarrow \infty, \text{ fixed }y.
\end{align}
That means that, although the asymptotic values of $\vec \zeta$ can change as we move to a different, parallel geodesic, the change can only occur in a direction that is perpendicular to the original geodesic. The generalization to an $n$-dimensional flat moduli spaces involves replacing $y$ with a vector $\vec y\in \mathbb R^{n-1}$.

The above covers the case where the impact parameter of the geodesic is varied while its direction is held fixed (i.e., we take different geodesics going to the same infinite-distance point). Suppose, instead, that we vary the direction of the geodesic, and let the new and old geodesics intersect at some point, which we take to be the origin. Then it is convenient to write $f=f(r,\theta)$ in polar coordinates. Now we have
\begin{align}
	\zeta_{\hat r}=\frac{\partial f}{\partial r}, \qquad \zeta_{\hat \theta}=\frac{1}{r}\frac{\partial f}{\partial \theta}.
\end{align}
Thus, in order for $\vec \zeta$ to approach a $\theta$-dependent constant as $r\rightarrow \infty$ at fixed $\theta$, we must have 
\begin{align}
	f(r,\theta)=rf_1(\theta)+\cdots,
\end{align}
up to terms that grow more slowly that $\mathcal O(r)$. Now we compute
\begin{align}
	\vec \zeta=f_1(\theta)\hat r+f_1'(\theta)\hat \theta.
\end{align}
Using $\partial_\theta \hat r=\hat \theta$ and $\partial_\theta \hat \theta=-\hat r$, we find:
\begin{align}
	\partial_\theta \vec \zeta =f_1'(\theta)\hat r+f_1(\theta)\hat\theta +f_1''(\theta)\hat \theta-f_1'(\theta)\hat r=[f_1(\theta)+f_1''(\theta)]\hat \theta,
\end{align}
so indeed the change in $\vec \zeta$ is always perpendicular to the direction of the geodesic $\vec r$.

In fact, this 2d analysis implies the sliding behaves in the same way in the higher-dimensional case. One could replace 2d polar coordinates with higher-dimensional spherical angular coordinates. A detailed analysis of a higher-dimensional spherical example reduces to a generalization of the 2d one.

\section{Geodesically-incomplete examples \label{a.GeodesicallyIncomplete}}

In this appendix, we discuss the applicability of our taxonomy rules in general geodesically incomplete moduli spaces, where not every direction leads to an infinite-distance singularity. Another example of our taxonomy rules in such spaces can be found in Appendix \ref{app:K3}.

As a broad class of geodesically incomplete moduli spaces, consider compactifications on Calabi-Yau threefolds $X$,\footnote{See \cite{ xiao2016positivity,lehmann2016convexity,Lanza:2021udy} for more details on these concepts in K\"ahler geometry and the associated cones.} with the K\"ahler form $J=s^a[D_a]$ expanded in an integral basis $[D_a]$ Poincar\'e-dual to a set of divisors $D_a$. In terms of the triple intersection numbers $\kappa_{abc}=D_a\cdot D_b\cdot D_c$, the volume of $X$ in string units is given by $V_X=\frac{1}{6}\kappa_{abc}s^a s^b s^c$, and the saxionic components of the moduli space metric are given by
\begin{equation}
	\mathsf{G}_{ab}=-\frac{1}{2}\partial_a\partial_b \log V_X,
\end{equation}
The K\"ahler cone $\mathcal{K}(X)\subseteq \mathbb{R}_{>0}^n$ is the set over which $J$ takes values, parameterized by $\{s^a\}_{a=1}^{h^{1,1}=n}$. In general the inclusion is strict, as these saxions measure the volume of the effective curves generating the dual Mori cone $\overline{\rm Eff}_1(X)$, whose relationship with $D^i$ is nontrivial. From the definition of $\mathcal{K}(X)$, it follows that the (saxionic) moduli space\footnote{This is at least true parameterizing this patch, there could be flops to other K\"ahler cones.
} might be asymptotically  flat\footnote{While it is easy to check that, up to $\alpha'$ and instanton corrections, which can be safely ignored in the asymptotic regime, the saxionic moduli space is Riemann flat for $n=2$, this is not generally the case for $n>2$, see \cite{Marchesano:2023thx} for more on the asymptotic curvature of CY moduli spaces.} but is never geodesically complete. One can compute that, for flat $\mathcal{K}(X)$, the subtended angle is given by 
\begin{equation}\label{e.solid angle}
	\Omega_{\mathcal{K}(X)}=\frac{4}{n}\lim_{r\to \infty} r^{-n}\int_{\mathcal{K}(X)\cap B(\mathbf{s}_0,r)}\sqrt{\mathsf{G}}{\rm d}s^1... {\rm d}s^n\;,
\end{equation}
where $B(\mathbf{s}_0,r)=\{\mathbf{s}:\|\mathbf{s},\mathbf{s}_0\|\leq r\}$ and $\mathbf{s}_0$ is some arbitrary fixed point. In general $\Omega_{\mathcal{K}(X)}<\frac{2\pi^{n/2}}{\Gamma(\frac{n}{2})}$, but as we will argue now, we expect these angles to be quantized by the taxonomy rules.

As described in \cite{Corvilain:2018lgw}, $\mathcal{K}(X)$ can be divided in several \emph{growth sectors} $R_{i_1...i_n}=\{s^{i_1}\gtrsim...\gtrsim s^{i_n}\}$ as we move to infinite-distance limits. Each growth sector is associated to a specific singularity type\footnote{This is determined by the behavior of the triple intersection numbers $\kappa_{abc}$ \cite{Corvilain:2018lgw} or a geometrical analysis \cite{Lee:2019wij}.}, with an associated fibration structure in $X$ and dominant leading tower becoming light. As limiting interfaces of the K\"ahler cone will correspond to the ``deepest'' regions of some growth sectors, the scalar charge-to-mass ratio vectors of the leading towers in these limits are expected to point along these these directions. This way, even if the moduli space is not geodesically complete (so that the polytope following the taxonomy rules closes), it still needs to ``fill up'' the solid angle $\Omega_{\mathcal{K}(X)}$ enclosed by these tower vectors.

The nuance here comes when realizing that these tower vectors can correspond to points that in the ``complete'' vectors appear in edges/facets/etc closest points to the origin, such as $\vec{\zeta}_{\rm pc}\equiv\vec{\zeta}_{\rm KK_2}$ along an edge spanned by $\vec{\zeta}_{\rm KK_1}$ vertices, see \eqref{e:zeta min}.

To illustrate this, we refer to the two 2-moduli $d=4$ examples appearing in Section 5 from \cite{Castellano:2023jjt}, corresponding to Type IIA string theory on $\mathbb{P}^{(1,1,1,6,9)}[18]$ and $\mathbb{P}^{(1,1,1,2,6)}[12]$. There is no sliding for any of these tower vectors (although there is for some of the subleading towers). This allows us to compute the subtended angles, either using \eqref{vertexangles again}, or directly from \eqref{e.solid angle}, as in both cases the explicit flat coordinates are known \cite{Castellano:2023jjt} and the integral can be computed. The different results appear in Table \ref{tab:subtended}.
\begin{table}
\centering
\begin{tabular}{|c|c|c|}
\hline 
$X$ & $\mathbb{P}^{(1,1,1,6,9)}[18]$ & $\mathbb{P}^{(1,1,1,2,6)}[12]$ \\ 
\hline 
$V_X$ & $\frac 32 (s^1)^3+\frac 32 (s^1)^2s^2+\frac{1}{2} s^2(s^2)^2$ & $\frac 43 (s^1)^3+2(s^1)^2s^2$ \\ 
$\Omega_{\mathcal{K}(X)}$ & $\frac{1}{2}\theta_{1,1}={\rm arccot}\sqrt{2}$ & $\theta_1=\arctan\sqrt{2}$ \\ 
Limiting points & $\vec{\zeta}_{\rm KK_1},\;\vec{\zeta}_{\rm KK_2}$ & $\vec{\zeta}_{\rm KK_1},\; \vec{\zeta}_{\rm osc}$ \\ 
\hline 
\end{tabular} 
\caption{Volume of the different CY$_3$ manifolds from Section 5 from \cite{Castellano:2023jjt}, as well as subtended angles of the two-saxion K\"ahler cones and limiting towers.\label{tab:subtended}}
\end{table}

The procedure could, in principle, be generalized for $n>2$ and performing flops to adjacent K\"ahler cones, but will not be further developed in this paper, and will be left for a future work \cite{TBAkahler}.

\bibliographystyle{JHEP}
\bibliography{ref}

\providecommand{\href}[2]{#2}\begingroup\raggedright\begin{thebibliography}{10}

\bibitem{Palti:2017elp}
E.~Palti, \emph{{The Weak Gravity Conjecture and Scalar Fields}},
  \href{http://dx.doi.org/10.1007/JHEP08(2017)034}{\emph{JHEP} {\bf 08} (2017)
  034}, [\href{http://arxiv.org/abs/1705.04328}{{\tt 1705.04328}}].

\bibitem{Lee:2018spm}
S.-J. Lee, W.~Lerche and T.~Weigand, \emph{{A Stringy Test of the Scalar Weak
  Gravity Conjecture}},
  \href{http://dx.doi.org/10.1016/j.nuclphysb.2018.11.001}{\emph{Nucl. Phys. B}
  {\bf 938} (2019) 321--350}, [\href{http://arxiv.org/abs/1810.05169}{{\tt
  1810.05169}}].

\bibitem{Gonzalo:2019gjp}
E.~Gonzalo and L.~E. Ib\'a\~nez, \emph{{A Strong Scalar Weak Gravity Conjecture
  and Some Implications}},
  \href{http://dx.doi.org/10.1007/JHEP08(2019)118}{\emph{JHEP} {\bf 08} (2019)
  118}, [\href{http://arxiv.org/abs/1903.08878}{{\tt 1903.08878}}].

\bibitem{DallAgata:2020ino}
G.~Dall'Agata and M.~Morittu, \emph{{Covariant formulation of BPS black holes
  and the scalar weak gravity conjecture}},
  \href{http://dx.doi.org/10.1007/JHEP03(2020)192}{\emph{JHEP} {\bf 03} (2020)
  192}, [\href{http://arxiv.org/abs/2001.10542}{{\tt 2001.10542}}].

\bibitem{Andriot:2020lea}
D.~Andriot, N.~Cribiori and D.~Erkinger, \emph{{The web of swampland
  conjectures and the TCC bound}},
  \href{http://dx.doi.org/10.1007/JHEP07(2020)162}{\emph{JHEP} {\bf 07} (2020)
  162}, [\href{http://arxiv.org/abs/2004.00030}{{\tt 2004.00030}}].

\bibitem{Benakli:2020pkm}
K.~Benakli, C.~Branchina and G.~Lafforgue-Marmet, \emph{{Revisiting the scalar
  weak gravity conjecture}},
  \href{http://dx.doi.org/10.1140/epjc/s10052-020-8268-0}{\emph{Eur. Phys. J.
  C} {\bf 80} (2020) 742}, [\href{http://arxiv.org/abs/2004.12476}{{\tt
  2004.12476}}].

\bibitem{Ooguri:2006in}
H.~Ooguri and C.~Vafa, \emph{{On the Geometry of the String Landscape and the
  Swampland}},
  \href{http://dx.doi.org/10.1016/j.nuclphysb.2006.10.033}{\emph{Nucl.Phys.}
  {\bf B766} (2007) 21--33}, [\href{http://arxiv.org/abs/hep-th/0605264}{{\tt
  hep-th/0605264}}].

\bibitem{Vafa:2005ui}
C.~Vafa, \emph{{The String landscape and the swampland}},
  \href{http://arxiv.org/abs/hep-th/0509212}{{\tt hep-th/0509212}}.

\bibitem{Brennan:2017rbf}
T.~D. Brennan, F.~Carta and C.~Vafa, \emph{{The String Landscape, the
  Swampland, and the Missing Corner}},
  \href{http://dx.doi.org/10.22323/1.305.0015}{\emph{PoS} {\bf TASI2017} (2017)
  015}, [\href{http://arxiv.org/abs/1711.00864}{{\tt 1711.00864}}].

\bibitem{Palti:2019pca}
E.~Palti, \emph{{The Swampland: Introduction and Review}},
  \href{http://dx.doi.org/10.1002/prop.201900037}{\emph{Fortsch. Phys.} {\bf
  67} (2019) 1900037}, [\href{http://arxiv.org/abs/1903.06239}{{\tt
  1903.06239}}].

\bibitem{vanBeest:2021lhn}
M.~van Beest, J.~Calder\'on-Infante, D.~Mirfendereski and I.~Valenzuela,
  \emph{{Lectures on the Swampland Program in String Compactifications}},
  \href{http://dx.doi.org/10.1016/j.physrep.2022.09.002}{\emph{Phys. Rept.}
  {\bf 989} (2022) 1--50}, [\href{http://arxiv.org/abs/2102.01111}{{\tt
  2102.01111}}].

\bibitem{Grana:2021zvf}
M.~Gra\~na and A.~Herr\'aez, \emph{{The Swampland Conjectures: A Bridge from
  Quantum Gravity to Particle Physics}},
  \href{http://dx.doi.org/10.3390/universe7080273}{\emph{Universe} {\bf 7}
  (2021) 273}, [\href{http://arxiv.org/abs/2107.00087}{{\tt 2107.00087}}].

\bibitem{Harlow:2022ich}
D.~Harlow, B.~Heidenreich, M.~Reece and T.~Rudelius, \emph{{Weak gravity
  conjecture}},
  \href{http://dx.doi.org/10.1103/RevModPhys.95.035003}{\emph{Rev. Mod. Phys.}
  {\bf 95} (2023) 035003}, [\href{http://arxiv.org/abs/2201.08380}{{\tt
  2201.08380}}].

\bibitem{Agmon:2022thq}
N.~B. Agmon, A.~Bedroya, M.~J. Kang and C.~Vafa, \emph{{Lectures on the string
  landscape and the Swampland}},  \href{http://arxiv.org/abs/2212.06187}{{\tt
  2212.06187}}.

\bibitem{VanRiet:2023pnx}
T.~Van~Riet and G.~Zoccarato, \emph{{Beginners lectures on flux
  compactifications and related Swampland topics}},
  \href{http://dx.doi.org/10.1016/j.physrep.2023.11.003}{\emph{Phys. Rept.}
  {\bf 1049} (2024) 1--51}, [\href{http://arxiv.org/abs/2305.01722}{{\tt
  2305.01722}}].

\bibitem{Baume:2016psm}
F.~Baume and E.~Palti, \emph{{Backreacted Axion Field Ranges in String
  Theory}}, \href{http://dx.doi.org/10.1007/JHEP08(2016)043}{\emph{JHEP} {\bf
  08} (2016) 043}, [\href{http://arxiv.org/abs/1602.06517}{{\tt 1602.06517}}].

\bibitem{Klaewer:2016kiy}
D.~Klaewer and E.~Palti, \emph{{Super-Planckian Spatial Field Variations and
  Quantum Gravity}},
  \href{http://dx.doi.org/10.1007/JHEP01(2017)088}{\emph{JHEP} {\bf 01} (2017)
  088}, [\href{http://arxiv.org/abs/1610.00010}{{\tt 1610.00010}}].

\bibitem{Blumenhagen:2017cxt}
R.~Blumenhagen, I.~Valenzuela and F.~Wolf, \emph{{The Swampland Conjecture and
  F-term Axion Monodromy Inflation}},
  \href{http://dx.doi.org/10.1007/JHEP07(2017)145}{\emph{JHEP} {\bf 07} (2017)
  145}, [\href{http://arxiv.org/abs/1703.05776}{{\tt 1703.05776}}].

\bibitem{Grimm:2018ohb}
T.~W. Grimm, E.~Palti and I.~Valenzuela, \emph{{Infinite Distances in Field
  Space and Massless Towers of States}},
  \href{http://dx.doi.org/10.1007/JHEP08(2018)143}{\emph{JHEP} {\bf 08} (2018)
  143}, [\href{http://arxiv.org/abs/1802.08264}{{\tt 1802.08264}}].

\bibitem{Heidenreich:2018kpg}
B.~Heidenreich, M.~Reece and T.~Rudelius, \emph{{Emergence of Weak Coupling at
  Large Distance in Quantum Gravity}},
  \href{http://dx.doi.org/10.1103/PhysRevLett.121.051601}{\emph{Phys. Rev.
  Lett.} {\bf 121} (2018) 051601}, [\href{http://arxiv.org/abs/1802.08698}{{\tt
  1802.08698}}].

\bibitem{Blumenhagen:2018nts}
R.~Blumenhagen, D.~Kl\"awer, L.~Schlechter and F.~Wolf, \emph{{The Refined
  Swampland Distance Conjecture in Calabi-Yau Moduli Spaces}},
  \href{http://dx.doi.org/10.1007/JHEP06(2018)052}{\emph{JHEP} {\bf 06} (2018)
  052}, [\href{http://arxiv.org/abs/1803.04989}{{\tt 1803.04989}}].

\bibitem{Grimm:2018cpv}
T.~W. Grimm, C.~Li and E.~Palti, \emph{{Infinite Distance Networks in Field
  Space and Charge Orbits}},
  \href{http://dx.doi.org/10.1007/JHEP03(2019)016}{\emph{JHEP} {\bf 03} (2019)
  016}, [\href{http://arxiv.org/abs/1811.02571}{{\tt 1811.02571}}].

\bibitem{Buratti:2018xjt}
G.~Buratti, J.~Calder\'on and A.~M. Uranga, \emph{{Transplanckian axion
  monodromy!?}}, \href{http://dx.doi.org/10.1007/JHEP05(2019)176}{\emph{JHEP}
  {\bf 05} (2019) 176}, [\href{http://arxiv.org/abs/1812.05016}{{\tt
  1812.05016}}].

\bibitem{Corvilain:2018lgw}
P.~Corvilain, T.~W. Grimm and I.~Valenzuela, \emph{{The Swampland Distance
  Conjecture for K\"ahler moduli}},
  \href{http://dx.doi.org/10.1007/JHEP08(2019)075}{\emph{JHEP} {\bf 08} (2019)
  075}, [\href{http://arxiv.org/abs/1812.07548}{{\tt 1812.07548}}].

\bibitem{Joshi:2019nzi}
A.~Joshi and A.~Klemm, \emph{{Swampland Distance Conjecture for One-Parameter
  Calabi-Yau Threefolds}},
  \href{http://dx.doi.org/10.1007/JHEP08(2019)086}{\emph{JHEP} {\bf 08} (2019)
  086}, [\href{http://arxiv.org/abs/1903.00596}{{\tt 1903.00596}}].

\bibitem{Erkinger:2019umg}
D.~Erkinger and J.~Knapp, \emph{{Refined swampland distance conjecture and
  exotic hybrid Calabi-Yaus}},
  \href{http://dx.doi.org/10.1007/JHEP07(2019)029}{\emph{JHEP} {\bf 07} (2019)
  029}, [\href{http://arxiv.org/abs/1905.05225}{{\tt 1905.05225}}].

\bibitem{Marchesano:2019ifh}
F.~Marchesano and M.~Wiesner, \emph{{Instantons and infinite distances}},
  \href{http://dx.doi.org/10.1007/JHEP08(2019)088}{\emph{JHEP} {\bf 08} (2019)
  088}, [\href{http://arxiv.org/abs/1904.04848}{{\tt 1904.04848}}].

\bibitem{Font:2019cxq}
A.~Font, A.~Herr\'aez and L.~E. Ib\'a\~nez, \emph{{The Swampland Distance
  Conjecture and Towers of Tensionless Branes}},
  \href{http://dx.doi.org/10.1007/JHEP08(2019)044}{\emph{JHEP} {\bf 08} (2019)
  044}, [\href{http://arxiv.org/abs/1904.05379}{{\tt 1904.05379}}].

\bibitem{Gendler:2020dfp}
N.~Gendler and I.~Valenzuela, \emph{{Merging the weak gravity and distance
  conjectures using BPS extremal black holes}},
  \href{http://dx.doi.org/10.1007/JHEP01(2021)176}{\emph{JHEP} {\bf 01} (2021)
  176}, [\href{http://arxiv.org/abs/2004.10768}{{\tt 2004.10768}}].

\bibitem{Lanza:2020qmt}
S.~Lanza, F.~Marchesano, L.~Martucci and I.~Valenzuela, \emph{{Swampland
  Conjectures for Strings and Membranes}},
  \href{http://dx.doi.org/10.1007/JHEP02(2021)006}{\emph{JHEP} {\bf 02} (2021)
  006}, [\href{http://arxiv.org/abs/2006.15154}{{\tt 2006.15154}}].

\bibitem{Klaewer:2020lfg}
D.~Klaewer, S.-J. Lee, T.~Weigand and M.~Wiesner, \emph{{Quantum Corrections in
  4d N=1 Infinite Distance Limits and the Weak Gravity Conjecture}},
  \href{http://arxiv.org/abs/2011.00024}{{\tt 2011.00024}}.

\bibitem{Rudelius:2023mjy}
T.~Rudelius, \emph{{Revisiting the Refined Distance Conjecture}},
  \href{http://arxiv.org/abs/2303.12103}{{\tt 2303.12103}}.

\bibitem{Ooguri:2024ofs}
H.~Ooguri and Y.~Wang, \emph{{Universal Bounds on CFT Distance Conjecture}},
  \href{http://arxiv.org/abs/2405.00674}{{\tt 2405.00674}}.

\bibitem{Aoufia:2024awo}
C.~Aoufia, I.~Basile and G.~Leone, \emph{{Species scale, worldsheet CFTs and
  emergent geometry}},  \href{http://arxiv.org/abs/2405.03683}{{\tt
  2405.03683}}.

\bibitem{Etheredge:2022opl}
M.~Etheredge, B.~Heidenreich, S.~Kaya, Y.~Qiu and T.~Rudelius,
  \emph{{Sharpening the Distance Conjecture in Diverse Dimensions}},
  \href{http://arxiv.org/abs/2206.04063}{{\tt 2206.04063}}.

\bibitem{Etheredge:2023odp}
M.~Etheredge, B.~Heidenreich, J.~McNamara, T.~Rudelius, I.~Ruiz and
  I.~Valenzuela, \emph{{Running Decompactification, Sliding Towers, and the
  Distance Conjecture}},  \href{http://arxiv.org/abs/2306.16440}{{\tt
  2306.16440}}.

\bibitem{Calderon-Infante:2023ler}
J.~Calder\'on-Infante, A.~Castellano, A.~Herr\'aez and L.~E. Ib\'a\~nez,
  \emph{{Entropy Bounds and the Species Scale Distance Conjecture}},
  \href{http://arxiv.org/abs/2306.16450}{{\tt 2306.16450}}.

\bibitem{Etheredge:2023usk}
M.~Etheredge, \emph{{Dense Geodesics, Tower Alignment, and the Sharpened
  Distance Conjecture}},  \href{http://arxiv.org/abs/2308.01331}{{\tt
  2308.01331}}.

\bibitem{Lee:2019xtm}
S.-J. Lee, W.~Lerche and T.~Weigand, \emph{{Emergent Strings, Duality and Weak
  Coupling Limits for Two-Form Fields}},
  \href{http://arxiv.org/abs/1904.06344}{{\tt 1904.06344}}.

\bibitem{Lee:2019wij}
S.-J. Lee, W.~Lerche and T.~Weigand, \emph{{Emergent Strings from Infinite
  Distance Limits}},  \href{http://arxiv.org/abs/1910.01135}{{\tt 1910.01135}}.

\bibitem{Baume:2020dqd}
F.~Baume and J.~Calder\'on~Infante, \emph{{Tackling the SDC in AdS with CFTs}},
  \href{http://dx.doi.org/10.1007/JHEP08(2021)057}{\emph{JHEP} {\bf 08} (2021)
  057}, [\href{http://arxiv.org/abs/2011.03583}{{\tt 2011.03583}}].

\bibitem{Perlmutter:2020buo}
E.~Perlmutter, L.~Rastelli, C.~Vafa and I.~Valenzuela, \emph{{A CFT distance
  conjecture}}, \href{http://dx.doi.org/10.1007/JHEP10(2021)070}{\emph{JHEP}
  {\bf 10} (2021) 070}, [\href{http://arxiv.org/abs/2011.10040}{{\tt
  2011.10040}}].

\bibitem{CalderonIV}
J.~Calder\'on-Infante and I.~Valenzuela, \emph{to appear}, .

\bibitem{Lee:2018urn}
S.-J. Lee, W.~Lerche and T.~Weigand, \emph{{Tensionless Strings and the Weak
  Gravity Conjecture}},
  \href{http://dx.doi.org/10.1007/JHEP10(2018)164}{\emph{JHEP} {\bf 10} (2018)
  164}, [\href{http://arxiv.org/abs/1808.05958}{{\tt 1808.05958}}].

\bibitem{Baume:2019sry}
F.~Baume, F.~Marchesano and M.~Wiesner, \emph{{Instanton Corrections and
  Emergent Strings}},
  \href{http://dx.doi.org/10.1007/JHEP04(2020)174}{\emph{JHEP} {\bf 04} (2020)
  174}, [\href{http://arxiv.org/abs/1912.02218}{{\tt 1912.02218}}].

\bibitem{Xu:2020nlh}
F.~Xu, \emph{{On TCS G$_{2}$ manifolds and 4D emergent strings}},
  \href{http://dx.doi.org/10.1007/JHEP10(2020)045}{\emph{JHEP} {\bf 10} (2020)
  045}, [\href{http://arxiv.org/abs/2006.02350}{{\tt 2006.02350}}].

\bibitem{Lanza:2021udy}
S.~Lanza, F.~Marchesano, L.~Martucci and I.~Valenzuela, \emph{{The EFT stringy
  viewpoint on large distances}},
  \href{http://dx.doi.org/10.1007/JHEP09(2021)197}{\emph{JHEP} {\bf 09} (2021)
  197}, [\href{http://arxiv.org/abs/2104.05726}{{\tt 2104.05726}}].

\bibitem{Castellano:2023jjt}
A.~Castellano, I.~Ruiz and I.~Valenzuela, \emph{{Stringy Evidence for a
  Universal Pattern at Infinite Distance}},
  \href{http://arxiv.org/abs/2311.01536}{{\tt 2311.01536}}.

\bibitem{Rudelius:2023odg}
T.~Rudelius, \emph{{Gopakumar-Vafa Invariants and the Emergent String
  Conjecture}},  \href{http://arxiv.org/abs/2309.10024}{{\tt 2309.10024}}.

\bibitem{Calderon-Infante:2020dhm}
J.~Calder\'on-Infante, A.~M. Uranga and I.~Valenzuela, \emph{{The Convex Hull
  Swampland Distance Conjecture and Bounds on Non-geodesics}},
  \href{http://dx.doi.org/10.1007/JHEP03(2021)299}{\emph{JHEP} {\bf 03} (2021)
  299}, [\href{http://arxiv.org/abs/2012.00034}{{\tt 2012.00034}}].

\bibitem{vandeHeisteeg:2023ubh}
D.~van~de Heisteeg, C.~Vafa and M.~Wiesner, \emph{{Bounds on Species Scale and
  the Distance Conjecture}},  \href{http://arxiv.org/abs/2303.13580}{{\tt
  2303.13580}}.

\bibitem{vandeHeisteeg:2023uxj}
D.~van~de Heisteeg, C.~Vafa, M.~Wiesner and D.~H. Wu, \emph{{Bounds on Field
  Range for Slowly Varying Positive Potentials}},
  \href{http://arxiv.org/abs/2305.07701}{{\tt 2305.07701}}.

\bibitem{Castellano:2023stg}
A.~Castellano, I.~Ruiz and I.~Valenzuela, \emph{{A Universal Pattern in Quantum
  Gravity at Infinite Distance}},  \href{http://arxiv.org/abs/2311.01501}{{\tt
  2311.01501}}.

\bibitem{Rudelius:2023spc}
T.~Rudelius, \emph{{Persistence of the Pattern in the Interior of 5d Moduli
  Spaces}},  \href{http://arxiv.org/abs/2312.00120}{{\tt 2312.00120}}.

\bibitem{Maldacena:2000mw}
J.~M. Maldacena and C.~Nunez, \emph{{Supergravity description of field theories
  on curved manifolds and a no go theorem}},
  \href{http://dx.doi.org/10.1142/S0217751X01003937}{\emph{Int. J. Mod. Phys.
  A} {\bf 16} (2001) 822--855},
  [\href{http://arxiv.org/abs/hep-th/0007018}{{\tt hep-th/0007018}}].

\bibitem{Castellano:2022bvr}
A.~Castellano, A.~Herr\'aez and L.~E. Ib\'a\~nez, \emph{{The emergence proposal
  in quantum gravity and the species scale}},
  \href{http://dx.doi.org/10.1007/JHEP06(2023)047}{\emph{JHEP} {\bf 06} (2023)
  047}, [\href{http://arxiv.org/abs/2212.03908}{{\tt 2212.03908}}].

\bibitem{Castellano:2021mmx}
A.~Castellano, A.~Herr\'aez and L.~E. Ib\'a\~nez, \emph{{IR/UV mixing, towers
  of species and swampland conjectures}},
  \href{http://dx.doi.org/10.1007/JHEP08(2022)217}{\emph{JHEP} {\bf 08} (2022)
  217}, [\href{http://arxiv.org/abs/2112.10796}{{\tt 2112.10796}}].

\bibitem{Alvarez-Garcia:2023qqj}
R.~\'Alvarez-Garc\'\i{}a, S.-J. Lee and T.~Weigand, \emph{{Non-minimal Elliptic
  Threefolds at Infinite Distance II: Asymptotic Physics}},
  \href{http://arxiv.org/abs/2312.11611}{{\tt 2312.11611}}.

\bibitem{Etheredge:2023zjk}
M.~Etheredge and B.~Heidenreich, \emph{{Geodesic Gradient Flows in Moduli
  Space}},  \href{http://arxiv.org/abs/2311.18693}{{\tt 2311.18693}}.

\bibitem{Herraez:2020tih}
A.~Herraez, \emph{{A Note on Membrane Interactions and the Scalar potential}},
  \href{http://dx.doi.org/10.1007/JHEP10(2020)009}{\emph{JHEP} {\bf 10} (2020)
  009}, [\href{http://arxiv.org/abs/2006.01160}{{\tt 2006.01160}}].

\bibitem{Alvarez-Garcia:2021pxo}
R.~\'Alvarez-Garc\'\i{}a, D.~Kl\"awer and T.~Weigand, \emph{{Membrane limits in
  quantum gravity}},
  \href{http://dx.doi.org/10.1103/PhysRevD.105.066024}{\emph{Phys. Rev. D} {\bf
  105} (2022) 066024}, [\href{http://arxiv.org/abs/2112.09136}{{\tt
  2112.09136}}].

\bibitem{Etheredge:BraneDC}
M.~Etheredge, B.~Heidenreich and T.~Rudelius, \emph{A distance conjecture for
  branes (to appear)}, .

\bibitem{Etheredge:BraneTaxonomy}
M.~Etheredge, \emph{Taxonomy of branes in infinite distance limits (to
  appear)}, .

\bibitem{Bedroya:2023xue}
A.~Bedroya and Y.~Hamada, \emph{{Dualities from Swampland principles}},
  \href{http://arxiv.org/abs/2303.14203}{{\tt 2303.14203}}.

\bibitem{vandeHeisteeg:2024lsa}
D.~van~de Heisteeg, \emph{{Charting the Complex Structure Landscape of
  F-theory}},  \href{http://arxiv.org/abs/2404.03456}{{\tt 2404.03456}}.

\bibitem{Brcker1985RepresentationsOC}
T.~Br{\"o}cker and T.~tom Dieck, \emph{Representations of compact lie groups},
  1985.

\bibitem{Moore:1998zu}
G.~W. Moore, \emph{{Attractors and arithmetic}},
  \href{http://arxiv.org/abs/hep-th/9807056}{{\tt hep-th/9807056}}.

\bibitem{xiao2016positivity}
J.~Xiao, \emph{Positivity in K{\"a}hler geometry}.
\newblock PhD thesis, Universit{\'e} de Fudan (Shanghai, Chine), 2016.

\bibitem{lehmann2016convexity}
B.~Lehmann and J.~Xiao, \emph{Convexity and zariski decomposition structure},
  {\emph{Geometric and Functional Analysis} {\bf 26} (2016) 1135--1189}.

\bibitem{Marchesano:2023thx}
F.~Marchesano, L.~Melotti and L.~Paoloni, \emph{{On the moduli space curvature
  at infinity}},  \href{http://arxiv.org/abs/2311.07979}{{\tt 2311.07979}}.

\bibitem{TBAkahler}
N.~Gendler, L.~Melotti, I.~Ruiz and I.~Valenzuela, \emph{to appear}, .

\end{thebibliography}\endgroup
\end{document}